\documentclass[fleqn,usenatbib]{mnras}

\usepackage{newtxtext,newtxmath}

\usepackage[T1]{fontenc}
\usepackage{ae,aecompl}
\usepackage{comment}
 \usepackage{booktabs}
 \usepackage{multi row}
 
\usepackage{epsfig}
\usepackage{amssymb}
\usepackage{natbib}

\usepackage{graphicx}	
\usepackage{amsmath}	
\usepackage{amssymb}	

\title[Properties bulges-discs in CANDELS]{The structural properties of classical bulges and discs from $z\sim2$}

\author[Dimauro et al.]{
Paola Dimauro$^{1}$, \thanks{E-mail: paola.dimauro@obspm.fr}
Marc Huertas-Company$^{2,1,3}$, Emanuele Daddi$^{4}$,
\newauthor
Pablo G. P\'erez-Gonz\'alez$^{5,20}$, Mariangela Bernardi$^{24}$, Fernando Caro$^{1}$,
\newauthor
Andrea Cattaneo$^{8}$, Boris H\"au\ss ler$^{10}$, Ulrike Kuchner$^{13}$,Francesco Shankar$^{17}$
\newauthor
Guillermo Barro$^{6}$, Fernando Buitrago$^{7}$, Sandra M. Faber$^{9}$, Dale D. Kocevski$^{11}$
\newauthor
Anton M. Koekemoer$^{12}$, David C. Koo$^{23}$, Simona Mei$^{21,1,22}$, Reynier Peletier$^{14}$,
 \newauthor
 Joel Primack$^{9}$,  Aldo Rodriguez-Puebla$^{15}$, Mara Salvato$^{16}$, Diego Tuccillo$^{18,19}$
\\\\
$^{1}$ Sorbonne Universit\'e, Observatoire de Paris, Universit\'e PSL, CNRS, LERMA, F-75014, Paris, France\\
$^{2}$ Instituto de Astrof\'sica de Canarias, c/ Via Láctea s/n, E-38205 La Laguna, Tenerife, Spain\\
$^{3}$ Departamento de Astrof\'sica, Universidad de La Laguna, E-38205 La Laguna, Tenerife, Spain\\
$^{4}$ CEA Saclay, Laboratoire AIM-CNRS-Universit\'e Paris Diderot, Irfu/SAp, Orme des Merisiers, 91191, Gif-sur-Yvette, France\\
$^{5}$ Centro de Astrobiolog\'ia (CAB, INTA-CSIC), Carretera de Ajalvir km 4, E-28850 Torrej\'on de Ardoz, Madrid, Spain\\
$^{6}$ University of the Pacific, Stockton, CA 90340 USA\\
$^{7}$ Instituto de Astrofisica e Ciencias do Espaco, Universidade de Lisboa, OAL, Tapada da Ajuda, P-1349-018 Lisbon, Portugal \\
$^{8}$ GEPI, Observatoire de Paris, 61 Avenue de l'Observatoire, 75014, Paris, France\\
$^{9}$ University of California, Santa Cruz, CA 95064, USA\\
$^{10}$ European Southern Observatory, Alonso de Cordova 3107, Vitacura, Casilla 19001, Santiago, Chile\\
$^{11}$ Department of Physics and Astronomy, Colby College, Waterville, ME 04961, USA \\
$^{12}$ Space Telescope Science Institute, 3700 San Martin Drive, Baltimore, MD 21218, USA \\
$^{13}$ School of Physics \& Astronomy, The University of Nottingham, University Park, Nottingham NG7 2RD, UK\\
$^{14}$ Kapteyn Astronomical Institute, University of Groningen, Postbus 800, NL-9700 Au Groningen, The Netherlands\\
$^{15}$ Instituto de Astronom\'ia, Universidad Nacional Aut\'onoma de M\'exico, A.P. 70-264, 04510 M\'exico, D.F., Mexico\\
$^{16}$ Max Planck Institut fur Plasma Physik and Excellence Cluster, 85748 Garching, Germany \\
$^{17}$ Department of Physics and Astronomy, University of Southampton, Highfield, SO17 1BJ, UK\\
$^{18}$ MINES Paristech, PSL Research University, Centre for Mathematical Morphology, Fontainebleau, France\\
$^{19}$ Instituto de Fisica de Cantabria (CSIC - Universidad de Cantabria), Av. de los Castros s/n, E-39005 Santander, Spain\\
$^{20}$ Departamento de Astrof\'isica, Facultad de CC. F\'isicas, Universidad Complutense de Madrid, E-28040 Madrid, Spain\\
$^{21}$ Universit\'{e} Paris Denis Diderot, Universit\'e Paris Sorbonne Cit\'e, 75205 Paris Cedex 13, France\\
$^{22}$ Jet Propulsion Laboratory, Cahill Center for Astronomy \& Astrophysics, California Institute of Technology, 4800 Oak Grove California, USA\\
$^{23}$ Department of Astronomy \& Astrophysics, University of California, Santa Cruz, 1156 High Street, Santa Cruz, CA 95064, USA\\
$^{24}$ Department of Physics and Astronomy, University of Pennsylvania, Philadelphia, PA 19104, USA
}

\date{Accepted XXX. Received YYY; in original form ZZZ}

\pubyear{2019}

\begin{document}
\label{firstpage}
\pagerange{\pageref{firstpage}--\pageref{lastpage}}
\maketitle

\def \aj {AJ}
\def \mnras {MNRAS}
\def \pasp {PASP}
\def \apj {ApJ}
\def \apjs {ApJS}
\def \apjl {ApJL}
\def \aap {A\&A}
\def \nat {Nature}
\def \araa {ARAA}
\def \iaucirc {IAUC}
\def \aaps {A\&A Suppl.}
\def \qjras {QJRAS}
\def \na {New Astronomy}
\def \aapr {A\&ARv}
\def\lesssim{\mathrel{\hbox{\rlap{\hbox{\lower4pt\hbox{$\sim$}}}\hbox{$<$}}}}
\def\gtrsim{\mathrel{\hbox{\rlap{\hbox{\lower4pt\hbox{$\sim$}}}\hbox{$>$}}}}

\newcommand{\Sersic}{S\'{e}rsic\:}

\begin{abstract}
We study the rest-frame optical mass-size relation of bulges and discs from z$\sim$2 to z$\sim$0 for a complete sample of massive galaxies in the CANDELS fields using 2 component \Sersic models (\citealp{Dimauro2018}).
Discs and star forming galaxies follow similar mass-size relations. The mass-size relation of bulges is less steep than the one of quiescent galaxies (best fit slope of $\sim$ 0.7 for quiescent galaxies against $\sim$ 0.4 for bulges). 
We find little dependence of the structural properties of massive bulges and discs with the global morphology of galaxies (disc vs. bulge dominated) and the star formation activity (star-forming vs. quiescent). This result suggests similar bulge formation mechanisms for most massive galaxies and also that the formation of the bulge component does not significantly affect the disc structure. 
Our findings pose a challenge to models envisioning multiple channels for massive bulge growth, such as disc instabilities and mergers. 
 \end{abstract}




\section{Introduction}

The life of a galaxy is a tight balance between processes that trigger the formation of stars and others that tend to reduce or halt its star-formation activity (e.g. \citealp{Carollo2013}). Understanding how this balance works and therefore how mass is assembled, is a fundamental question for our understanding of galaxy formation and evolution. 

One of the main results of the last decade has been the discovery (since $z\sim3$ at least) of a tight correlation between the star-formation rate and the stellar masses of galaxies (Main Sequence of star-formation, MS, e.g \citealp{Elbaz2007,Brinchmann2004,Whitaker2012}). This empirical correlation suggests that the star formation history of a galaxy is preferentially driven by regular mass dependent processes, instead of stochastic events such as mergers. 
Within this picture, an additional key observational fact is the presence of a population of galaxies which do not form stars (quenched galaxies) and therefore are somehow decoupled from the underlying cosmic structure. This population already exists at $z\sim3$ (\citealp{Huertas2016}, \citealp{Whitaker2012}), or even earlier (\citealp{Glazebrook2017}) and dominates the high-mass end of the stellar mass function at $z<1$ (e.g. \citealp{Ilbert2013}, \citealp{Muzzin2013}). The coexistence of the two populations creates a bimodal distribution in the $SFR-M_*$ plane. Galaxies following the main sequence (MS) of star-forming galaxies within a scatter of $\sim0.5$ dex, are typically found at lower masses, while very low specific star-formation rates (sSFR) dominate the high mass end. This behavior is often interpreted as galaxies growing along the MS at a rather constant sSFR until they reach a critical stellar mass (or halo mass) threshold, above which they likely quench (e.g. \citealp{Peng2010}, although see \citealp{Abramson2015} for a different interpretation). 
Several groups have measured a bending of the SF-MS at high stellar masses. This may be due to galaxies that are in the process of leaving the MS, i.e. quenching. The exact physical processes that cause quenching are still a matter of discussions. However, some of these galaxies might be in the inverse process. After a passive phase, they are starting a new star formation activity (called rejuvenation phase), thanks to the accretion of the surrounding gas (e.g. \citealp{Fang2013,Mancini2015}). 

The bimodality also exists in the stellar mass-size plane. Independently of classification (e.g. star-forming/passive early/late type), the two main classes of galaxies indeed occupy different regions in the stellar mass-size plane, in the local universe (\citealp{Shen2003}), as well as at high redshift (\citealp{vanderWel2014,Whitaker2012,Whitaker2015}). Early-type galaxies follow a steeper relation than late-type galaxies, i.e. early-type galaxies appear more compact than late-type ones at fixed stellar mass (\citealp{Daddi2005,Trujillo2006}). In general, all galaxies appear progressively smaller at earlier epochs, especially early-type spheroidals (e.g., \citealp{Trujillo2007}).

There is also a well-known correlation between galaxy morphology and star-formation rate, with early-type galaxies tending to be redder and more passive, and late-type galaxies bluer and starforming (e.g \citealp{Baldry2004,Mei2009,Huertas2015, Huertas2016}).
The quenched population is mainly composed of bulge-dominated galaxies, while star-forming galaxies tend to have lower bulge fractions. The origin of this dichotomy is still unknown. Despite this consensus, there is still an open debate in the literature on how the growth of a central stellar over density (bulge) relates to quenching.
 A variety of mechanisms have been proposed to link the growth of a bulge with the quenching of the star-formation rate in the galaxy. The classical picture is that bulges grow mainly through mergers (e.g. \citealp{Toomre1977,Hammer2005,Hopkins2010}) that provoke a loss of angular momentum and a burst of star formation. If the resulting halo is massive enough the infalling gas is shock heated and prevented from a further accreation through the center, hence the galaxy will eventually quench (\citealp{Dekel2006}, \citealp{Cattaneo2006,Cattaneo2009}). More recent works have also suggested that bulges can grow in-situ within the disc. High-redshift discs are indeed known to be turbulent and gas rich (e.g.~\citealp{Genzel2015, Wisnioski2015}). Violent disc instabilities can generate starforming clumps in the disc that can survive and migrate towards the centre to successively build up a bulge (e.g. \citealp{Bournaud2014,Bournaud2016}). The bulge itself further stabilizes the gas within the disc avoiding new clump formation. Consequently the galaxy quenches (e.g. \citealp{Martig2009}). This is also observationally supported by the increasing abundance of clumpy galaxies observed at high-redshift (e.g.~\citealp{Huertas2016,Shibuya2016,Guo2015}). An alternative or complementary mechanism is the re-growth of a disc component after gas-rich mergers or through accretion of gas from the cosmic web (\citealp{Mancini2015, Mancini2019,Fang2013}).

Recent high-resolution cosmological simulations have shown that main-sequence galaxies can experience multiple gas-compaction events that create an overdensity of stars and a loss of angular momentum (e.g. \citealp{Dekel2014,Zolotov2015,Tacchella2016}). These compaction events are expected to be frequent in main-sequence galaxies (e.g. \citealp{Tacchella2017}) and can lead to quenching if the halo reaches a critical mass that prevents accretion from the outer regions. This is supported by recent observational works that report a tight correlation between stellar mass density within 1 Kpc radii and the star formation activity (e.g. \citeauthor{Barro2015} \citeyear{Barro2015}, \citeyear{Barro2016}, \citealp{Whitaker2017}). Additionally, a population of dense star-forming galaxies (\emph{blue nuggets}) with similar structural properties and abundances comparable to their passive counterparts have been observed \citep{Barro2013,Huertas2018}. Molecular gas observations of some of these dense star-forming galaxies, carried out with ALMA, also confirm a large reservoir of gas in the central regions \citep{Barro2016,Popping2017}. These dense star-forming galaxies are possible candidates to be in the compaction phase that precedes the quenching. Signatures of an inside-out quenching have also been detected by inspecting the SFR gradients of massive galaxies at $z\sim2$  \citep{Tacchella2015,Tacchella2017}.
Moreover, using the bulge-disc catalog from \cite{Simard2011}, \cite{Morselli2017} show the existence of a link between the position of a galaxy on the MS and the presence of a central component. 

These views are challenged by several works that do not find strong signatures of a morphological transformation in the quenching process (e.g \citealp{Carollo2014}). Indeed, \cite{Abramson2016} did not find any evidence of an excess of dense galaxies among the passive population, suggesting that the observed relation between mass density and star formation might be essentially driven by observational biases (i.e. progenitor bias) instead of being a consequence of physical processes. \cite{Abramson2016} suggest that galaxies which are quenched at a given epoch look denser because their progenitors were denser.  A simple toy model presented by \cite{Lilly2016} is able to reproduce the observed trends, namely the distribution of passive and star-forming galaxies in the mass-size plane (see also~\citealp{Stringer2014}). 

In order to get a deeper insight into the different bulge growth mechanisms and their relation to quenching, we move from integrated properties to resolved quantities within galaxies. 
In \cite{Dimauro2018} (hereafter DM18), we released a catalog of bulge-disc decompositions of $\sim 17.600$ galaxies at $z<2$ in the CANDELS fields. Taking advantage of the large wavelength coverage at high spatial resolution in this area of the sky, we fit the light profiles of galaxies with 2 components in 4-7 filters spanning the spectral range $400-1600$ nm using the \textsc{GalfitM} and \textsc{Galapagos-2} codes from the \textsc{MEGAMORPH} project \citep{haeussler2013,Barden2012,Vika2014}. Additionally, we use  a novel method based on unsupervised feature learning (deep-learning) to select the best model to be fitted to the surface brightness profile of a galaxy (namely one or two components). This allows us to derive stellar masses as well as rest-frame colors and star-formation rates separately for the bulge and disc components. 

This paper is the first of a series using this catalog to dig into the properties of embedded bulges and discs to provide new clues on the link between morphological transformations and quenching. We focus on the mass-size relation as a mean to  link bulge growth and quenching. With few exceptions (e.g., \citealp{Kuchner2017}), in the past this scaling relation has generally been studied using integrated light. Using detailed bulge and disc decomposition enables us to add additional constraints on the processes of bulge formation and disc growth. The paper proceeds as follows. The data are presented in section~\ref{sec:data}.  We first analyze the scaling relations of passive and star-forming galaxies in section~\ref{sec:SF_Q}. We then focus on the analysis of the mass-size relation of the bulge and disc components separately in section~\ref{sec:results}. Sections~\ref{sec:morph} and \ref{sec:SF} show the structural properties of discs and bulges hosted by galaxies with different morphologies and star-formation activities respectively. The implications for the quenching and bulge formation mechanisms are discussed in section~\ref{sec:discussion}. All magnitudes are measured in the AB system. We adopt a Planck 2013 cosmology~\citep{Planck2014}.

\section{Data}
\label{sec:data}
We use the bulge-disc catalog presented in DM18 in the CANDELS fields \citep{Grogin2011,Koekemoer2011}. We refer the reader to that paper for more details on the methods. Briefly, the catalog contains multi-wavelength fits (7 bands for the GOODS-N and GOODS-S fields, and 4 bands for UDS, COSMOS and AEGIS) to the surface brightness profiles of $\sim17.600$ galaxies using the code \textsc{galfitM}. Galaxies are H band (F160W) selected down to a magnitude of 23 from the official CANDELS catalogs (\citealp{Galametz2013} for UDS, \citealp{Guo2013} for GOODS-S, Barro et al. 2017  for GOODS-N and \citealp{Stefanon2017} for COSMOS and AEGIS). Each galaxy is fitted with a 1-component \Sersic model and with a 2-component \Sersic+Exponential model. Several setups are provided in the catalog in which different wavelength dependences of the model are tested. The main quantities used in this paper are taken from the main setup (setups 1 and 4 from Table 1 in DM18) where sizes are modelled with a second order polynomial function over the wavelength. 
Other setups are used to correct ambiguous cases (according with the deep-learning classification: Setups 4 and 6) and also to estimate random uncertainties on the structural parameters derived from the fits following a similar approach to \cite{vanderWel2014}. In DM18 we show that the typical random uncertainties on the sizes of different components (the main quantity used in this work) are below $20\%$ and that no bias are introduced in the measurements. 


The catalog also contains a set of four probabilities for each galaxy, estimated with a convolutional neural network, that measures how well a given model describes the F160W surface brightness profile of a galaxy. This allows us to then select the optimal model to be used for every galaxy: the probabilities $P_{B}$ and $P_{D}$ measure how well a 1-component \Sersic model with high ($>2.5$) and low ($<1.5$) \Sersic index respectively matches the light distribution. In practice this means that for objects with large $P_{B}$ or $P_{D}$ a 1-component \Sersic model should be preferred. The two additional probabilities $P_{BD}$ and $P_{PB}$ measure how well a 2-component \Sersic+Exponential model, with a high and low \Sersic index bulge respectively, matches the surface brightness distribution. High values indicate that a 2-component model should be used. 

In the following analysis, we use the structural properties of galaxies (and their internal components) measured in the optical rest-frame band. Thanks to the multi-wavelength fitting method used to build the catalog, i.e. the Chebyshev polynomials functions, we interpolate values at the 5000 \AA \, rest frame. This is the best choice considering the redshift range and the wavelength coverage probed by our data.

Stellar masses and stellar populations properties of bulges and discs are derived through the fitting of Spectral Energy Distribution using the FAST code~\citep{Kriek2009}. As detailed in DM18, we use BC03 stellar population models and a \cite{Chabrier2003} initial mass function. This allows us to estimate a stellar mass bulge-to-total ratio that will be used in the forthcoming analysis. The typical error on the stellar mass bulge-over-total ratio ($B/T$) is below 0.2. This error does not include uncertainties due to stellar population models. 
 

Additionally, we use photometric redshifts derived through SED fitting from the CANDELS collaboration (spectroscopic when available) by combining different available codes. More details can be found in \cite{Dahlen2013}. Integrated rest-frame magnitudes (U,V,J) are also derived as part of the SED fitting procedure and are used to divide galaxies between star-forming and quiescent systems using the widely used cuts in the UVJ plane (see \citealp{Whitaker2012}).

The stellar mass completeness of the sample is computed with two different methods as explained in DM18. We report in table~\ref{tbl:comp} the values at different redshifts. We estimate a stellar mass completeness limit of $\sim 10^{10.7}$ solar masses at $z\sim2$.

\begin{table}
\centering
\begin{tabular}{|c |c |c| c|}
\hline
z        &All&  $Q $& $SF$ \\
\hline
\hline
0-0.5   &   9.0    &  9.16 &  8.98  \\
\hline
0.5-1.0&  9.75   &  9.91 &  9.79 \\
\hline
1.0-1.4 &  10.3   & 10.38 & 10.28 \\
\hline
1.4-2.0 &  10.7  & 10.72 & 10.69\\
\hline
\end{tabular}
\caption{Stellar mass completeness limits of the sample used in this work. We show the values for all galaxies, quiescent (Q) and star-forming (SF). The table is taken from DM18 but reproduced here for clarity.}
\label{tbl:comp}
\end{table}

\section{Mass-size relation of passive and star-forming galaxies}
\label{sec:SF_Q}
We first start by analyzing the mass-size relation of star-forming and quenched galaxies. This relation has been extensively studied in the literature, thus we will use it as a benchmark for our analysis in this paper. We use the half-light semi-major axis (obtained from \Sersic model, done to reconstruct the surface brightness distribution of the entire galaxy), in the optical 5000 \AA \ rest frame band, as main size indicator (Setup 1 from DM18, Table 1).  

\begin{figure*}
\begin{center}
$\begin{array}{c c}
	\includegraphics[width=0.44\textwidth]{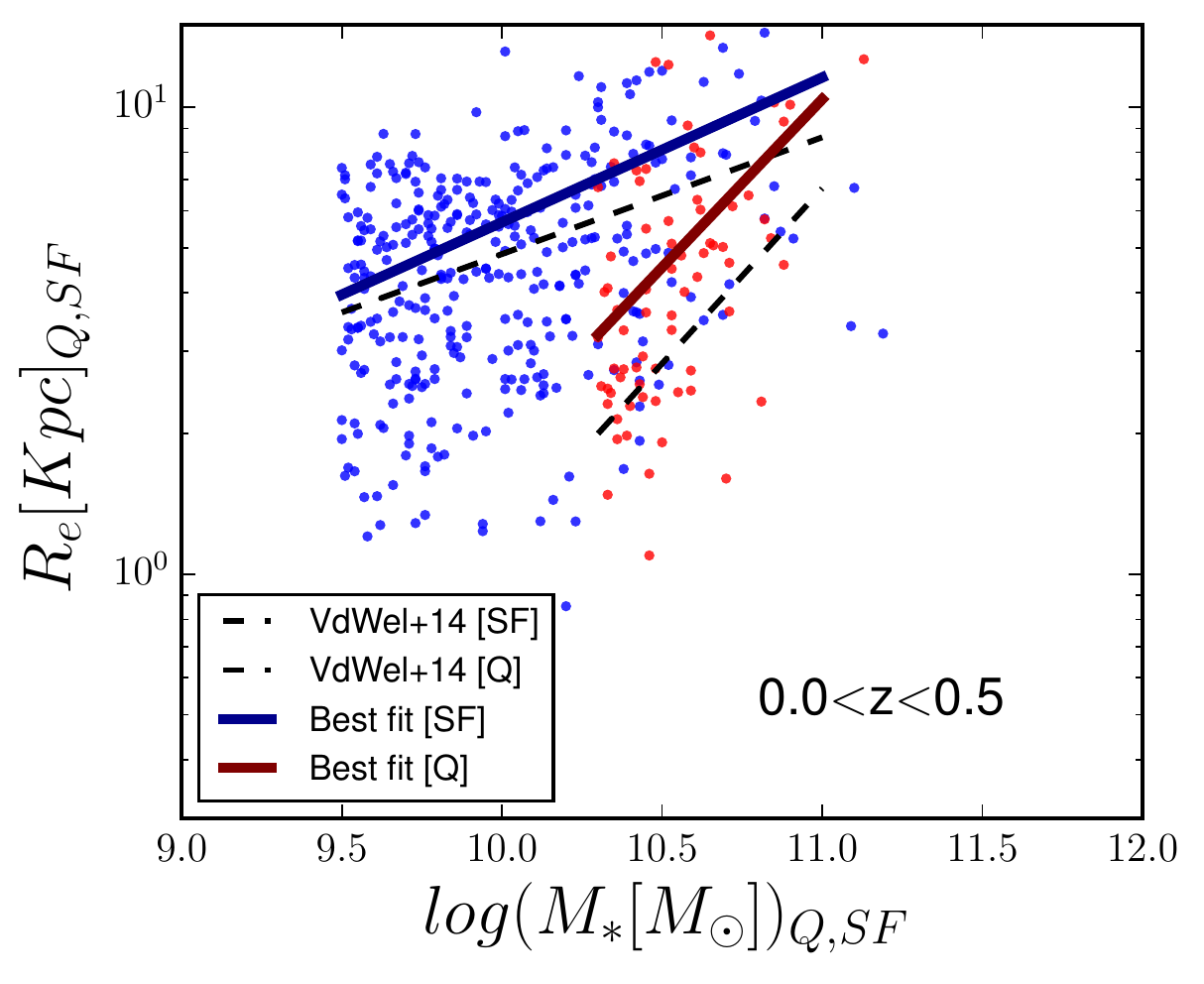} & 
	\includegraphics[width=0.44\textwidth]{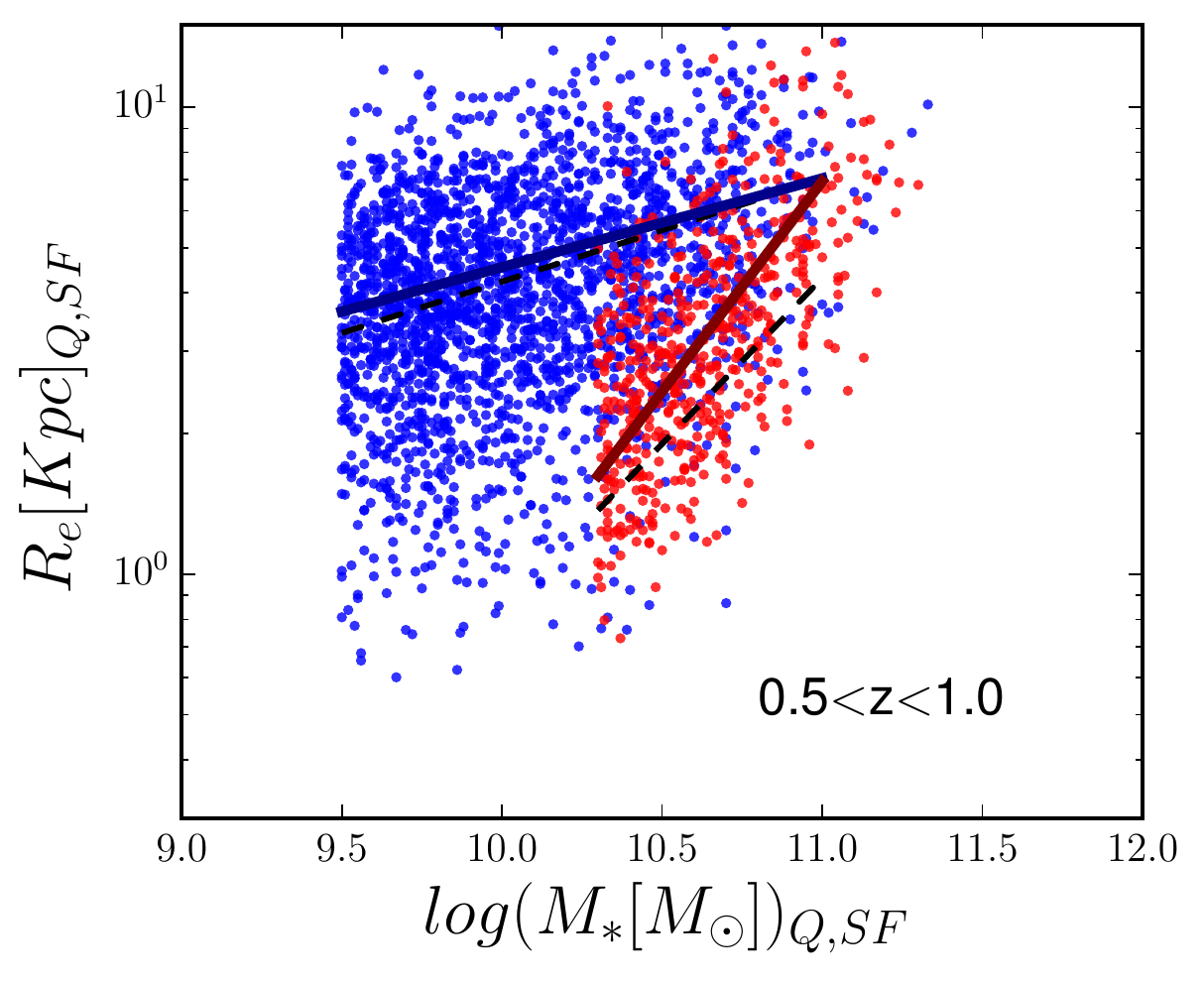} \\
	\includegraphics[width=0.44\textwidth]{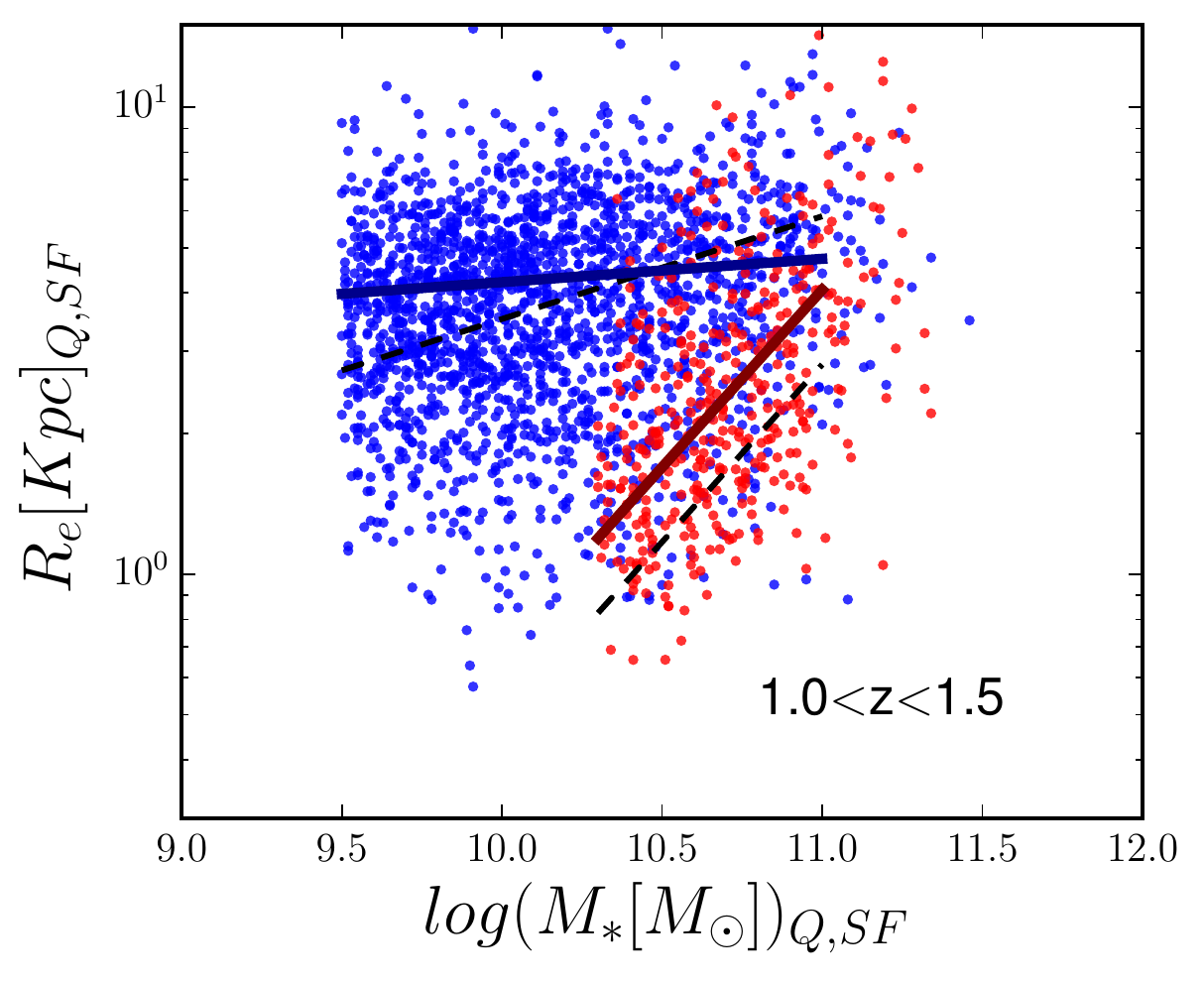} & 
	\includegraphics[width=0.44\textwidth]{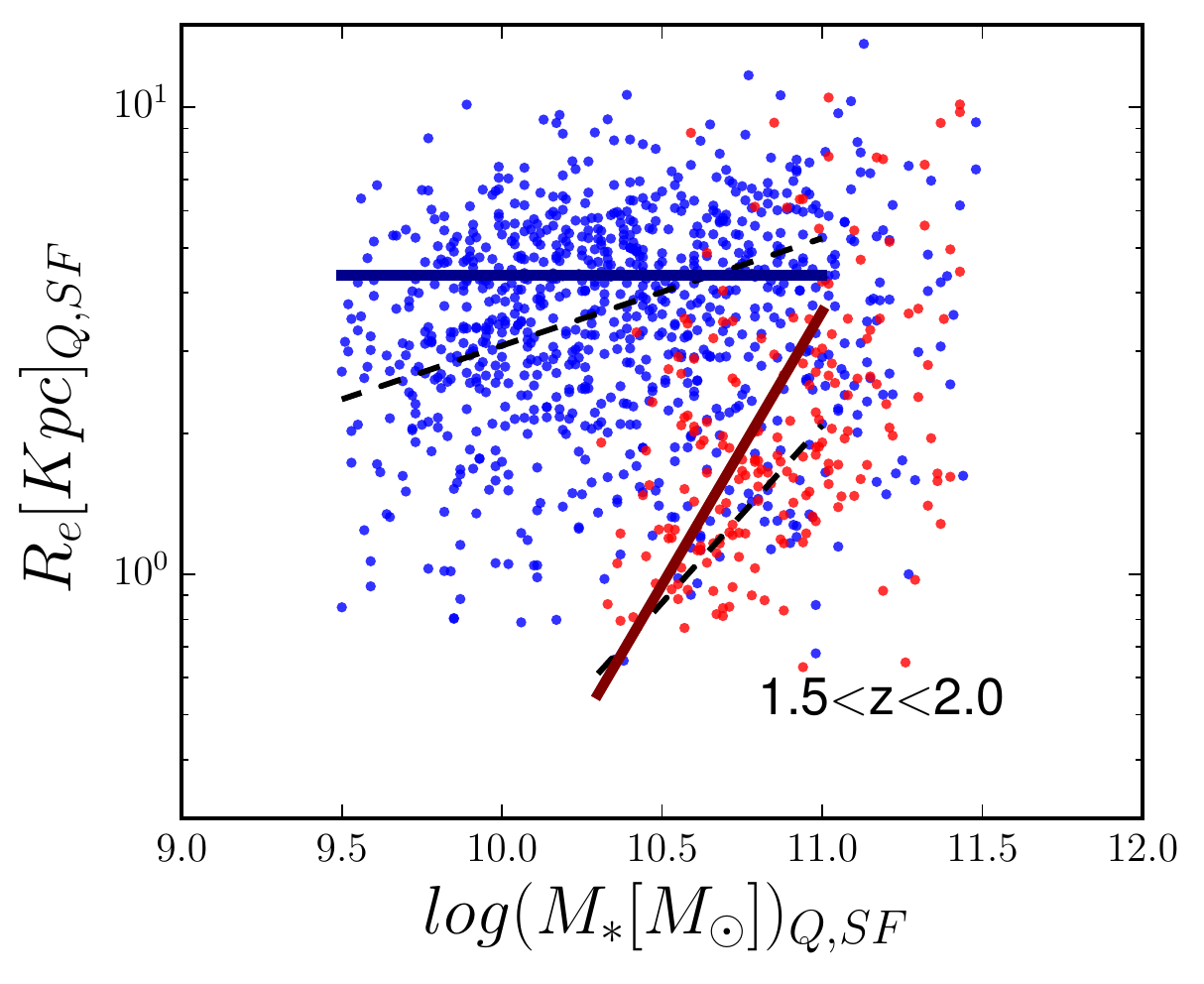} \\

\end{array}$
\caption{Mass-size relation of passive (red points) and star-forming (blue points) galaxies in our sample divided in four redshift bins as labeled. Only star forming/quiescent galaxies with $M_*$ in the range of ($9.5-11.5 M_{\odot}$)/($10.3-11.5 M_{\odot}$) are used for best fits and represented here. Half light sizes are computed in the 5000 \AA \, rest frame band, using a 1-component \Sersic model for all objects. Red (blue) solid lines show the best fit relations for passive and star-forming galaxies respectively. Black dashed lines are the best fits for star-forming (passive) galaxies from van der Wel (2014), reported here as a comparison.} 
 \label{fig:mass_size_SF_Q}
\end{center}
\end{figure*}

Star-forming and quenched populations are selected according to the rest-frame UVJ colors using standard cuts (e.g. \citealp{Whitaker2012}). The distribution of the sample in the mass-size plane is shown in figure~\ref{fig:mass_size_SF_Q}. As previously reported in the literature, star-forming and passive galaxies follow different relations in the mass-size plane (eg. \citealp{Shen2003,vanderWel2014}).

We quantify this result by fitting the two distributions with power-law models. We wish to compare the fits with the reference work by ~\cite{vanderWel2014} and thus we follow an analogous  fitting procedure for consistency. 
We assume a log-normal distribution for the size $N(log\,r,\sigma_{log\,r})$, where $log\,r$ is the mean and $\sigma_{log\,r}$ is the intrinsic dispersion. The semi-major axis $r$ is parametrized as a function of the stellar mass using the following relation:
$$ r(m_*)  = A.m_*^\alpha$$
where $m_*=M_*/5\times 10^{10}$. 
The model distribution $N(log\,r,\sigma_{log\,r})$ provides thus the probability for observing a galaxy with size $r$ given its mass $m_*$. We then use our measured sizes ($R$) together with the computed uncertainties, $\sigma\,log(R)$ (DM18), which are assumed to be Gaussian, and compute the probability of observation $P$:
$$ P=\langle\,N(log\,r(m_*),\sigma_{log\,r}),N(log\,R,\sigma_{log\,R})\rangle$$

The probability is computed for both passive ($P_{Q}$) and star-forming ($P_{SF}$) galaxies. Additionally, we include a random uncertainty of $0.2$ dex on the stellar mass (see DM18). To keep the probability $P$ of one dimension, we assume that errors on stellar mass and size are proportional: $\sigma\,log\,m_*=\alpha\times \sigma\,log\,R$. We adopt a constant $\alpha=0.2$ (0.5) for star-forming and quiescent galaxies respectively which are close to the expected slope of the mass-size relation. Finally, in order to not to be dominated by low mass galaxies which are more numerous, we weight the probability value for each galaxy by the inverse of the measured number density ($W=1/n(z,m_*)$) for a given mass and redshift (\citealp{Muzzin2013}). We also allow for $1\%$ of outliers. 

The final likelihood function that is maximized to estimate the six parameters of the model ($A$, $\alpha$, $\sigma_{log\,r}$) is:

$$L=\Sigma\,ln[W_Q.P_{Q}+0.01] + \Sigma\,ln[W_{SF}.P_{SF}+0.01]$$ 

\cite{vanderWel2014} included a contamination factor related to possible misclassifications. We tested the contribution of this factor since we want to apply a similar approach for the bulge and disc populations. We found that the choice of not consider the contamination factor does not drastically affect the final results (see appendix:  \ref{sec:appendix1}).  
The fits are performed in the stellar mass range [10.3,11.5] $log(M_*/M_\odot)$ / [9.5,11.5] $log(M_*/M_\odot)$ respectively for quiescent and star-forming galaxies. This was done in order to be above the completeness limits shown in table \ref{tbl:comp}, (except for the highest redshift bin) and to compare with the previous results by \cite{vanderWel2014}.

The best fit relations are shown in figure~\ref{fig:mass_size_SF_Q} and the resulting best fit values are reported in table~\ref{tab:mass_size_SF_Q}. Passive galaxies follow a relation with a steeper slope and lower normalization values than star-forming systems. 
We find that values for our sample differ from the ones estimated by \cite{vanderWel2014}, especially at high redshifts. In particular, we measure a shallower slope for star-forming galaxies in the last two redshift bins and a larger zero point for the quiescent population.  Additionally, we do find an evolution of the slope which is not reported by ~\cite{vanderWel2014}. There are several possible reasons to explain these discrepancies. First it might be a consequence of the different magnitude cut applied in the two works (see appendix \ref{sec:appendix2}).  Also, even though we were careful to repeat the same steps detailed in ~\cite{vanderWel2014} for a fair comparison, there is one difference that we would like to remind the reader of: Van der Wel used the 5000 \AA  \, rest-frame sizes and applied a correction, to take into account of the wavelength dependence of the size, while we directly extracted the values from the polynomial functions. In that regards our measurements are more robust and the two results are not directly comparable. To better highlight this point, figure  \ref{fig:test_mass_size_VdW} shows the best fits done using the ~\cite{vanderWel2014}, following the same selecting for the sample and applying the same corrections, in order to test the reliability of the fitting method. Our results appear in closer agreement with the best fits estimated by ~\cite{vanderWel2014}. 

As a consequence, we can conclude that the difference found in the literature are likely due to several factors, i.e., the brightness profile fitting, error estimation, galaxy selection. 

We will now extend the methodology outlined in this section to probe the mass-size relation separately for bulges and discs. We will therefore use our established mass-size relations for starforming and quiescent galaxies as a benchmark to compare to.
 
\begin{table*}
\centering
\begin{tabular}{c c c c c c c c}
   \hline
   \hline
   \multicolumn{5}{c}{Quiescent} & \multicolumn{3}{c}{Star-forming} \\
    \cline{2-4}
    \cline{6 -8}
    $z$ & $log(A)$ & $\alpha$ &$ \sigma log(R)$ & &$log(A)$ & $\alpha$& $ \sigma log(R)$ \\
    \hline
     0.25 & 0.80 & 0.72 & 0.05 & & 0.97 & 0.31 & 0.01  \\
     0.75 & 0.58 & 0.84 & 0.01 & & 0.78 & 0.19 & 0.03  \\
     1.2   & 0.37 & 0.71 & 0.01 & & 0.66 & 0.08 & 0.01  \\
     1.8   & 0.24 & 0.64 & 0.01 & & 0.63 & 0.03 & 0.01 \\                
    \hline
     \hline
 \end{tabular}
\caption{Best-fit results for the mass-size relations of star-forming and quiescent populations more massive than $M_*/M_\odot>2\times10^{10}$.}
\label{tab:mass_size_SF_Q}
\end{table*}

\section{Properties of bulges and discs populations}
\label{sec:results}

In following sections, we present in a systematic way the selections of bulges and discs and the result on the mass-size fitting. We take advantage of the bulge-disc catalog, released in DM18, that is the largest and complete catalog up to z$~2$ currently available. Here onwards, we will always classify galaxies via their B/T mass ratios, rather than relying on more traditional definitions such as "ellipticals", "spirals" or "lenticulars" strictly more appropriate to local galaxy morphologies. 

\subsection{Sample selection}
\label{sec:sel}
Our goal is to study a complete sample of bulges and discs. We thus select bulges and discs that are more massive than $2\times10^{10} M_{\odot}$. This stellar mass limit was chosen as a trade-off between completeness and robust statistics. We make the assumption that discs and bulges have the same completeness limit of star-forming and passive galaxies respectively (see table~\ref{tab:mass_size_SF_Q}). As a result, the highest redshift bin suffers from incompleteness effects. However, the main results do not change significantly if we apply a more conservative mass cut, as further discussed below. Considering the redshift range and wavelength coverage probed by our data, we believe that, using the 5000 \AA \, rest-frame half-light radius as main size estimator is the best choice considering the redshift range and the wavelength coverage probed by our data. 

\begin{figure*}
\begin{center}
\includegraphics[width=\textwidth]{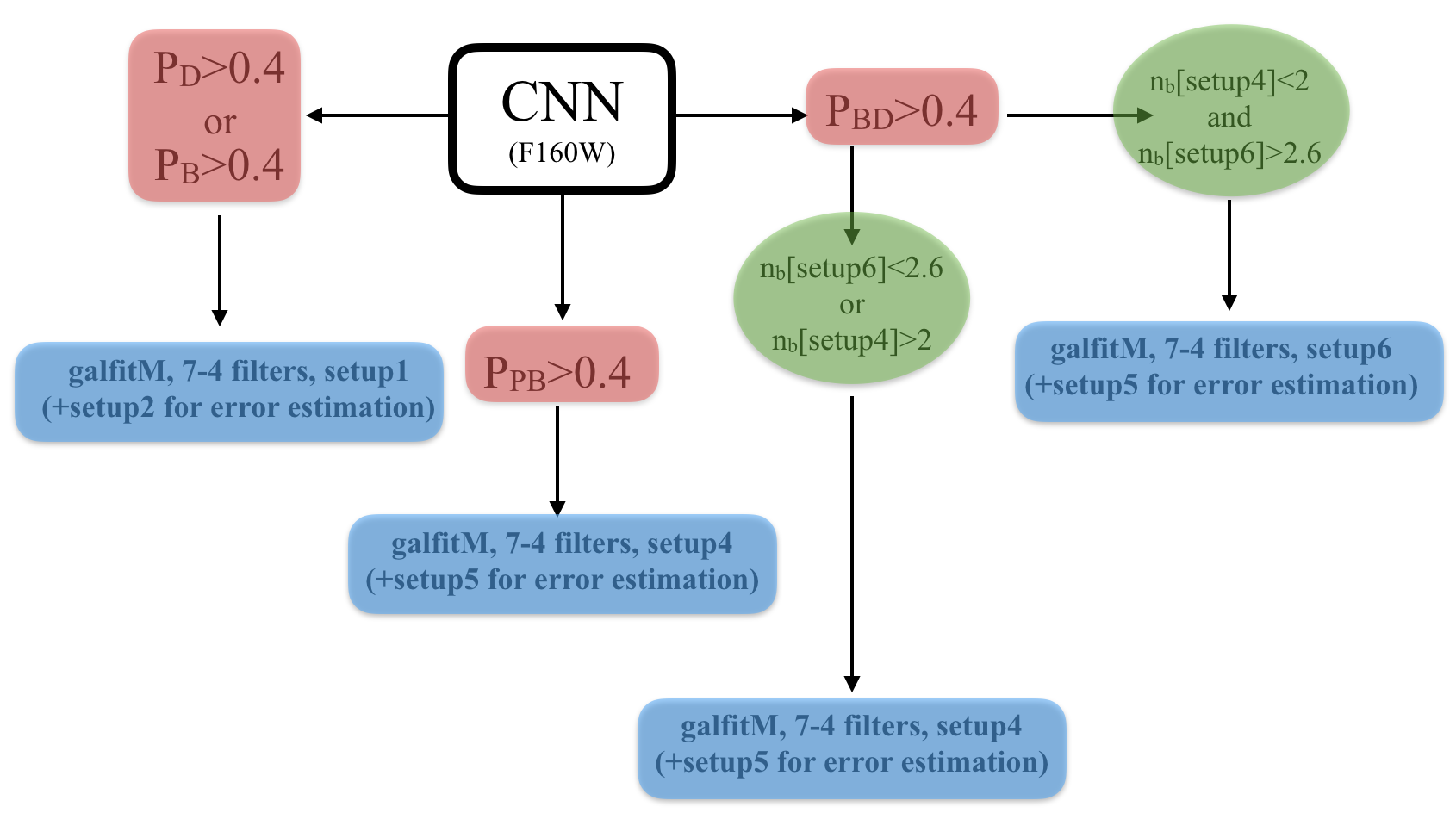} 
\caption{ Flow chart illustrating the selection applied to our sample. The figure is taken from DM18 and shown here for completeness. We refer the reader to the aforementioned work for more details. } 
\label{fig:flow_chart}
\end{center}
\end{figure*}

The choice of the best light-profile model for each galaxy (\Sersic or \Sersic+Exponential profile) is made using the classification based on a convolutional neural network (CNN). Galaxies identified as 2-component systems (i.e. selected because the probabilities $P_{BD}>0.4$ or $P_{PB}>0.4$, where $P_{BD}$ and $P_{PB}$ are the probability that a galaxy is preferentially described by \Sersic (n>2)+Exponential profile or \Sersic (n<2)+Exponential profile respectively) are represented by the \Sersic+Exponential model. The bulge-to-total ratio is then estimated by dividing the mass of the bulge component by the stellar mass of the whole galaxy, both estimated through SED fitting (see DM18). For galaxies that have $P_{B}>0.4$ or $P_{D}>0.4$ (probability to be described by a spheroidal or a disc  profile respectively), we use a 1-component \Sersic fit. The $B/T$ is then set to 1 or to 0 respectively and  the effective radius of the galaxy is used as primary estimator of the size in this cases. 
As shown in DM18, this selection leads to a purity and completeness around $\sim85\%$. The final selection algorithm is summarized in the flow chart of figure~\ref{fig:flow_chart}, taken from DM18 and reproduced here for clarity.

\subsection{Mass-size relation of bulges and discs}
\label{sec:mass_size}

The two populations of bulges and discs are shown in figure~\ref{fig:mass_size_B_D}. Red points represent galaxies that we classified as pure bulges (galaxy described by spheroidal profile i.e. 1-component model: see section~\ref{sec:sel}) as well as bulges of 2-component galaxies. Blue points include pure discs and disc components. As a consequence, most galaxies are represented by two points in this figure, one for the disc and another for the bulge. Galaxies with pseudo-bulges are not considered in this work (i.e. galaxies that host bulges for which a model with two low \Sersic index components is preferred - $P_{PB}>0.4$, see figure~\ref{fig:flow_chart}). However, most of these objects have low stellar masses ($<10^{10}M_\odot$), so the bulk of the results presented in the following are not affected by the inclusion or exclusion of this population. 

We find that the regions occupied by bulges and discs are similar to the ones of quiescent and  star forming galaxies respectively.
In order to quantify the extent of the agreement, we model the distributions with a power law relation using the same method described in section~\ref{sec:SF_Q}. Instead of computing $P_Q$ and $P_{SF}$, we compute now $P_{BULGE}$ and $P_{DISC}$. The best-fits are shown by solid lines, while the dashed lines refer to the best fits of the star-forming and passive populations. The best-fit parameters are summarized in tables~\ref{tbl:fit_mass_size_bulge} and~\ref{tbl:fit_mass_size_disc}. Bulges and discs have different slopes, and the normalizations of the mass-size relations are different for discs and bulges.
At all redshifts, the bulge mass-size relation has a steeper slope ($\alpha\sim0.3-0.4$) than the one of discs ($\alpha\sim0.2-0.1$) and a smaller normalization value at all redshifts. We also measure a slight decrease of the slope of both mass-size relations, especially in the highest redshift bin. 

These results confirm that disc components and star-forming galaxies follow very similar mass-size relations. This may be explained by the fact that we use light-weighted radii. The size of star-forming galaxies is driven by the star formation, which most probably comes from the disc component and therefore drives the relation. This result also reflects that the majority of star-forming galaxies are disc dominated.

Bulges follow a flatter relation compared to passive systems. The size of quiescent galaxies is mostly driven by the mass distribution which is more concentrated in the center, i.e., the bulge component. Indeed, bulge properties tend to drive the mass-size relations for pure bulge systems ($B/T>0.8$), while for double component systems the presence of the disc enlarges the total half light size. Consequently, the decrease of the slope can be due to the intermediate $B/T$ population (0.5 < $B/T$<0.8) which moves from the top right of the mass size plane to smaller sizes in the bulge-disc mass size plane (see appendix \ref{sec:appendix3}). 
Our finding are compatible with previous observations in the nearby universe. Purple lines in figure \ref{fig:mass_size_B_D} show the best fit for bulges and discs components done by \cite{Lange2016}, a recent reference in the local universe. The agreement indicates that, at least in the lowest redshift bin, we are not introducing any bias in our measurement. 

However, our result disagrees with the results reported by \citealp{Bernardi2014}. They find no significant difference between the mass-size relations of bulges and passive galaxies in SDSS. The main reason for this difference is that in all the analysis presented in this paper we use the semi-major axis as size estimator while \citealp{Bernardi2014} use circularized sizes (i.e. $R_{circ}=R_e*\sqrt{b/a}$). If our sizes are circularized both results are in agreement. Note that the fact that circularizing the sizes reduces the differences between the sizes of passive and bulges at fixes mass, supports our previous claim that the difference we measure can be explained by the presence of the disc component.

\begin{figure}
\centering
\includegraphics[width=0.35\textwidth]{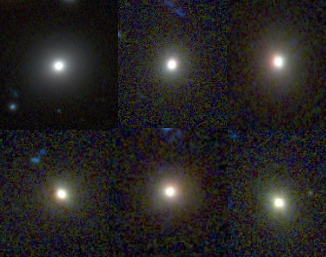}
\caption{Examples of pure bulges populating the massive end of the mass-size plane. They appear with an extended surrounding halo that might affect the profile selection. }
\label{fig:examples_Q}
\end{figure}
\begin{figure*}
\begin{center}
$\begin{array}{c c}
\includegraphics[width=0.44\textwidth]{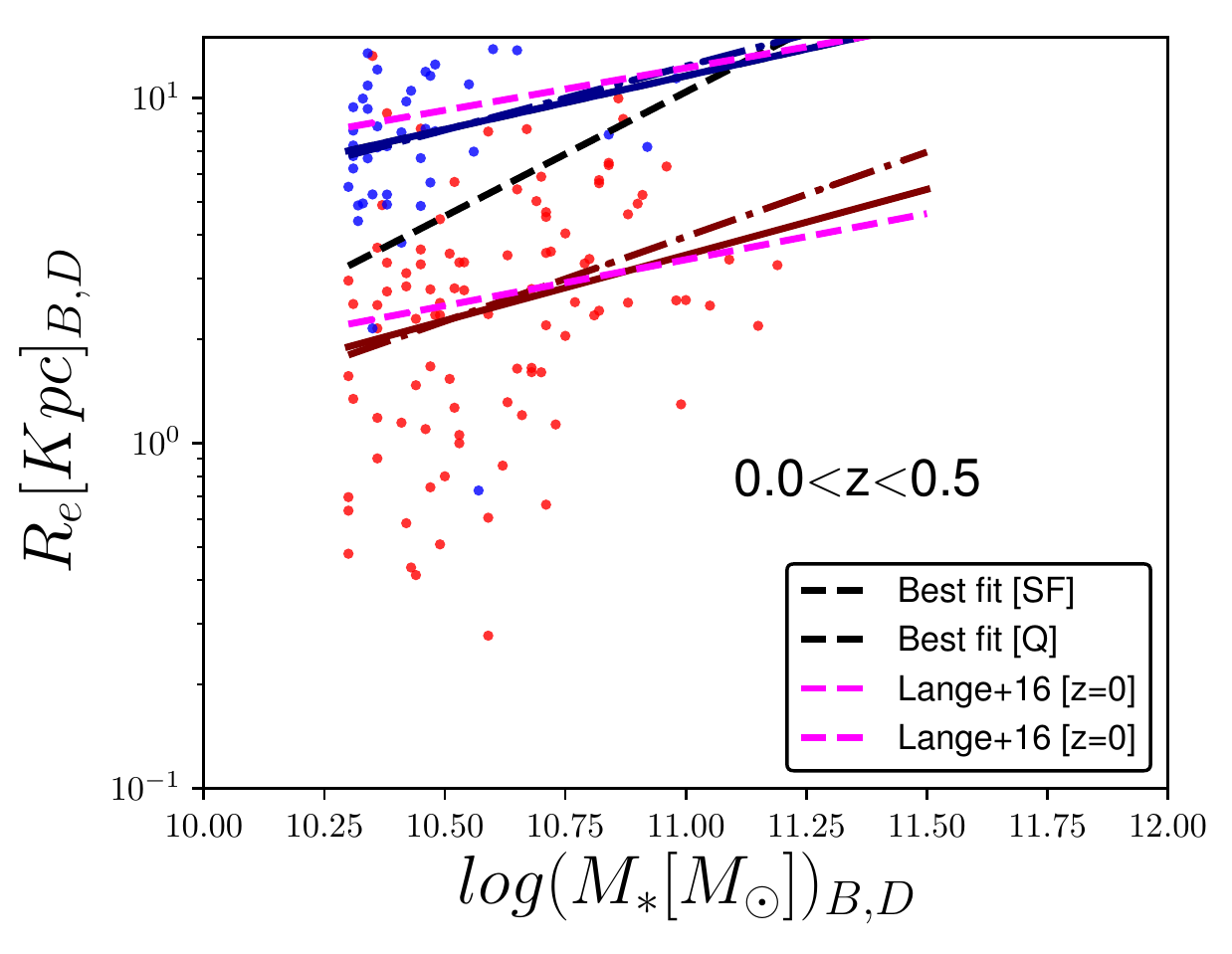} & 
\includegraphics[width=0.44\textwidth]{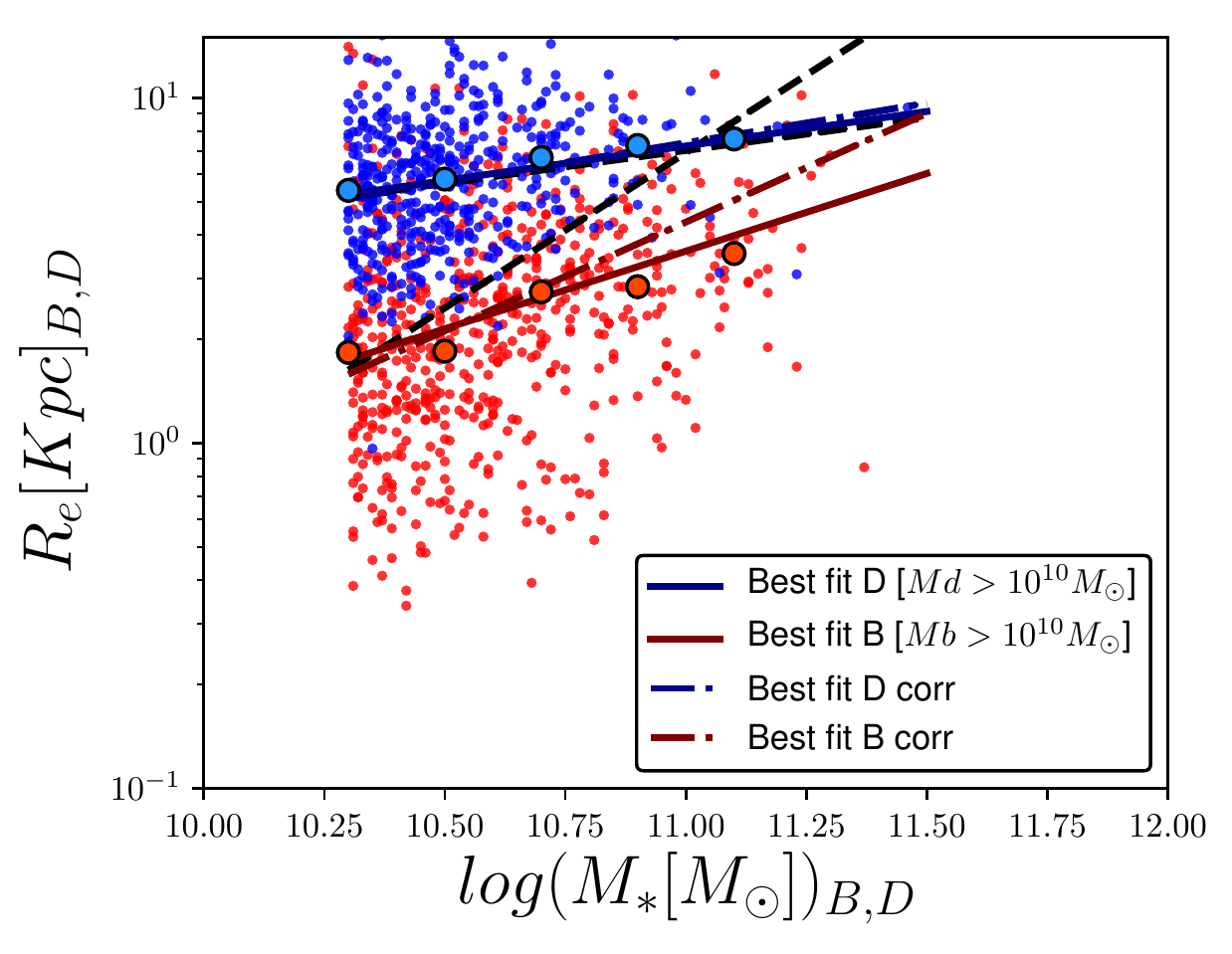} \\
\includegraphics[width=0.44\textwidth]{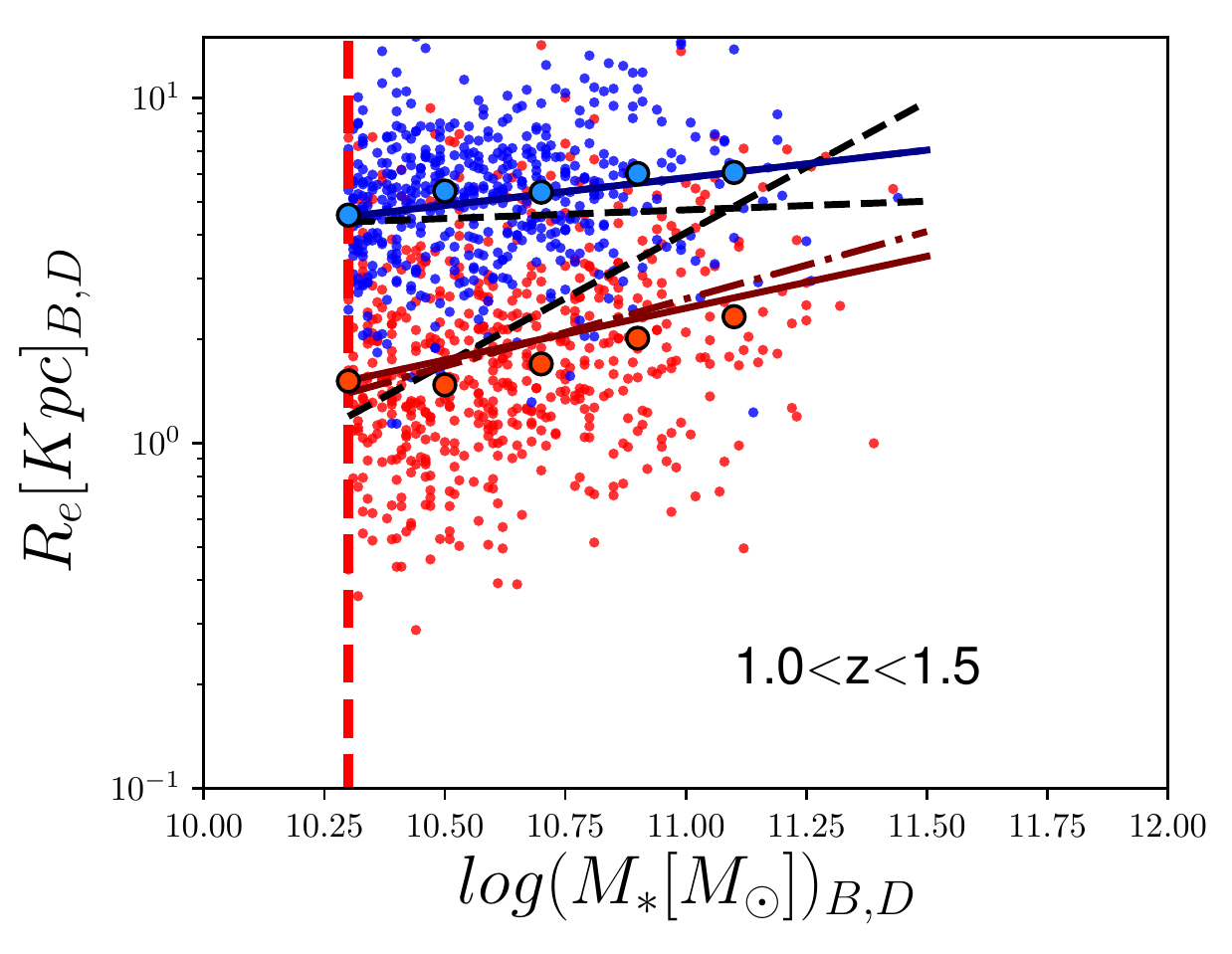} & 
\includegraphics[width=0.44\textwidth]{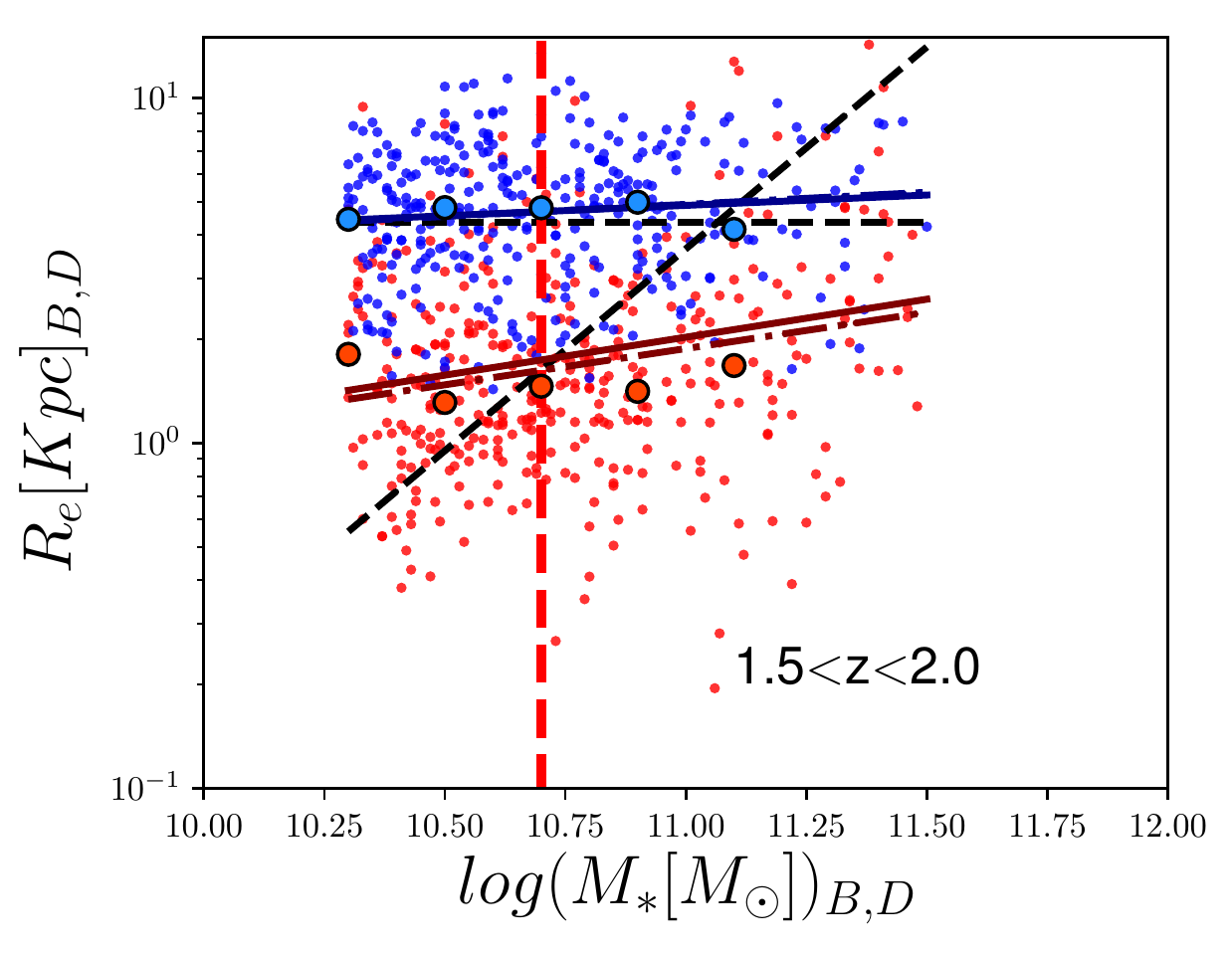} \\
\end{array}$
\caption{Mass-size relation of bulges and discs from our sample divided in four redshift bins. Red points are bulges (pure bulges and also bulges within two component systems). Blue points are discs (pure discs and also discs components). Only bulges and discs between 10.3 $M_{\odot} <M_{*,B,D}< 11.5 M_{\odot}$ are used for the best fit and are shown in the plots. Dark red/blue points are the median value of the sizes in each mass bins. Red (blue) solid lines show the best fit relations for bulges and discs respectively.  Dotted red and blues lines are the best fit done after applying the correction (see the text for more details). Purple lines are the best fit from Lange et al. (2016). Black lines show the best fit for star forming and quiescent galaxies (see figure: \ref{fig:mass_size_SF_Q}). Red vertical dashed lines show the stellar mass completeness.} 
\label{fig:mass_size_B_D}
\end{center}
\end{figure*}
To better investigate whether this effect is related to a wrong classification/model selection, we visually inspected most of the bulges in galaxies located in the massive end of the mass-size plane. We found a small number of galaxies for which the best profile classification does not agree with the visual morphology (\citealp{Huertas2015}). They are classified as 2-components systems while visually they appear as massive ellipticals with an extended surrounding halo (examples are shown in fig. \ref{fig:examples_Q}). Since in these cases a classification is ambiguos, we made an ulterior best-fit, applying a second level of correction. Namely, massive ($Log(M_{*}/M_\odot)>10.7 M_\odot$) passive 2-components galaxies, which have larger probabilities to be classified as a spheroid than a disc (from the visual morphology : $P_{sph}$>0.8 and $P_{disc}$ <0.5), are corrected changing their classification to 1-component model. Results are shown in fig: \ref{fig:mass_size_B_D}. The difference between the two fits does not change the main result. We conclude then that bulges and passive galaxies follow a different mass-size relation and this is mainly due to the existence of passive galaxies with discs even at the high mass end. We explore this further in the following section.

\begin{table*}
\centering
\begin{tabular}{cccccccccccc}
\toprule
\toprule
   \multicolumn{12}{c}{Bulge } \\
    \hline

    \multicolumn{4}{c}{$M_{b}>2\times10 M_{\odot}$ }& \multicolumn{4}{c}{$Q_{Bulge}$ } &\multicolumn{4}{c}{$B/T>0.8$ } \\
     \cline{2-4}
     \cline{6-8}
     \cline{10-12}
   $z$ & $log(A)$ & $\alpha$ &$ \sigma log(R)$ & & $log(A)$ & $\alpha$& $ \sigma log(R)$\ &&  $log(A)$ & $\alpha$ &$ \sigma log(R)$  \\
    \hline
    0.25 & 0.43 & 0.38 & 0.31 & & 0.42 & 0.47 & 0.29 & & 0.44 & 0.48 & 0.18\\
    0.75 & 0.42 & 0.45 & 0.27 & & 0.38 & 0.61 & 0.24 & & 0.40 & 0.43 & 0.16 \\
    1.25 & 0.29 & 0.30 & 0.25 & & 0.26 & 0.39 & 0.23 & & 0.29 & 0.34 & 0.17 \\
    1.75 & 0.21 & 0.22 & 0.25 & & 0.16 & 0.29 & 0.22 & & 0.21 & 0.14 & 016\\
\bottomrule
\bottomrule
\end{tabular}
\caption[Mass size fits results for bulges]{Results from the parametrized fit on the mass size relation for bulges with different selections. 1: all bulges with $M_{B}>2\times10 M_{\odot}$. 2: Bulges more massive than $M_{B}>2\times10^{10} M_{\odot}$ embedded in quiescent galaxies. 3: Bulges more massive than $M_{B}>2\times10^{10} M_{\odot}$ living in galaxies with $B/T >0.8$. }
\label{tbl:fit_mass_size_bulge}
\end{table*}
\begin{table*}
\centering
\begin{tabular}{cccccccccccc}
\toprule
\toprule
   \multicolumn{12}{c}{Disc } \\
    \hline

    \multicolumn{4}{c}{ $M_{d}>2\times10 M_{\odot}$ }& \multicolumn{4}{c}{$SF_{Disc}$ } &\multicolumn{4}{c}{ $B/T<0.2$ } \\
     \cline{2-4}
     \cline{6-8}
     \cline{10-12}
   $z$ & $log(A)$ & $\alpha$ &$ \sigma log(R)$ & & $log(A)$ & $\alpha$& $ \sigma log(R)$\ &&  $log(A)$ & $\alpha$ &$ \sigma log(R)$  \\
    \hline
     0.25 & 0.97 & 0.31 & 0.17 & & 0.98 & 0.38 & 0.18 && 0.91 & 0.25 & 0.18 \\
     0.75 & 0.80 & 0.20 & 0.18 & & 0.81 & 0.24 & 0.17 && 0.78 & 0.33 & 0.16 \\
     1.20 & 0.72 & 0.16 & 0.17 & & 0.72 & 0.12 & 0.16 && 0.69 & 0.08 & 0.17 \\
     1.80 & 0.67 & 0.06 & 0.17 & & 0.68 & 0.07 & 0.16 && 0.67 & 0.10 & 0.16 \\   
\bottomrule
\bottomrule
\end{tabular}
\caption[Mass size fits results for discs]{Results from the parametrized fit on the mass size relation for discs with different selections. 1: all discs with $M_{d}>2\times10 M_{\odot}$. 2: Discs more massive than $M_{D}>2\times10^{10}M_{\odot}$ embedded in star forming galaxies. 3: Discs more massive than $M_{D}>2\times10^{10} M_{\odot}$ living in galaxies with $B/T<0.2$.}
\label{tbl:fit_mass_size_disc}
\end{table*}
 
\section{Where do massive bulges and discs live?}
\label{sec:morph}

Following the results of the previous sections, we now aim to explore the dependence of the bulge and disc structure on the morphology of the galaxy that they reside in, using the bulge-to-total ratio.  
The top panel of figure~\ref{fig:nd} shows the morphological mix of galaxies hosting bulges more massive than $M_{*,B}>2\times10^{10}M_\odot$. About $\sim 60\%$ of massive bulges are in pure bulge dominated systems ($B/T >0.8$), while the remaining $40\%$ are embedded in discs ($B/T<0.8$). Moreover, at high redshift, $\sim60\%$ of massive discs ($Log(M_{*,D}/M_\odot)>10.3$) do not have a noticeable overdensity in the centre ($B/T<0.2$), while $\sim40\%$ of them host a relevant bulge component. At low redshift, the number of pure discs decrease. Indeed, $40\%$ of galaxies show no bulge. This morphological mix allows us to study how bulge and disc structural properties depend on the global morphology of the galaxy (we use here the $B/T$ as a proxy of the morphology) with the aim of putting some constraints on the formation mechanisms.

\subsection{Bulges}
If bulges are formed via different physical mechanisms, different processes should leave signatures on the galaxies internal structure i.e. on the bulge and disc components. We expect, then, to be able to find a statistical differences between pure bulges and bulges embedded in discs, at fixed stellar mass.

\begin{figure*}
\begin{center}
$\begin{array}{c c}
\includegraphics[width=0.44\textwidth]{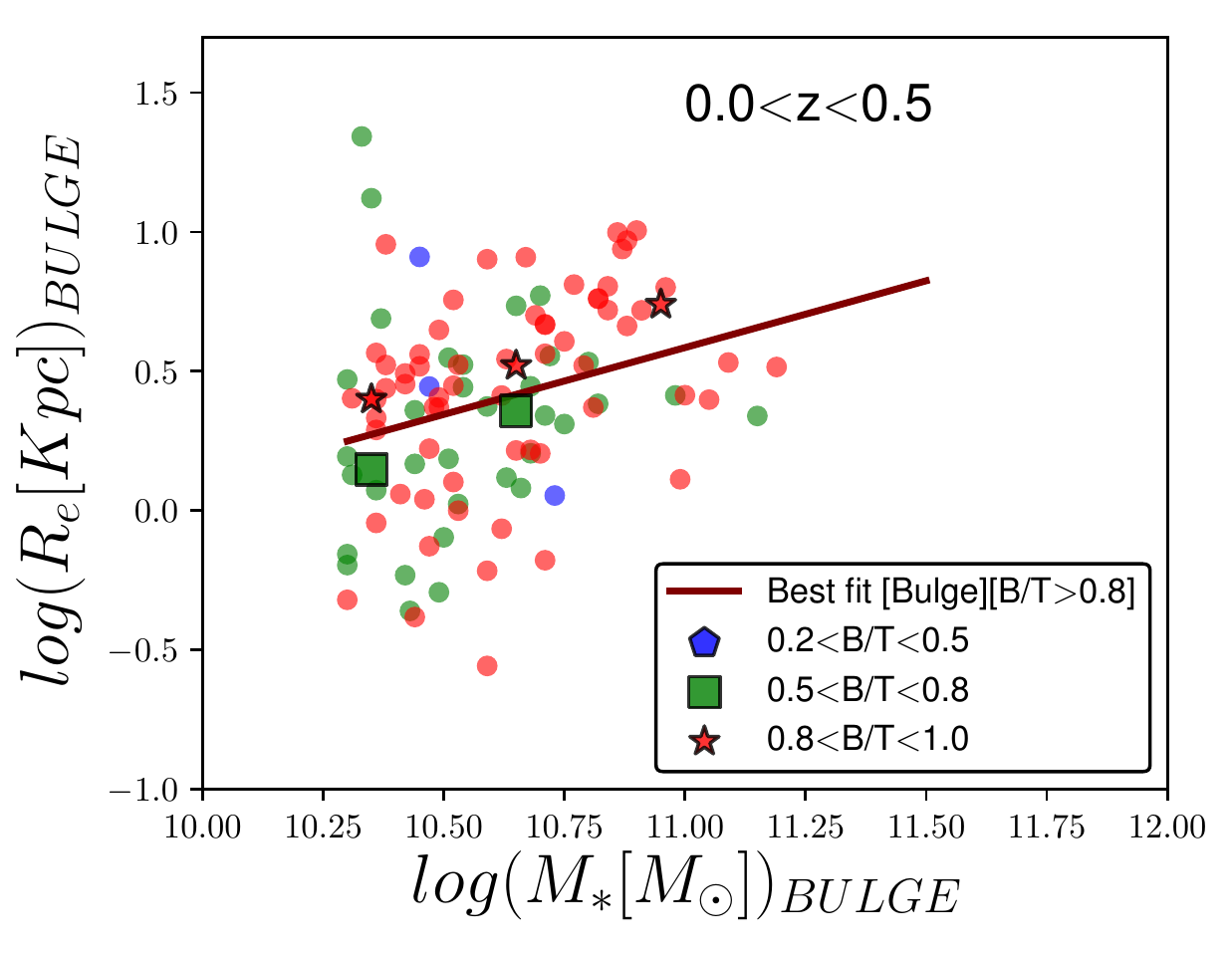} &
 \includegraphics[width=0.44\textwidth]{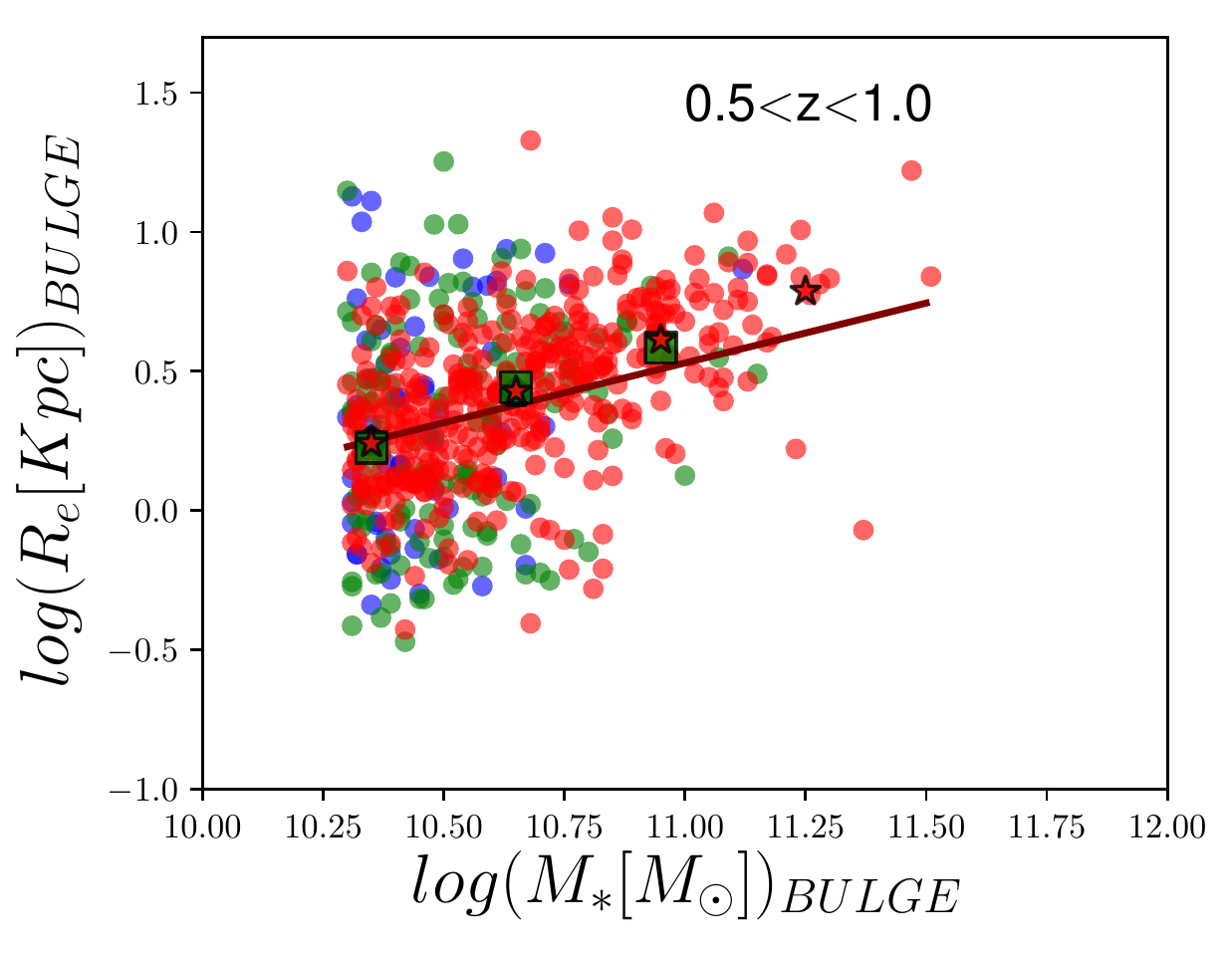} \\
\includegraphics[width=0.44\textwidth]{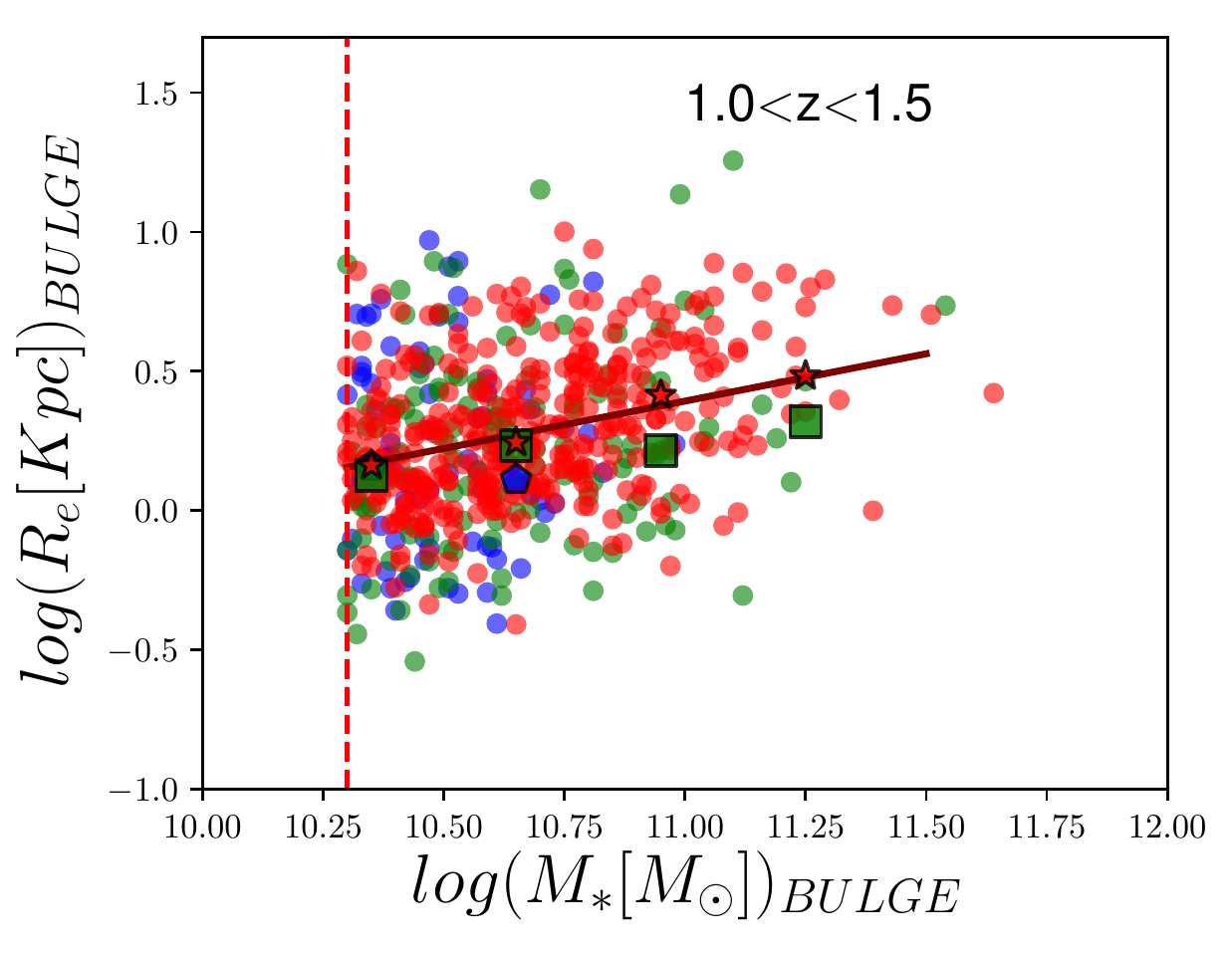} & 
\includegraphics[width=0.44\textwidth]{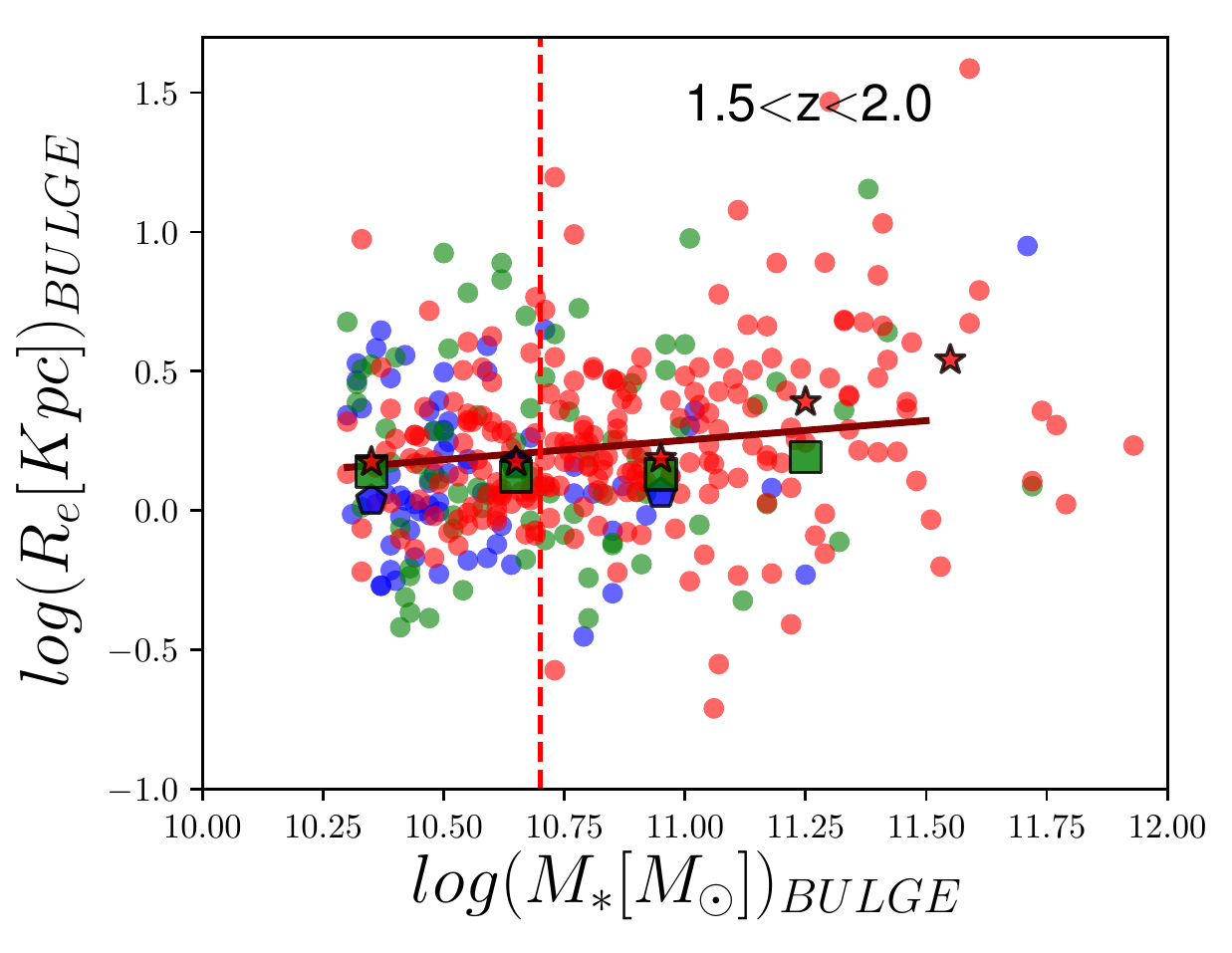} \\
\end{array}$
\caption{Mass-size relation of bulges embedded in galaxies with different bulge-to-total ratios, as labelled. The solid red line is the best fit model to the mass-size distribution of elliptical galaxies ($B/T>0.8$). Points with different colors and shapes are the median values of sizes in mass bins for different values of B/T as shown in the legend. The vertical dashed red lines are the stellar mass completeness limit for bulges. No strong correlation is revealed  between the size of the bulge and the morphology ($B/T$) of the host galaxy.} 
\label{fig:mass_size_B_BT}
\end{center}
\end{figure*}

\begin{figure*}
$\begin{array}{c c}
\includegraphics[width=0.4\textwidth]{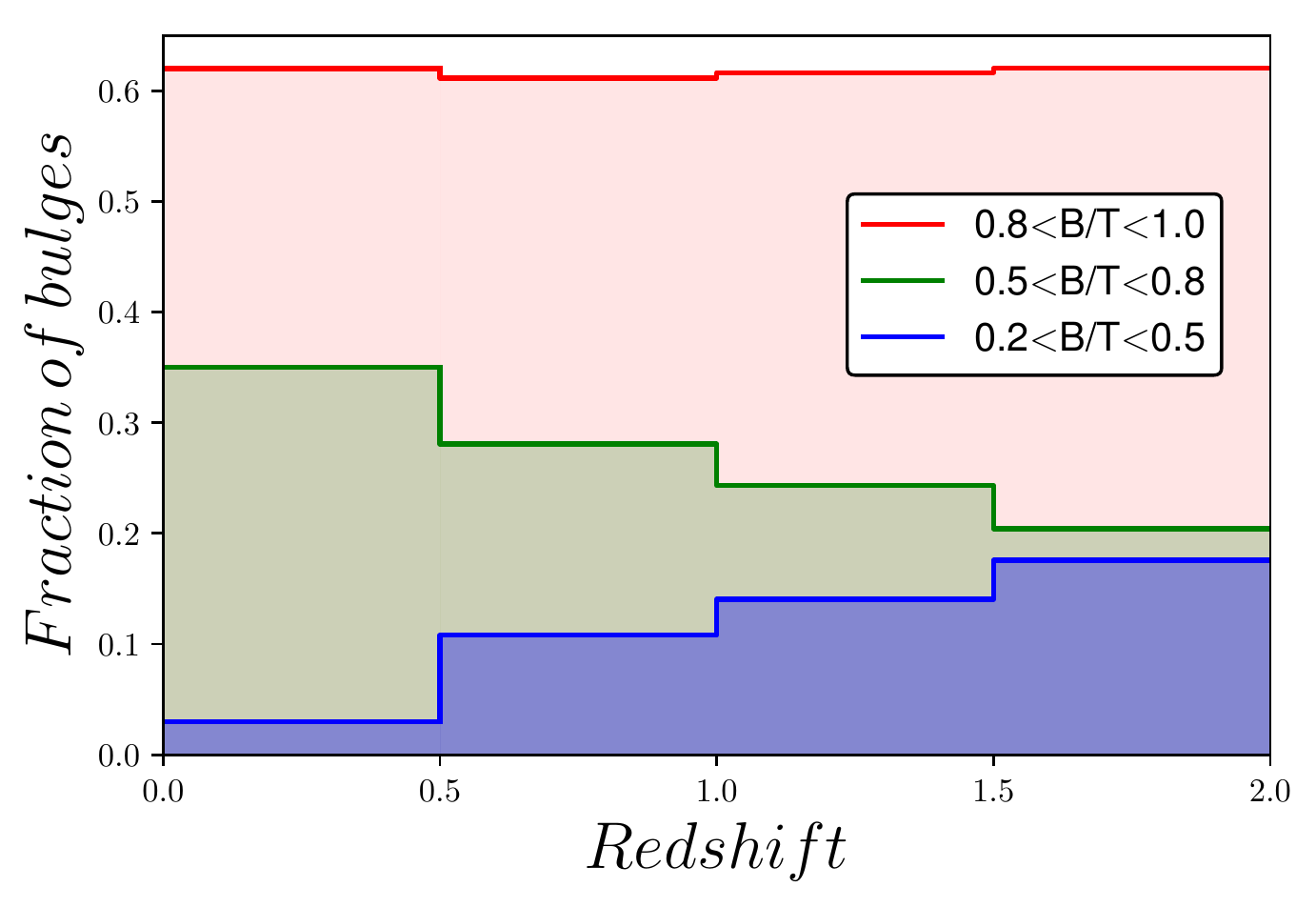} &
\includegraphics[width=0.4\textwidth]{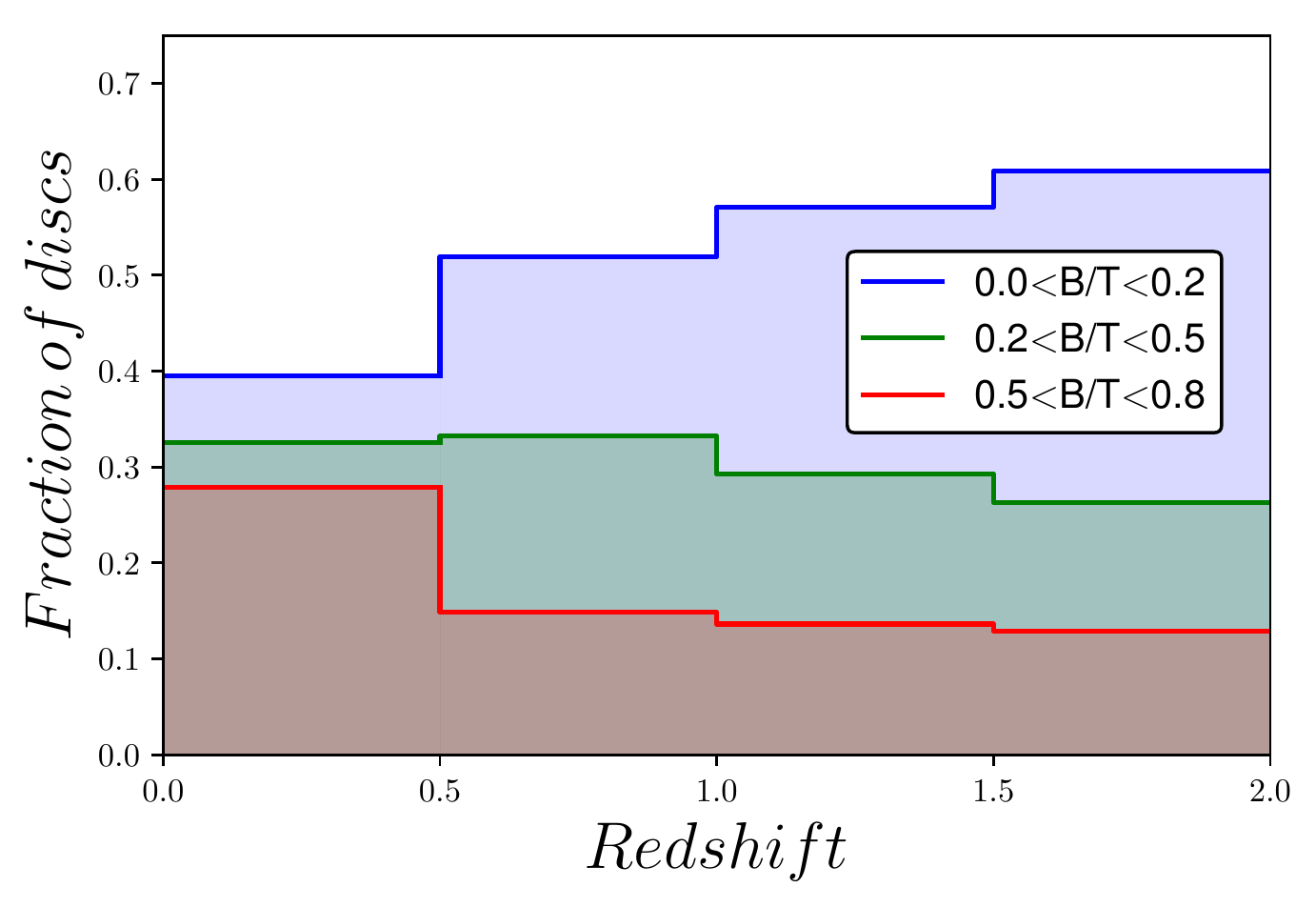} 
\end{array}$
\caption{Left: Fraction of massive bulges ($log(M_{*,B}/M_\odot)>10.3$) embedded in different morphologies as labeled as a function of redshift. Right: Same for massive discs ($log(M_{*,D}/M_\odot)>10.3$)}
\label{fig:nd}
\end{figure*}

 Figure~\ref{fig:mass_size_B_BT} shows the distribution of bulges in the mass-size plane color coded by the bulge-to-total ratio of the galaxy. Bulges in galaxies with $B/T<0.2$ are excluded from the plot since, as shown in DM18, their structural properties have large errors. 

\begin{figure}
\includegraphics[width=0.45\textwidth]{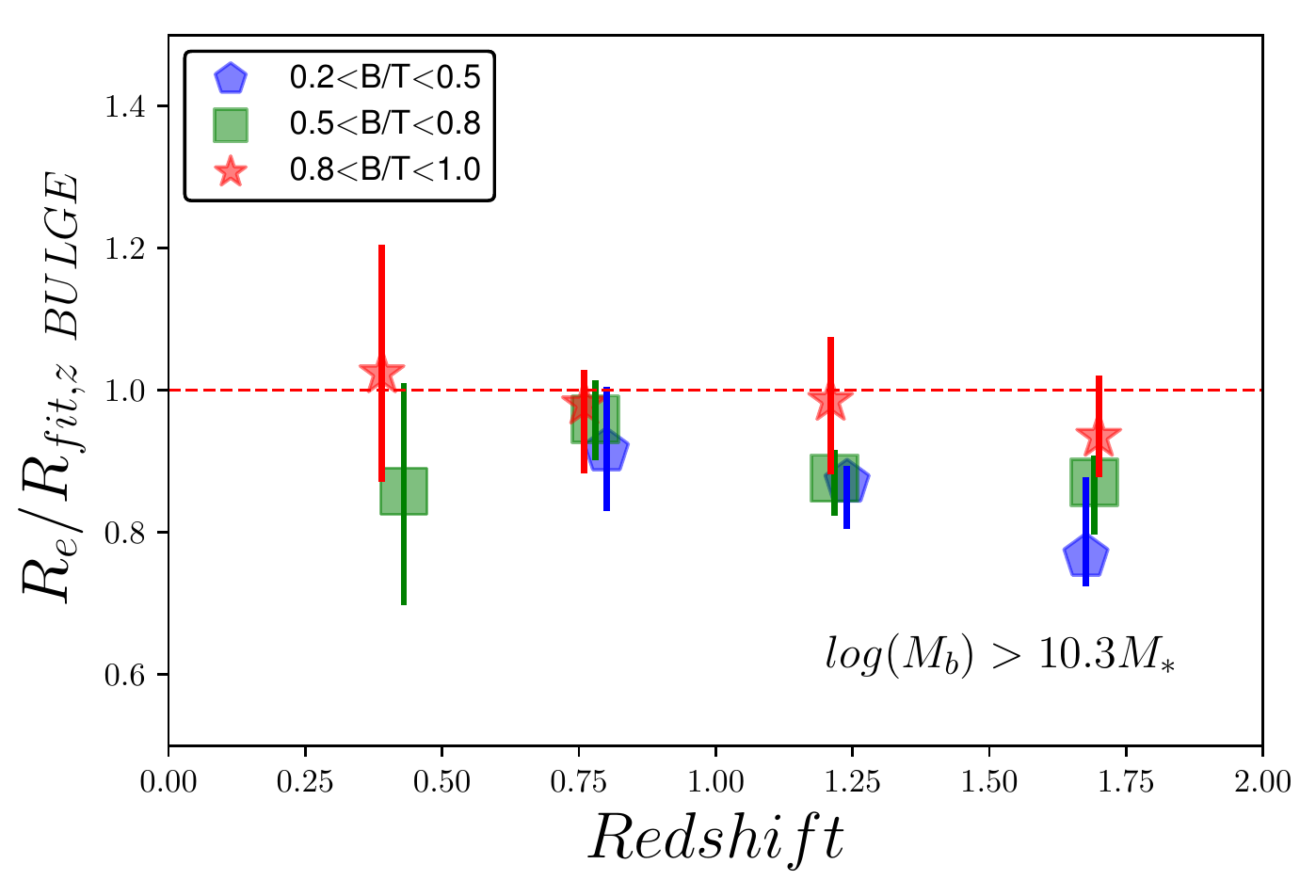} 
\caption{Median mass-normalized sizes of bulges  embedded in different morphologies as a function of redshift. For each redshift bin, the effective radius of every bulge is divided by the expected size from the best fit model to the population of elliptical galaxies ($B/T>0.8$). The median of the ratios is then reported. Errors bars are $68\%$ confidence levels estimated through bootstrapping $1000$ times. Bulges sizes in different morphologies are compatible. No systematic shift is observed. } \label{fig:gamma_evol}
\end{figure}

In order to quantify the results and measure possible systematic differences between bulge properties hosted in different systems, we fit the mass-size relation for pure bulges (B/T>0.8) using the same method as in section~\ref{sec:SF_Q}. (The best fit models are reported in table~\ref{tbl:fit_mass_size_bulge}). We then compute deviations from the best-fit line for every bulge by dividing each measured semi-major axis by the expected values computed from the best fit model at the same mass and redshift. Finally, the median of the ratios for different redshifts and $B/T$ bin is computed and reported in figure~\ref{fig:gamma_evol}. By definition, bulges in $B/T >0.8$ galaxies are expected to lie on average in the constant unity line. Every deviation indicates a statistical difference between populations. This methodology limits the impact of different mass distributions and allows an unbiased quantification of the difference between pure bulges and those living in more disky systems. We find that all the median values of the size ratios of bulges in different morphologies and at fixed stellar mass, are in agreement within a $\pm$ 2-$\sigma$ range of 20\%.  
The size of the bulge is independent of the mass of the disc component.
As a double check, we applied the Kolmogorov-Smirnov test (hereafter K-S test). The main aim of this test is to statistically compare the two samples in order to understand if 
they can be described by the same model. It results that the three populations of bulges are compatible with arising from the same distribution.
 We discuss the implications of these results on the bulge formation mechanisms in section~\ref{sec:discussion}.

\subsection{Discs}
\begin{figure*}
\begin{center}
$\begin{array}{c c}
\includegraphics[width=0.45\textwidth]{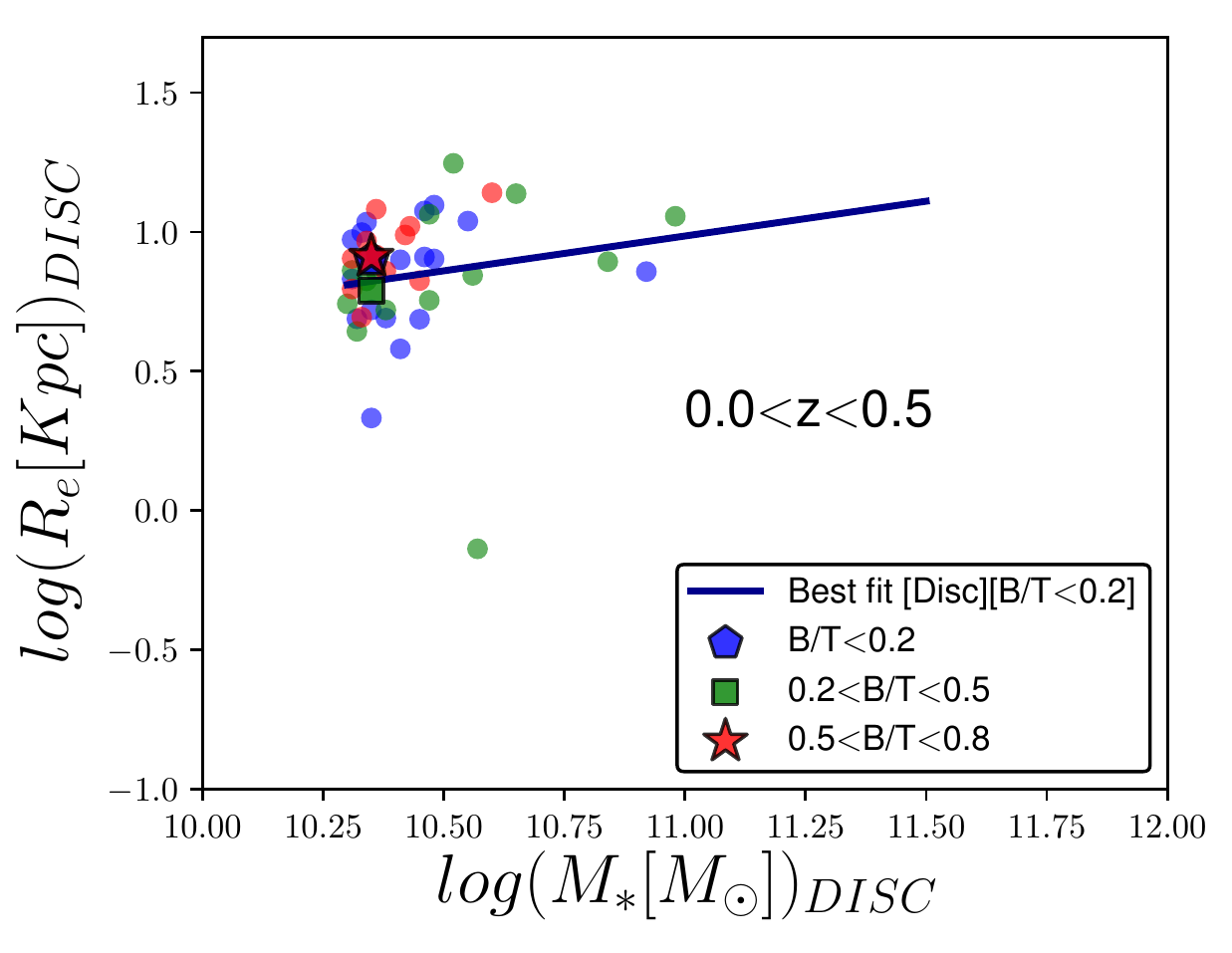} &
 \includegraphics[width=0.45\textwidth]{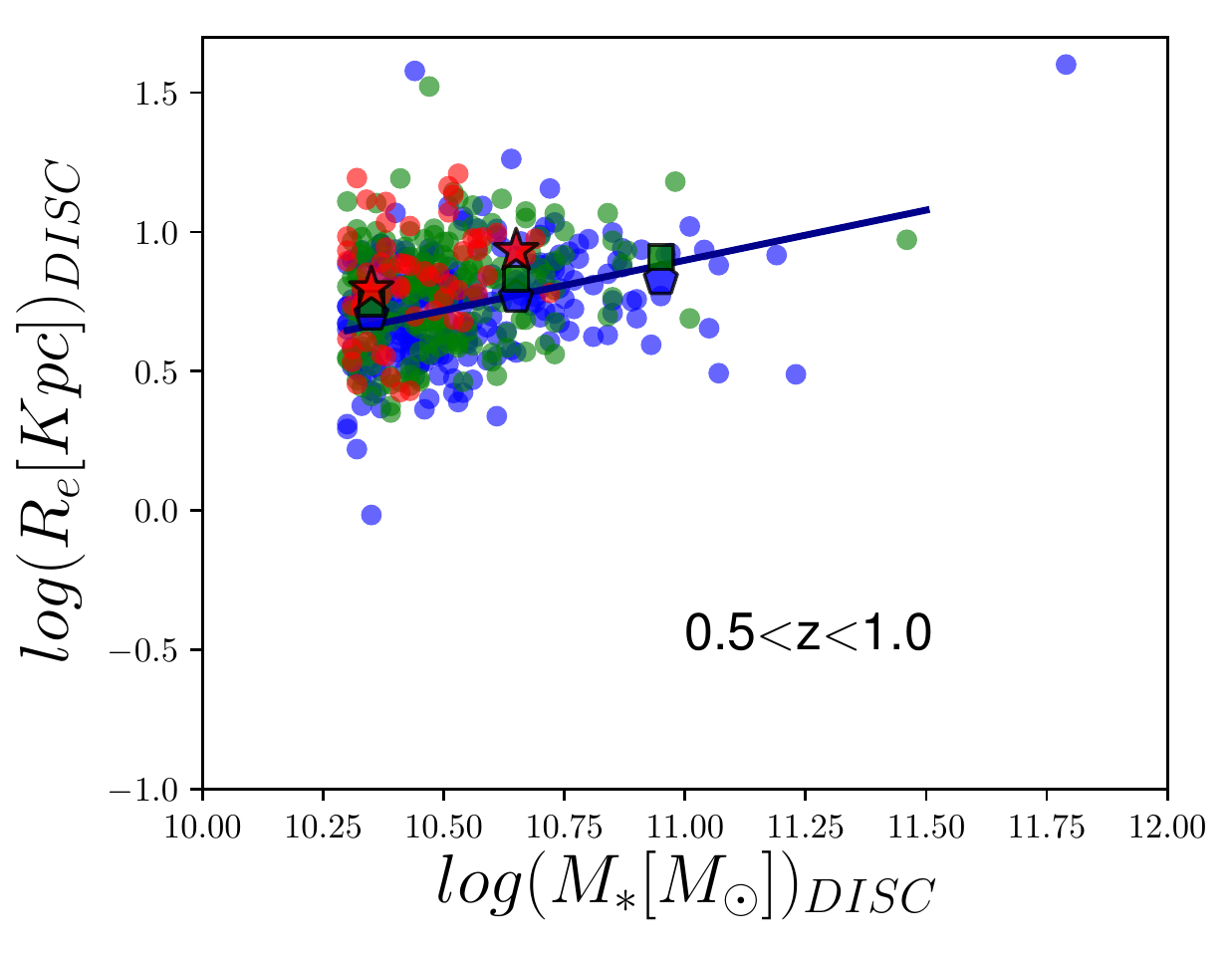} \\
\includegraphics[width=0.45\textwidth]{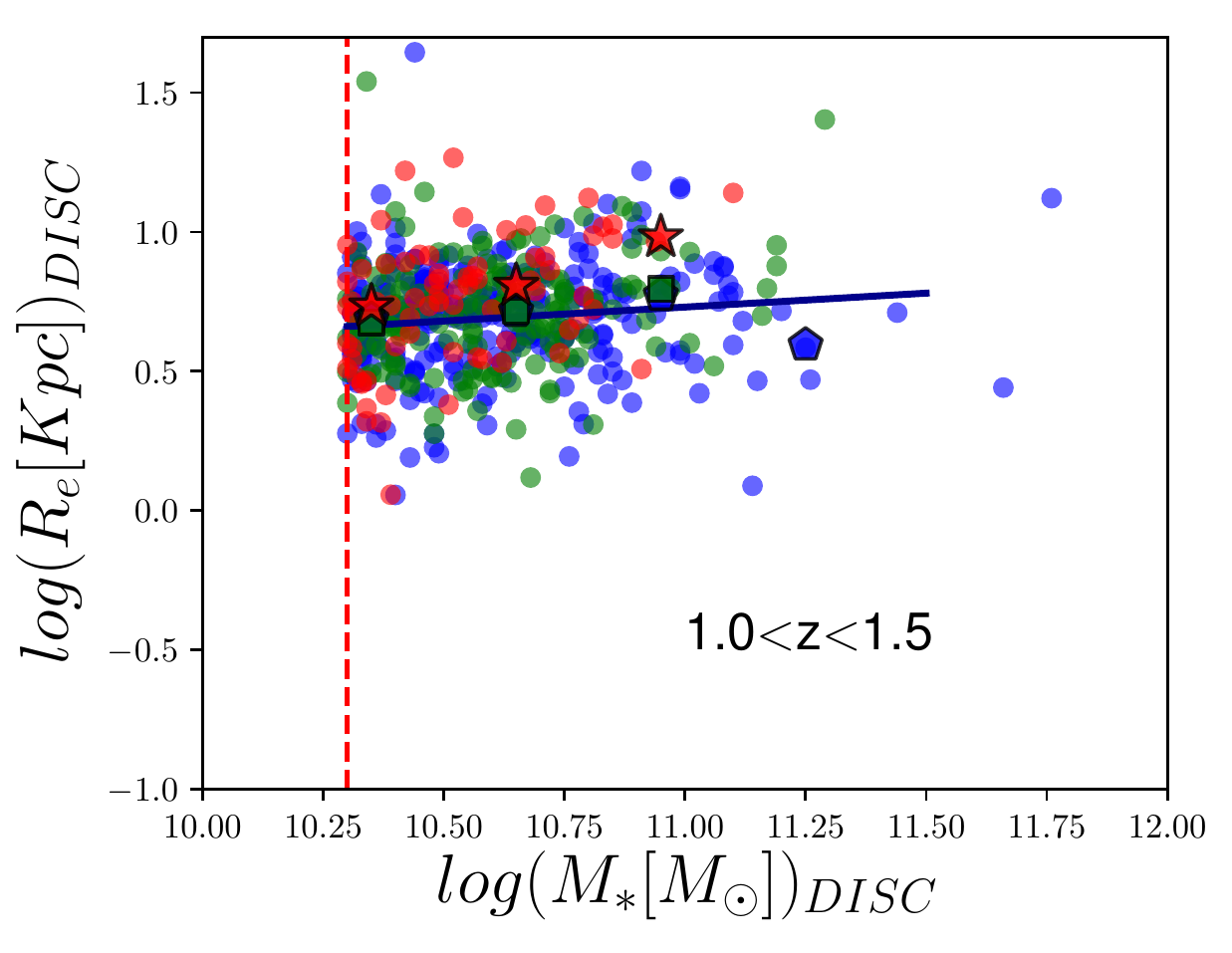} &
 \includegraphics[width=0.45\textwidth]{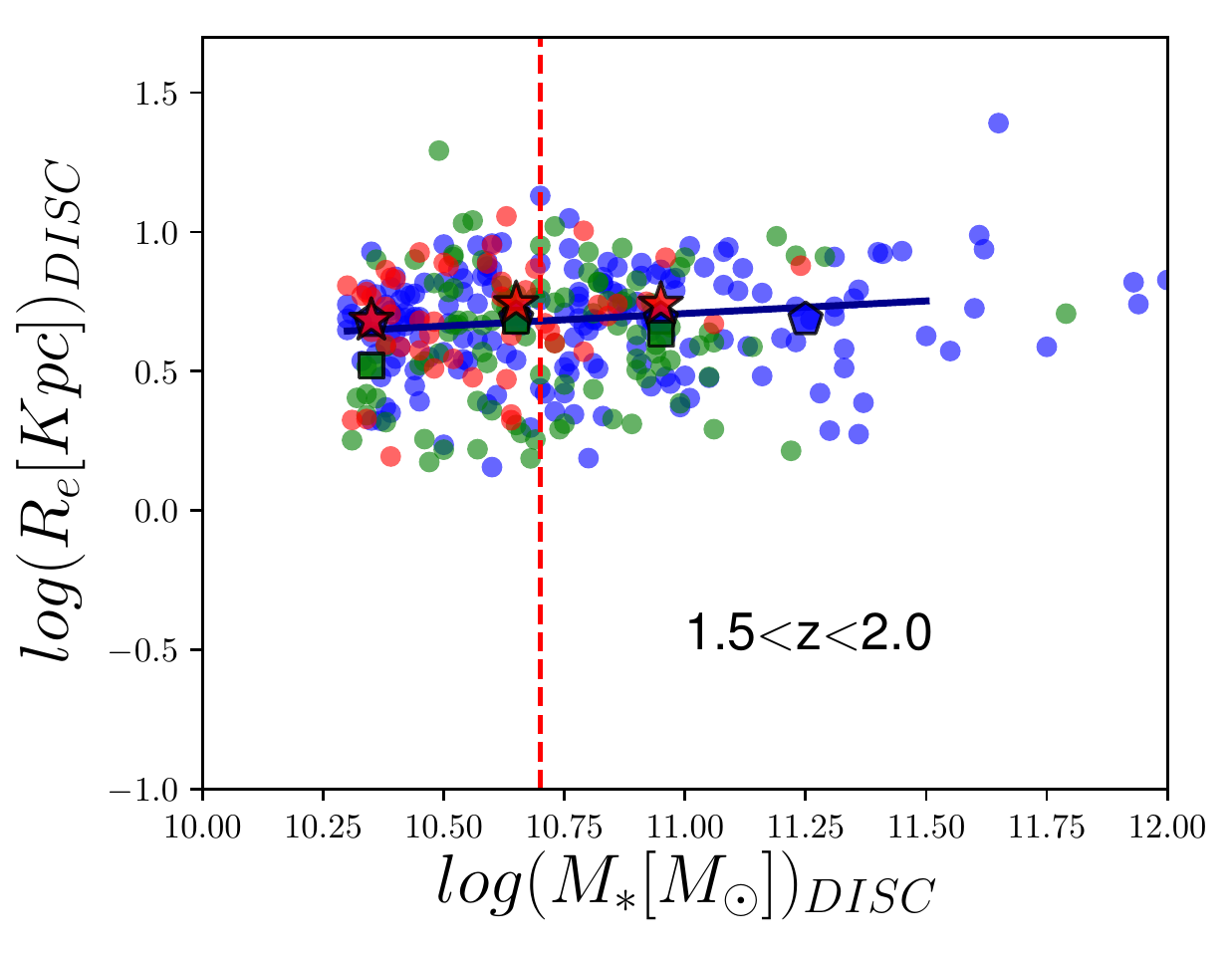} \\
\end{array}$
\caption{Mass-size relation of discs embedded in galaxies with different bulge-to-total ratios, as labelled. The solid blue line is the best fit model to the mass-size distribution of pure disc galaxies ($B/T<0.2$). Points with different colors and shapes are the median values of sizes in mass bins for different values of B/T as shown in the legend. The vertical dashed red lines are the stellar mass completeness limit for star forming galaxies. No clear correlation is observed between the size of discs and the morphology (B/T) of the hosting systems.} \label{fig:mass_size_D_BT}
\end{center}
\end{figure*}
\begin{figure*}
\includegraphics[width=0.45\textwidth]{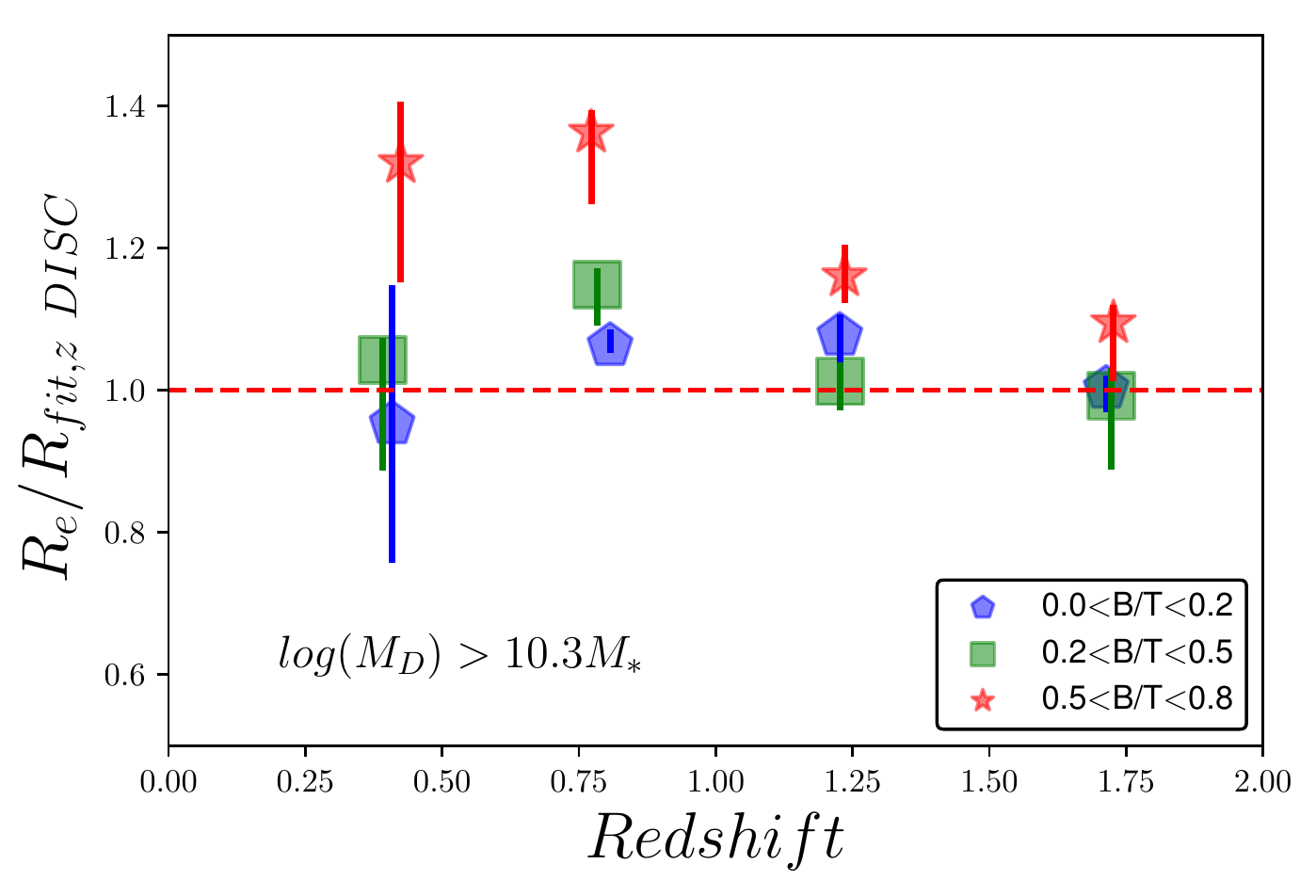} 
\caption{Median sizes of discs ($Log(M_{*,D}/M_\odot)>10.3$) embedded in different morphologies as labelled as a function of redshift. For each redshift bin, the effective radius of each disc is divided by the expected size from the best fit model to the population of pure disc galaxies ($B/T<0.2$). The median of the ratios is then reported. Errors bars are $68\%$ confidence levels estimated through bootstrapping $1000$ times.} \label{fig:gamma_evol_D}
\end{figure*}

 Following the same approach as for bulges, we now analyze the disc structural properties in different morphologies (B/T). Objects with $B/T>0.8$ are removed from the final disc sample to avoid possible bias. 
Figure~\ref{fig:mass_size_D_BT} shows the distribution of discs in the mass-size plane divided in three bins of $B/T$. No clear correlation between the size and the morphology, at fixed stellar mass is measured. This trend is better quantified in figure~\ref{fig:gamma_evol_D}. As done for bulges, we compare sizes of discs in different morphologies, using as a reference for the mass-size relation of pure disc galaxies (B/T<0.2). The mass dependence is removed by dividing each disc size by the expected value from the best fit (see table~\ref{tbl:fit_mass_size_disc}). 

Galaxies with low $B/T$ are close to being pure discs, therefore blue and green points are close to unity by construction. These results seem to suggest that in these class of galaxies, the presence of the bulge does not have a strong impact on the properties of the discs. 
However, discs in systems with values of $B/T>0.5$, appear slightly larger. This trend could be a consequence of statistical fluctuations (only $\sim10\%$ of discs live in these systems). Despite that, the fact that the red points are always above the others in all redshift bins allows us to claim that there is a slight difference between pure discs and discs surrounding massive bulges.  For the same disc mass, higher B/T implies a larger total galaxy mass which corresponds to a larger halo mass, therefore a higher virial radius. As disc sizes are expected to scale as $R_D \sim \lambda Rvir$ (where $\lambda$ is the spin parameter; \citealp{Mao1998}), galaxies with larger B/T have larger radii for the same disc mass. 

\section{Bulges and discs of star-forming and quiescent galaxies}
\label{sec:SF}
\begin{figure*}
$\begin{array}{c c}
\includegraphics[width=0.4\textwidth]{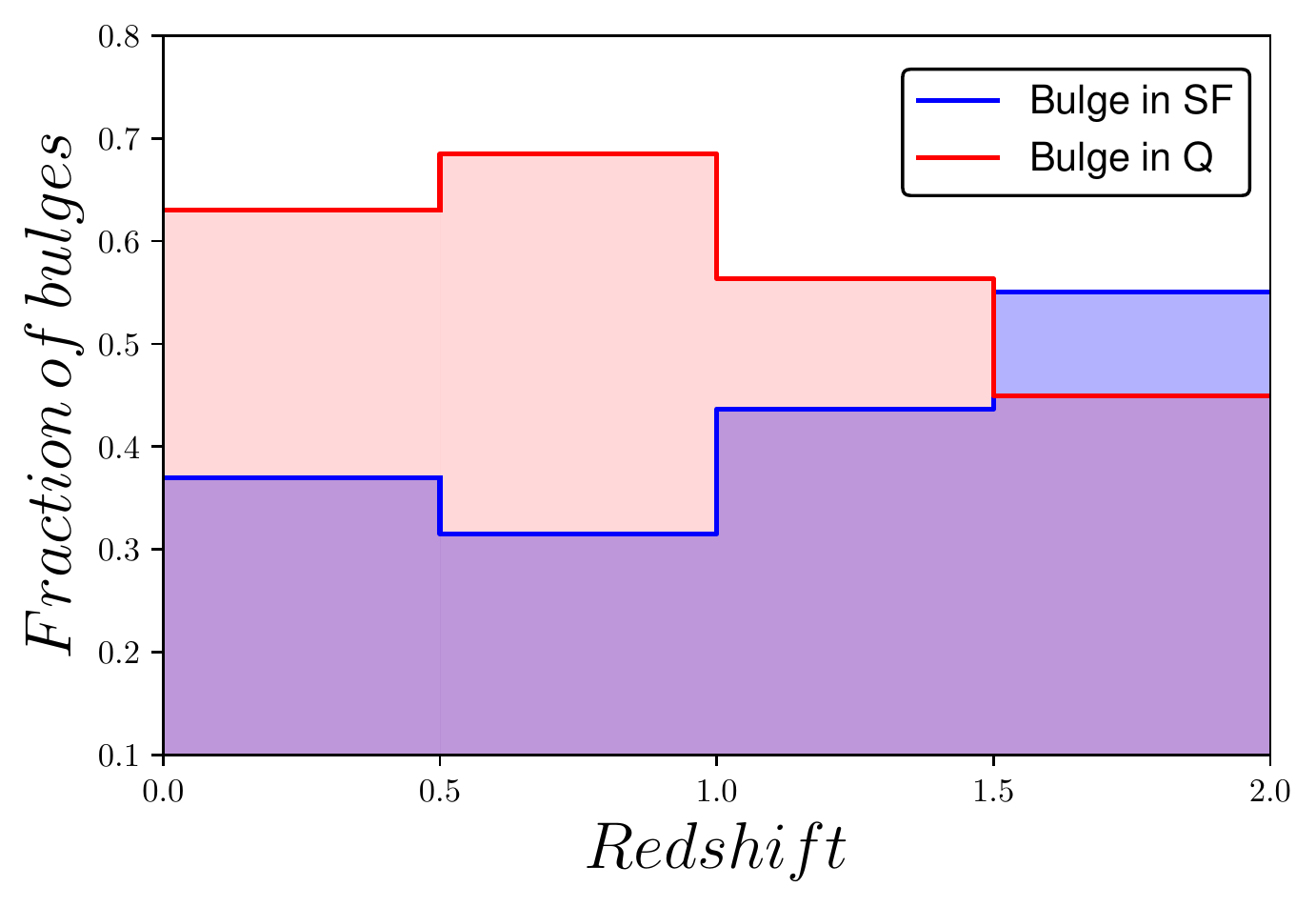} &
\includegraphics[width=0.4\textwidth]{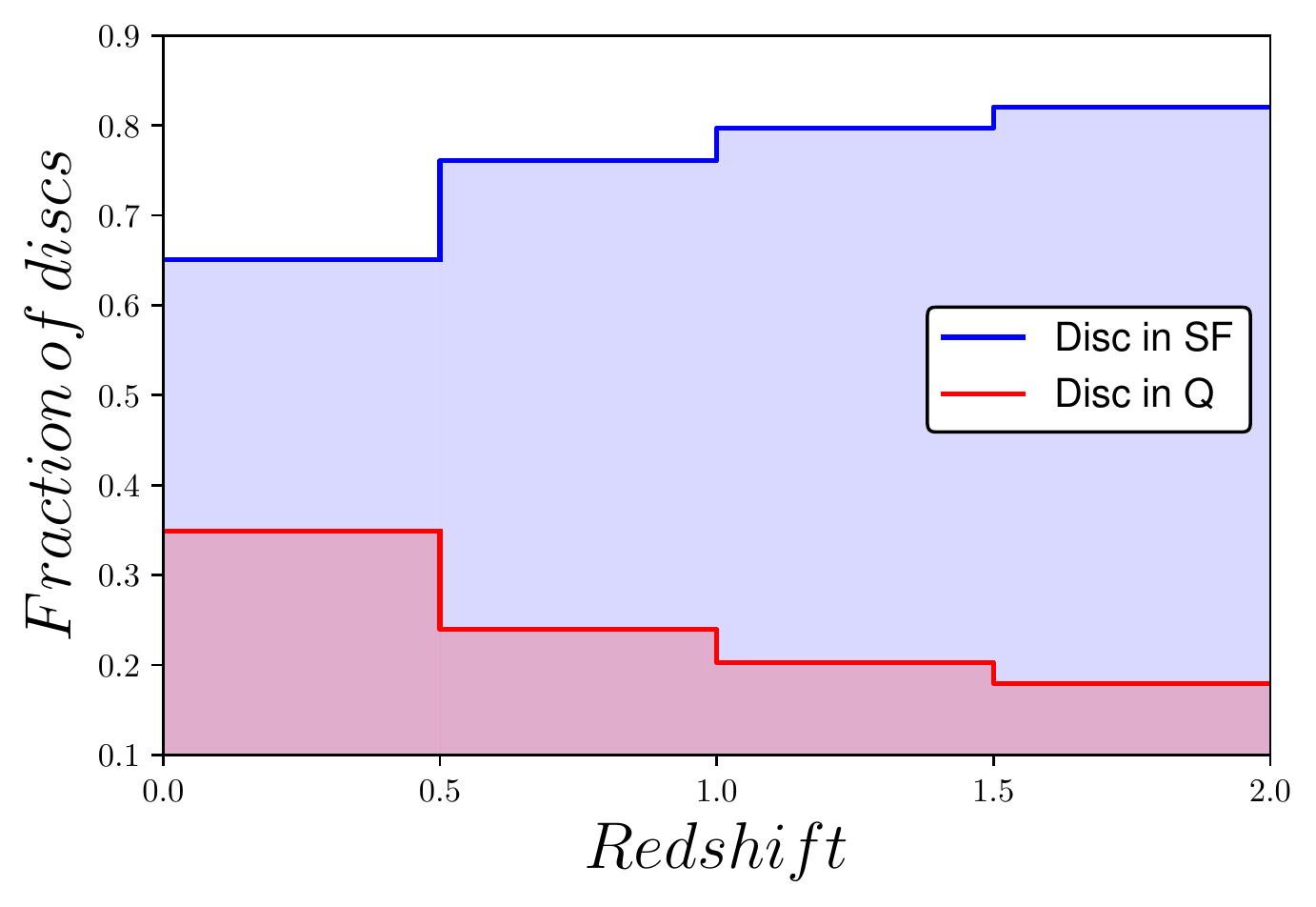} 
\end{array}$
\caption{Fraction of massive ($M_{*,B,D}/M_\odot>2\times10^{10}$) bulges (top panel) and discs (bottom panel) of passive (red line) and star-forming galaxies (blue line). } \label{fig:n_SF_Q}
\end{figure*}
The following sections focus on the analysis of the structural properties of bulges and discs as a function of the integrated star-formation activity of the host galaxy. From the literature it is known that most of the star-forming galaxies tend to have a disc dominated structure while quiescent systems are bulge-dominated (e.g. \citealp{Wuyts2011,Whitaker2017}). 
These results are in agreement with our analysis shown in figure~\ref{fig:n_SF_Q}. The two panels present the fractions of how many bulges and discs are hosted in quiescent or starforming galaxies respectively. The classification is done here using the UVJ rest-frame color of the host galaxies (eg. \citealp{Whitaker2012}).
About $\sim60\%$ of massive bulges are located in passive galaxies while the remaining $\sim40\%$ live in star-forming galaxies. The numbers are inverted at $z>1.5$. Furthermore, a fraction of > $\sim60\%$ of massive discs is hosted in star forming galaxies. We will explore this in more detail in a forthcoming work. However, in the following sections we explore the structural properties of internal components to investigate if and how the buildup of bulges and discs is linked to the the star formation history of the hosting galaxy.

\subsection{Bulges}

\begin{figure*}
\begin{center}
$\begin{array}{c c}
\includegraphics[width=0.45\textwidth]{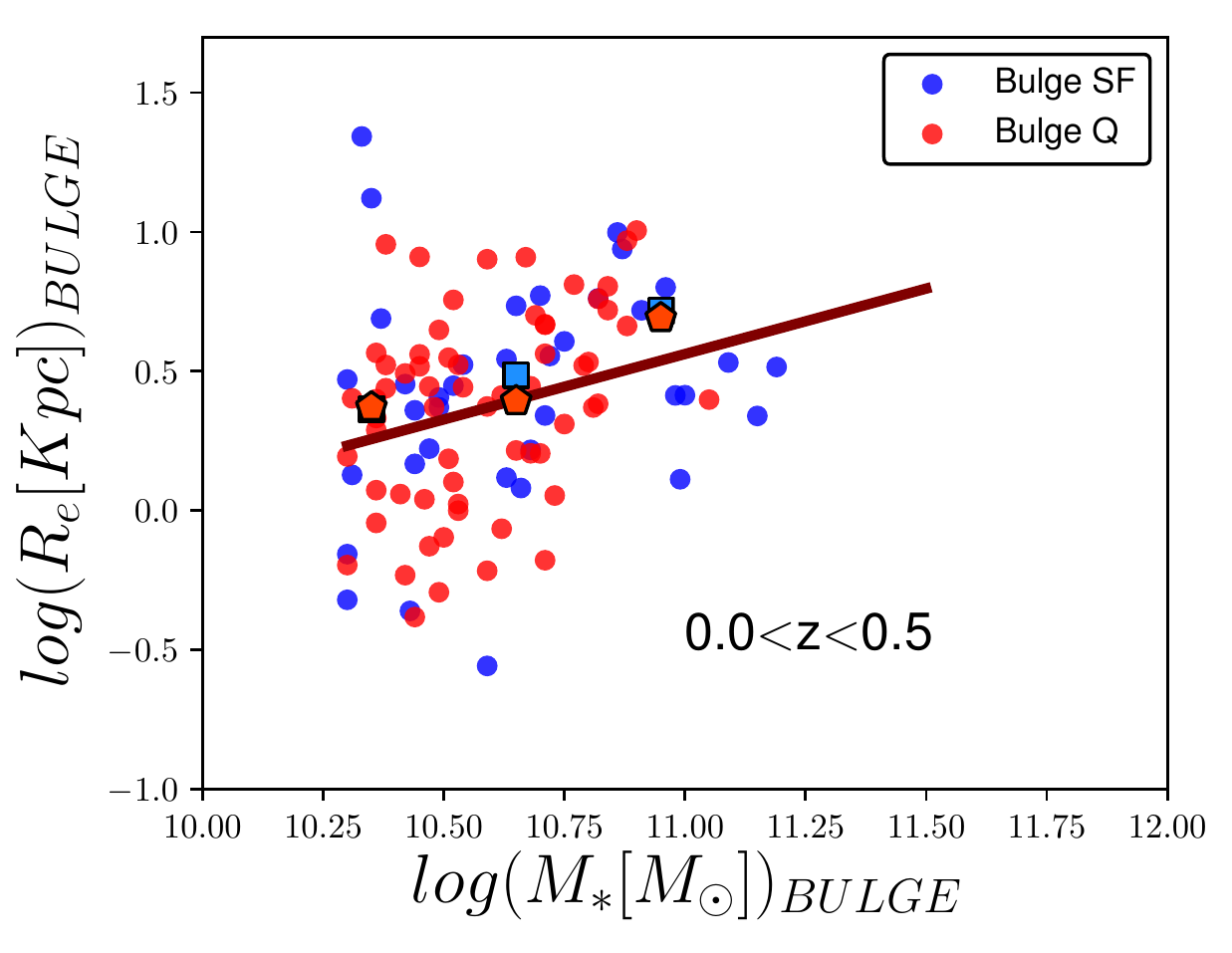} &
 \includegraphics[width=0.45\textwidth]{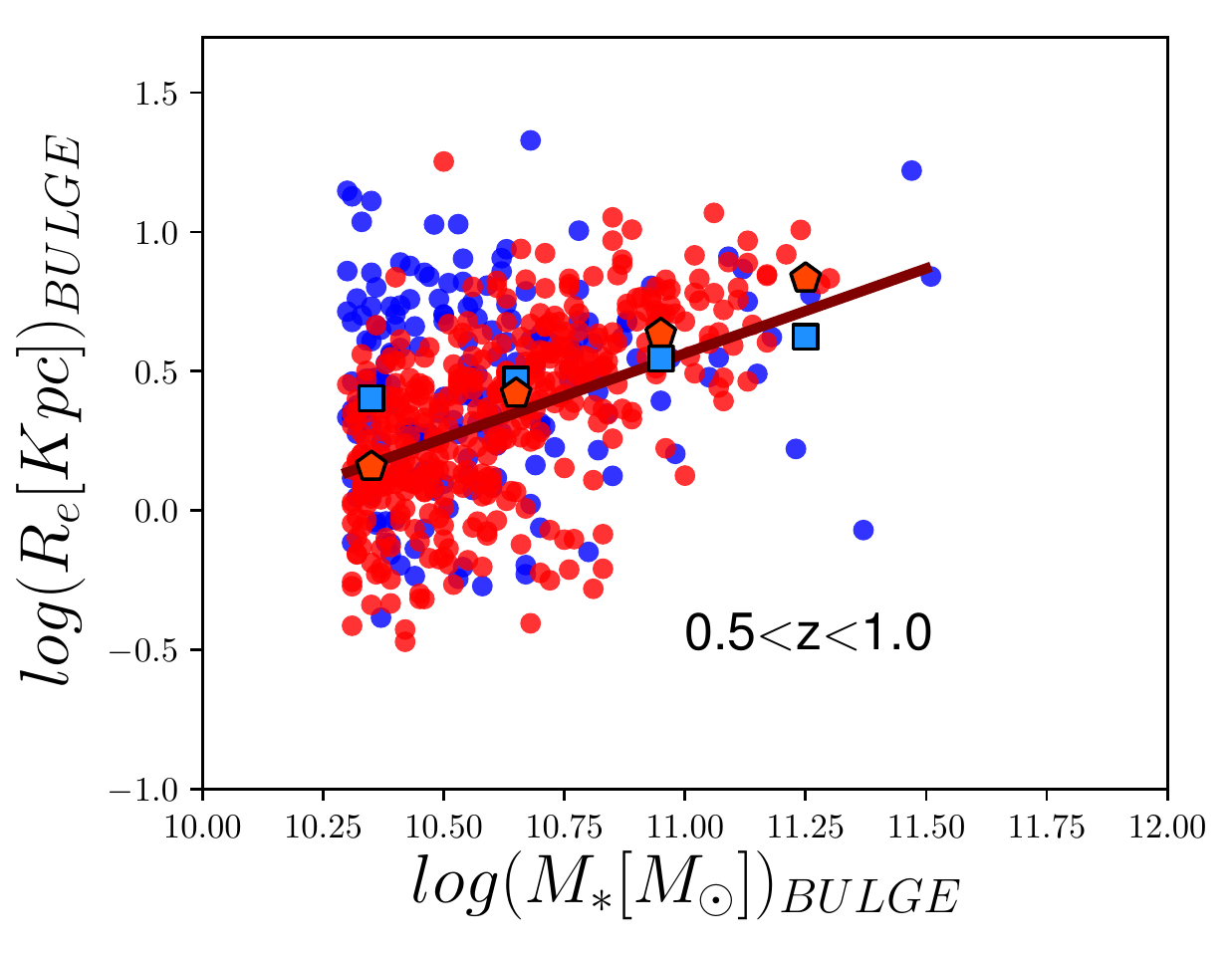} \\
\includegraphics[width=0.45\textwidth]{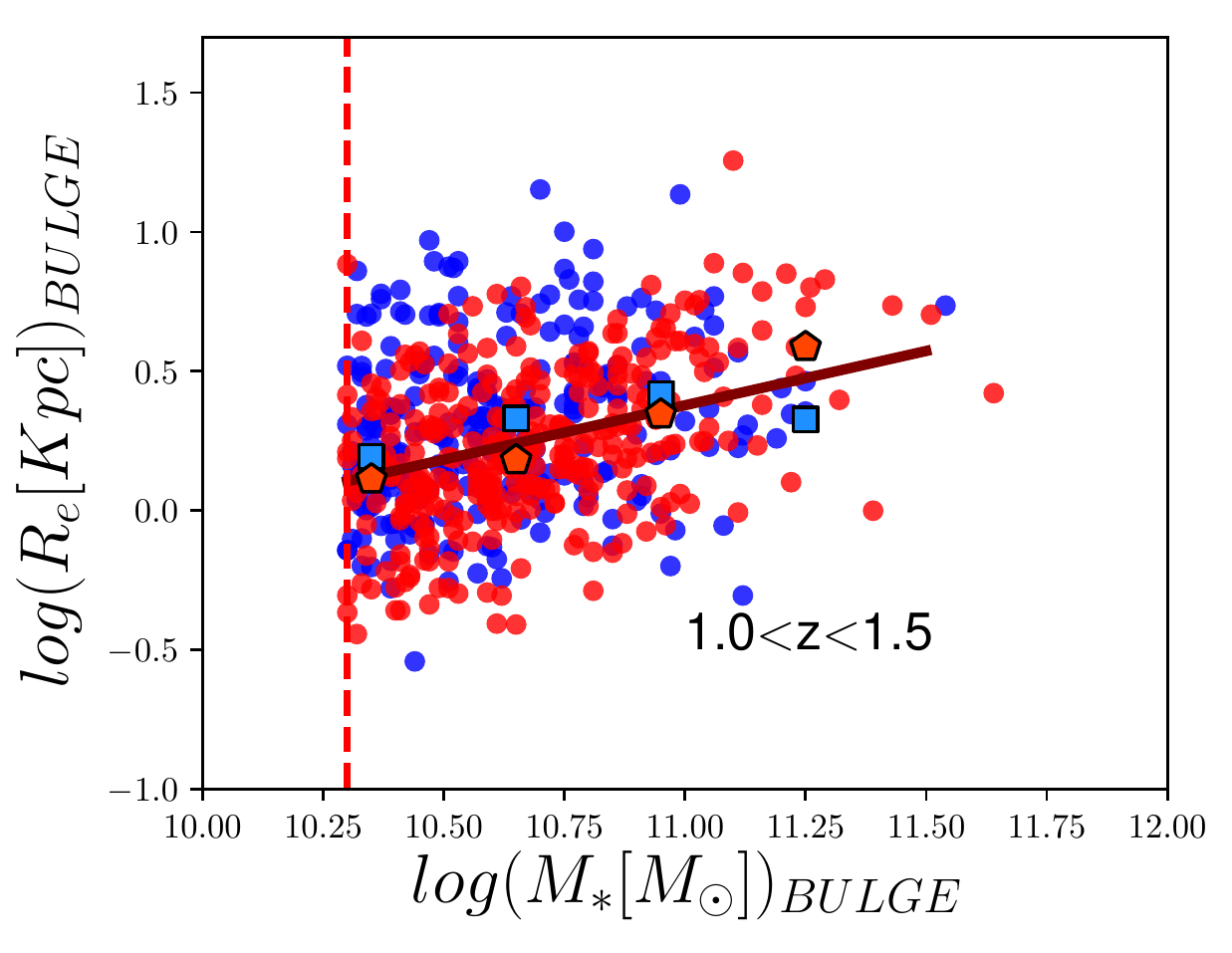} &
 \includegraphics[width=0.45\textwidth]{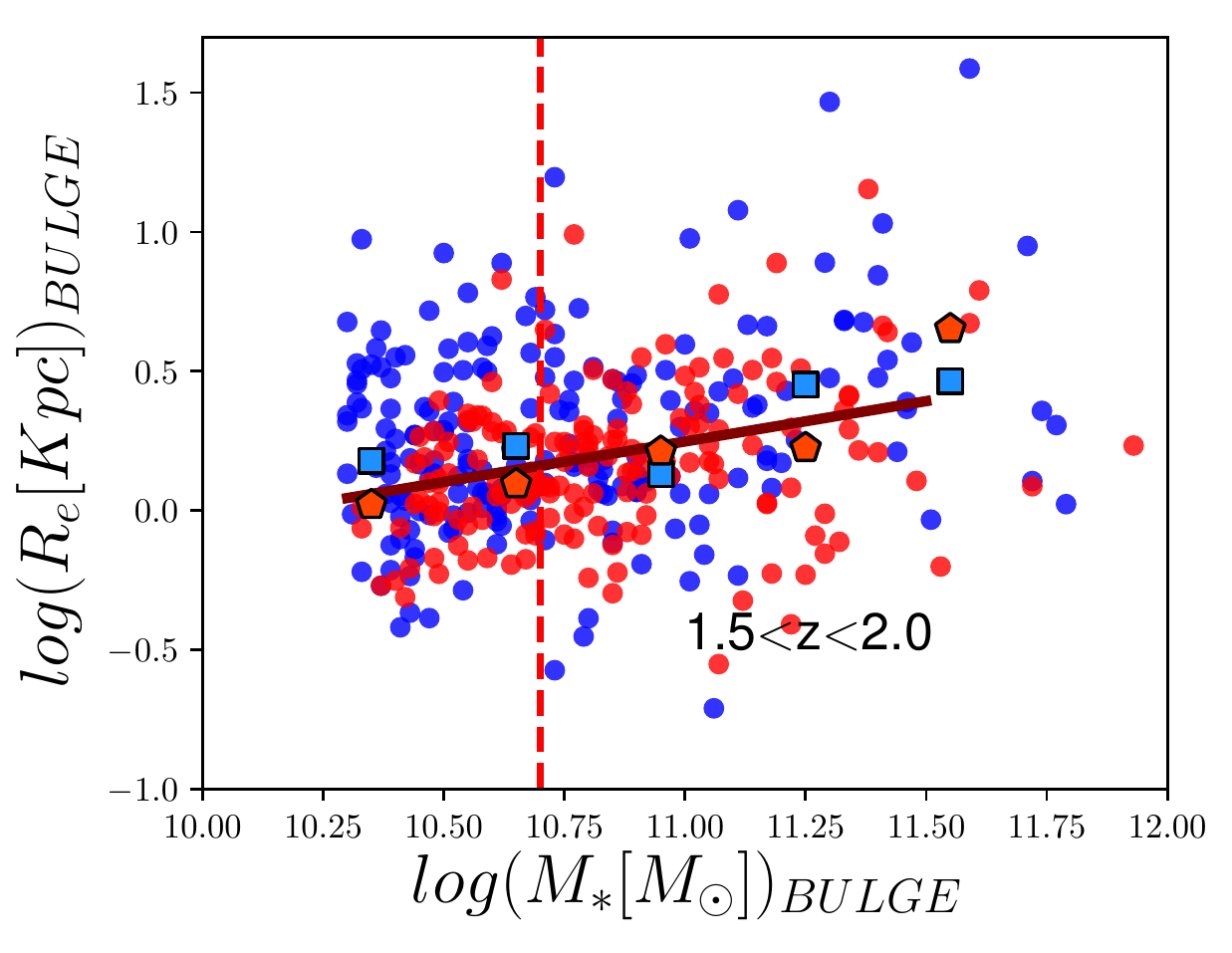} \\
\end{array}$
\caption{Mass-size relation of bulges  in star-forming (blue points) and passive (red points) galaxies. The solid red line is the best fit model to the mass-size distribution of bulges in passive galaxies. Dark red and blue points are the median values of sizes in mass bins for the two populations. The vertical dashed red lines are the stellar mass completeness limit for bulges. As can be seen from the median values a slight difference between bulges in star-forming/passive galaxies is measured and better analyzed in figure \ref{fig:gamma_evol_B_SF}. }
\label{fig:mass_size_B_SQ}
\end{center}
\end{figure*}

\begin{figure}
\includegraphics[width=0.44\textwidth]{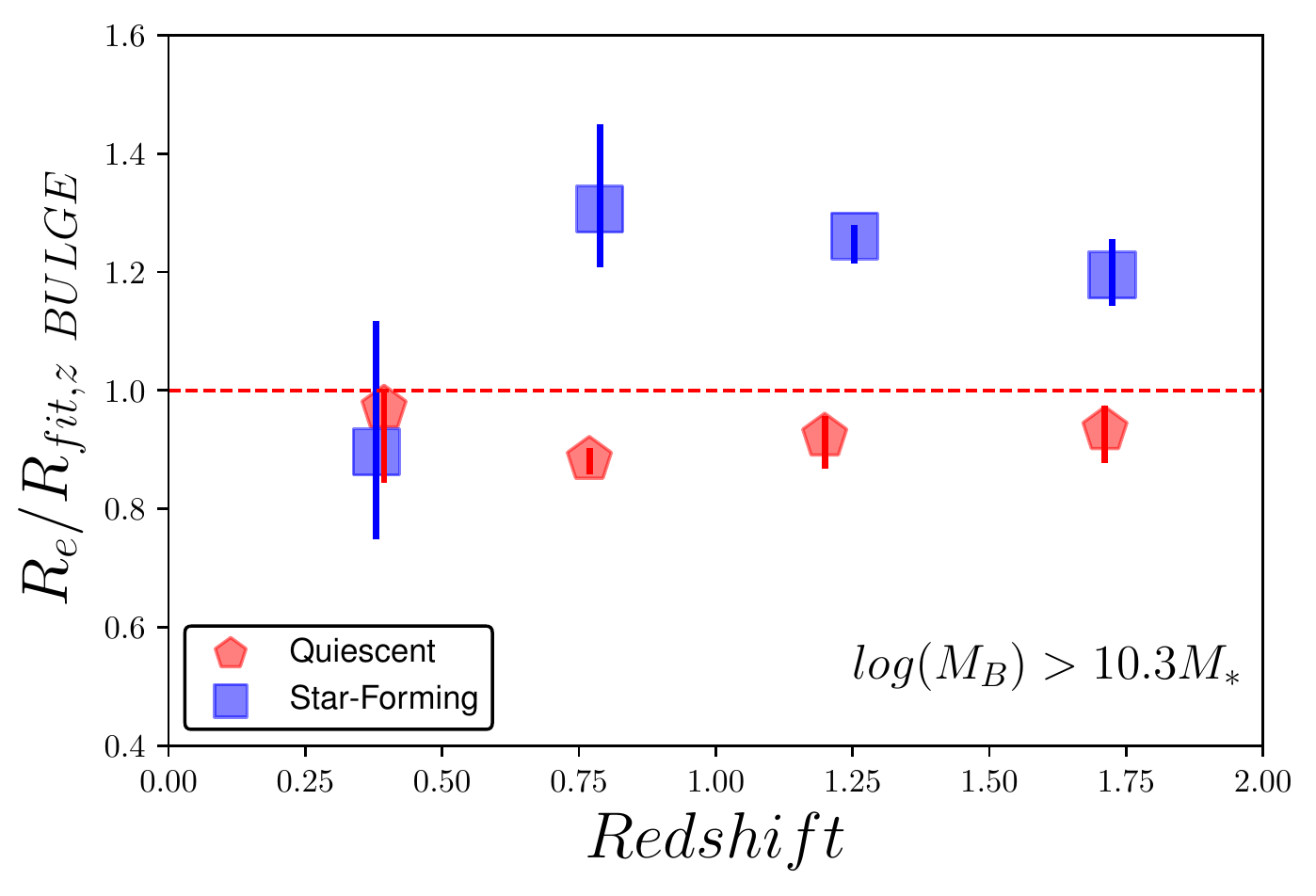} 
\includegraphics[width=0.44\textwidth]{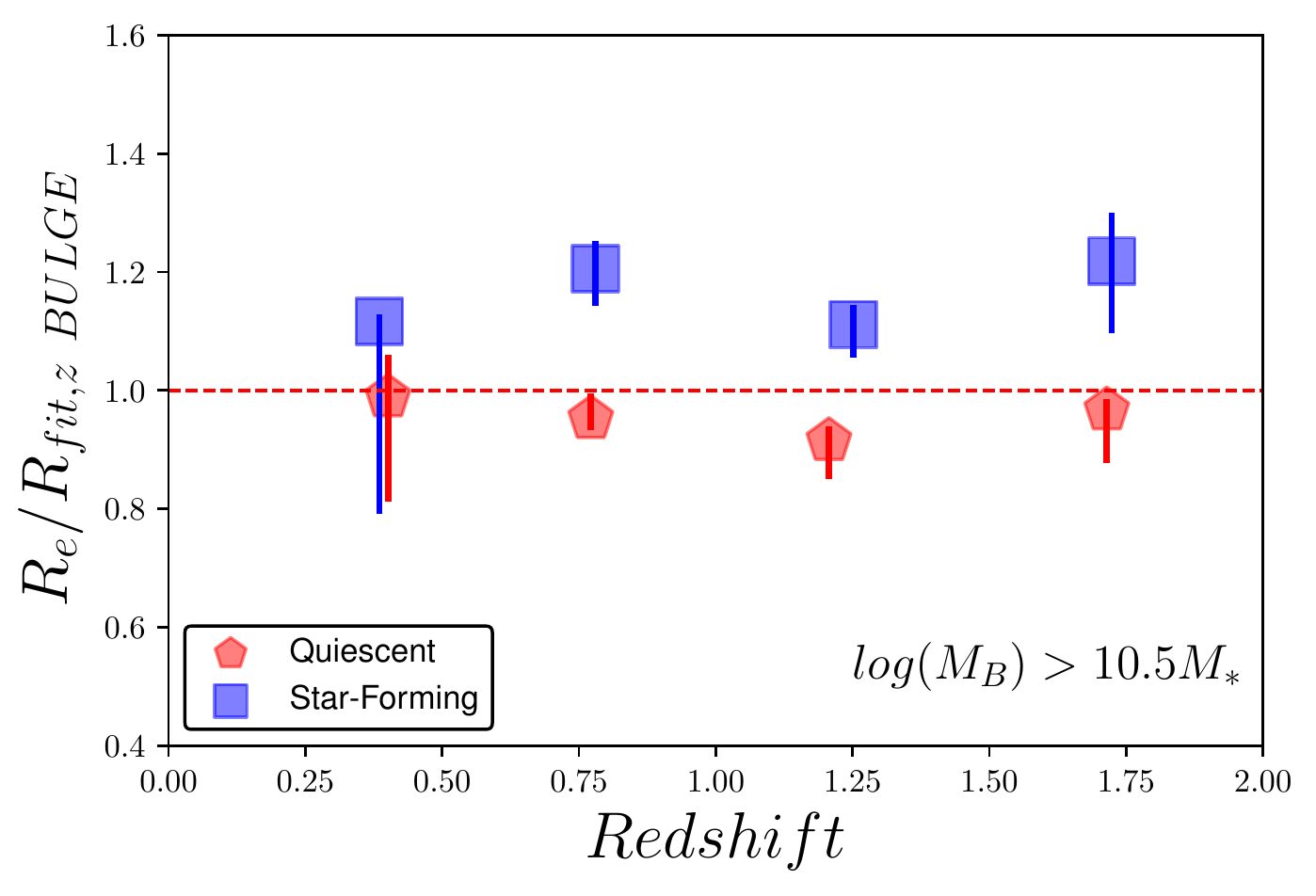} 
\includegraphics[width=0.44\textwidth]{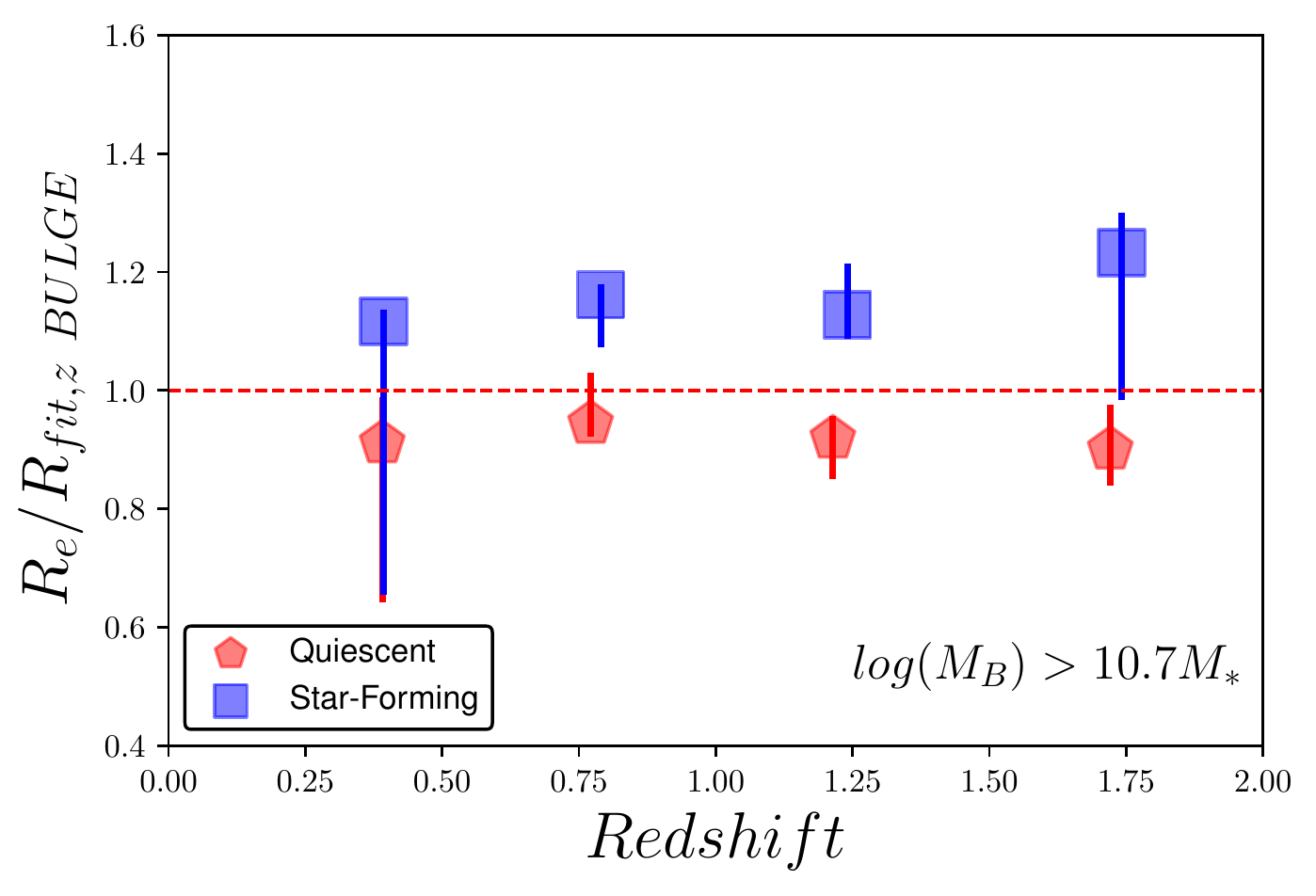} 

\caption{Median sizes of bulges of passive (red squares) and star-forming galaxies (blue squares). For each redshift bin, the effective radius of each bulge is divided by the expected size from the best fit model to the population of bulges of passive galaxies. The median of the ratios is then reported. Errors bars are $68\%$ confidence levels estimated through bootstrapping $1000$ times. The three panels show three different stellar mass thresholds as labeled. }
\label{fig:gamma_evol_B_SF}
\end{figure}

 Figure~\ref{fig:mass_size_B_SQ} shows the mass-size relation for bulges with a color code that indicates whether the galaxy is globally forming stars or not (blue and red points respectively). Blue squares and red circles represents the median values in mass bins. We measure a slight difference in sizes between bulges in star-forming/passive galaxies, at fixed stellar mass. We quantify this measurement, as done in the previous sections. As a reminder, we compute size deviations from the linear best-fit of the mass-size relations (the red line in figure \ref{fig:mass_size_B_SQ}), using the ratio between each measured size and the linear fit. The results are shown in figure~\ref{fig:gamma_evol_B_SF}. Bulges in star-forming galaxies are ~20\% larger than bulges of similar mass in passive galaxies. Systematic measurement errors can amplify the observed difference in size since discs in star-forming galaxies tend to be brighter (even more at high redshift). The separation between the two components cases becomes harder in cases in which the bulge is the fainter component since it component can be artificially enlarged because contaminated by disc light. Even though in DM18 we performed extensive simulations (including a star-forming disc and a passive bulge) to test our fitting procedure, we did not include a self-consistent model to account for the increase in star-formation efficiency of the disc. However, the signal is significant at $2-3\sigma$ and it is systematic at all redshifts except for the lowest redshift bin in which the statistics are poor. Furthermore we do not observe any increase of the trend at higher redshifts which could be expected if our measurements were biased towards bright discs. 
Another source of bias can be the different mass distributions even if partially corrected using our normalization procedure. To estimate the impact of this effect, we perform the same analysis applying an increasing mass threshold. Even though the difference is reduced at large masses ($Log(M_{*,B}/M_\odot>10.7)$), the trend remains. As previously done, we applied the K-S test. The output of the test confirms that sizes of star forming and quiescent bulges follow two different distributions. We discuss the interpretation of these results in section~\ref{sec:disc_Q}. 

\subsection{Discs}
\begin{figure*}
\begin{center}
$\begin{array}{c c}
\includegraphics[width=0.45\textwidth]{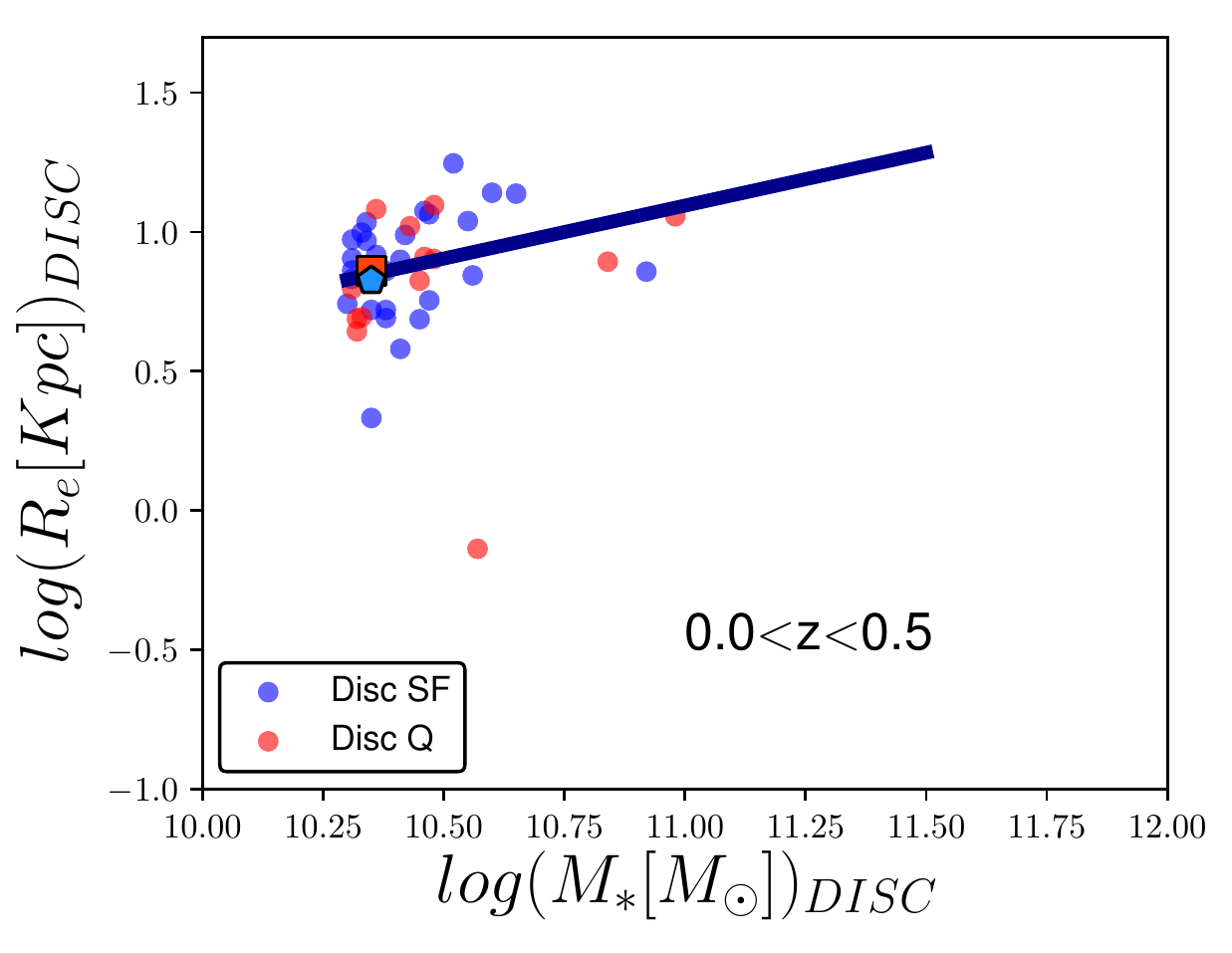} & 
\includegraphics[width=0.45\textwidth]{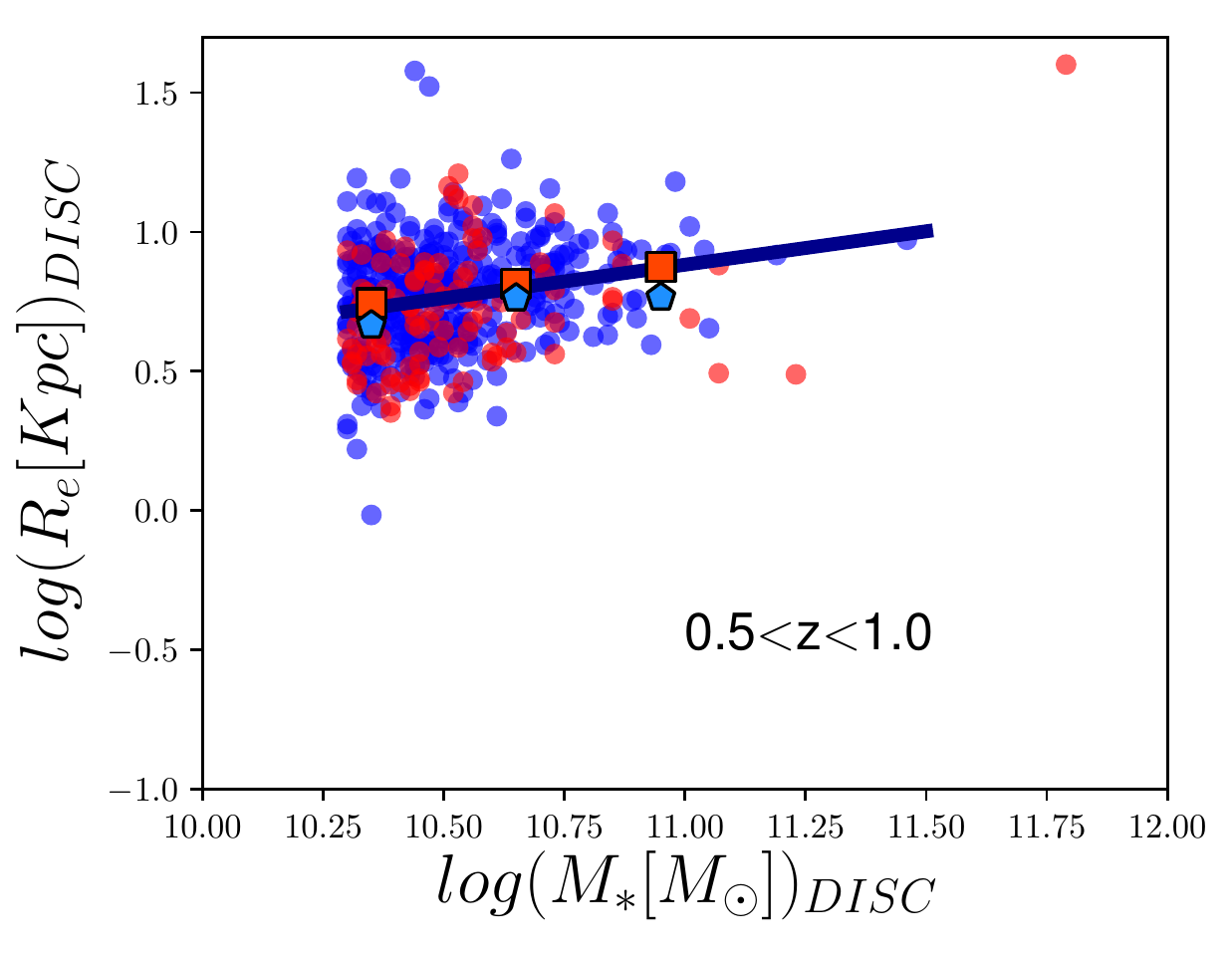} \\
\includegraphics[width=0.45\textwidth]{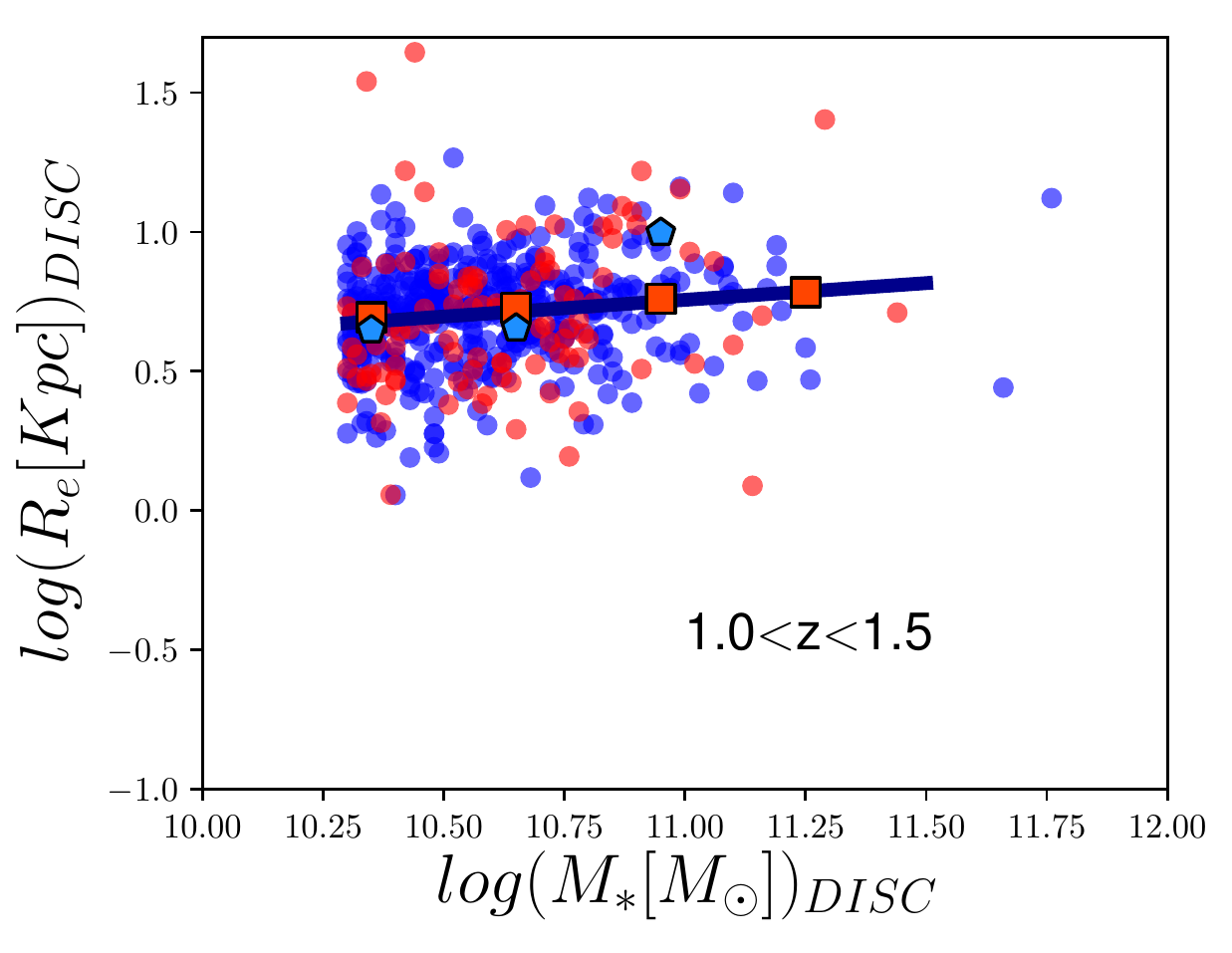} & 
\includegraphics[width=0.45\textwidth]{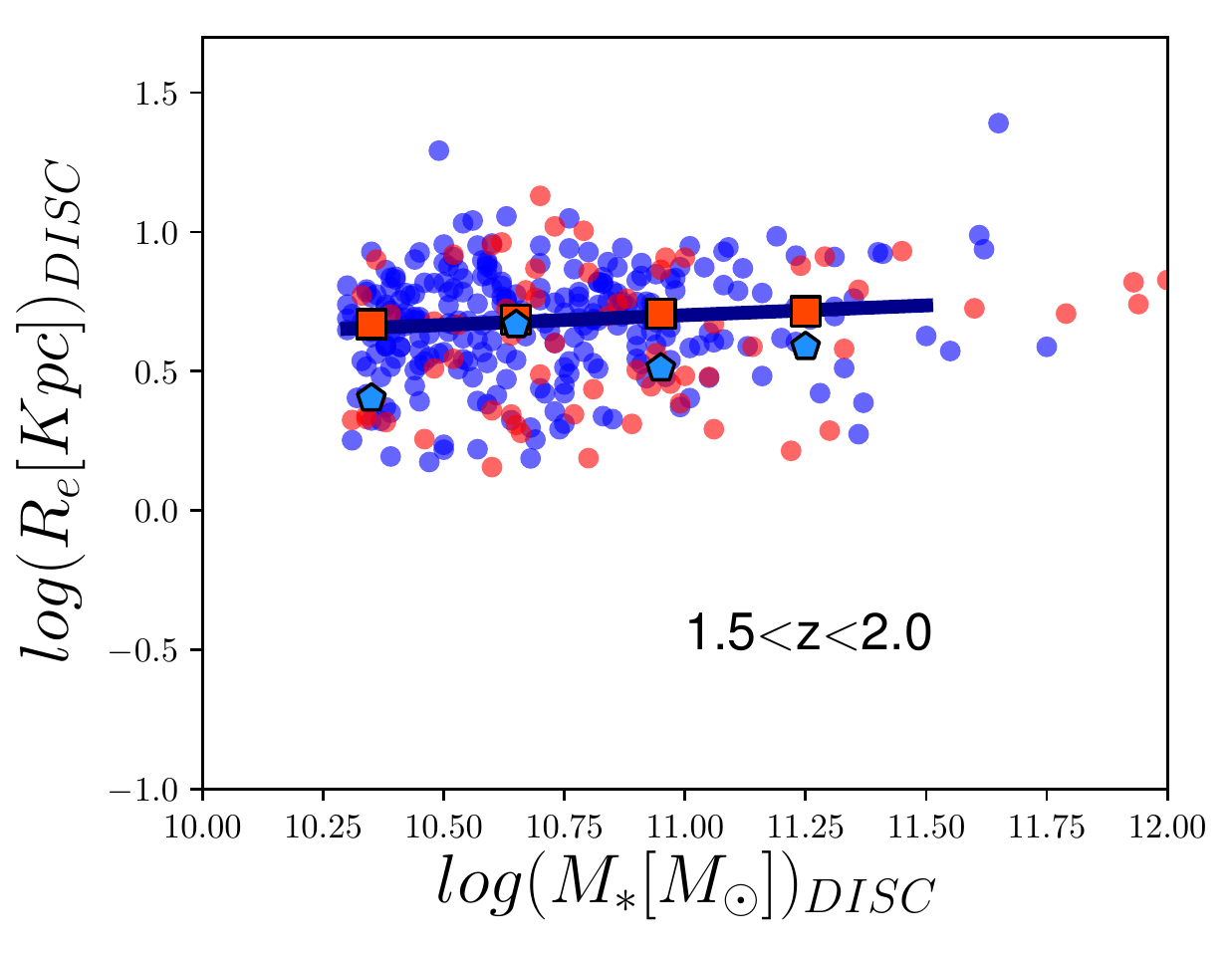} \\
\end{array}$
\caption{Mass-size relation of discs embedded in star-forming (blue points) or passive (red points) galaxies. The solid red line is the best fit model to the mass-size distribution of discs in star-forming galaxies. The vertical dashed red lines are the stellar mass completeness limit for bulges. Dark red and blue points shown the median sizes values of discs hosted star forming and passive galaxies respectively. No clear systematic difference is measured in the size median values for discs in the two populations. Result confirmed in figure \ref{fig:gamma_evol_D_SF}} 
\label{fig:mass_size_D_SQ}
\end{center}

\end{figure*}
\begin{figure}
\centering
\includegraphics[width=0.44\textwidth]{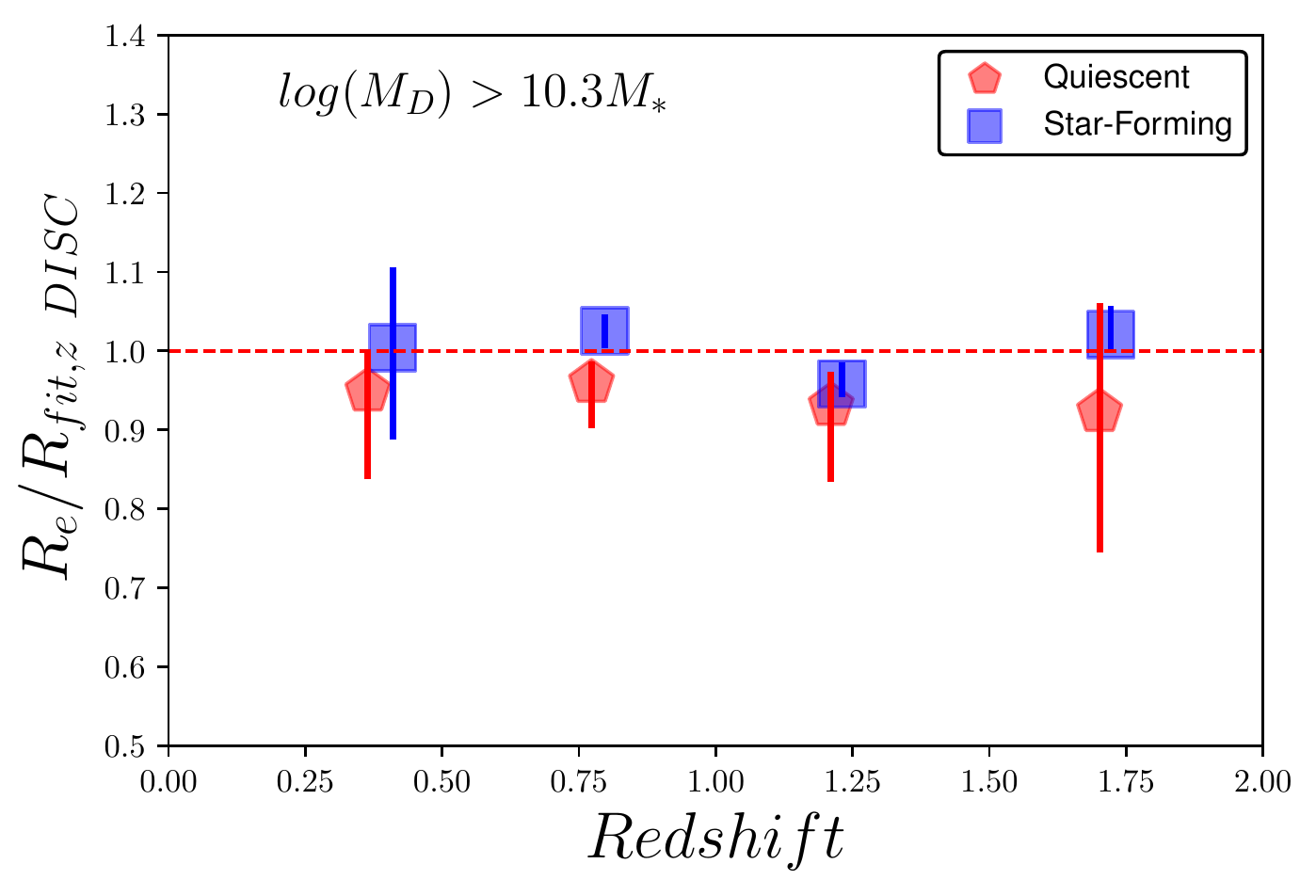} 
\caption{Median sizes of discs ($Log(M_*/M_\odot)>10$) embedded in passive (red squares) and star-forming galaxies (blue squares). For each redshift bin, the effective radius of each disc is divided by the expected size from the best fit model to the population of discs of star-forming galaxies. The median of the ratios is then reported. Errors bars are $68\%$ confidence levels estimated through bootstrapping $1000$ times. Interestingly, disc sizes in both populations are similar. } \label{fig:gamma_evol_D_SF}
\end{figure}

We repeat the same exercise for the disc population. Figures~\ref{fig:mass_size_D_SQ} and~\ref{fig:gamma_evol_D_SF} show the mass-size relations of discs and the size difference between discs of passive and star-forming galaxies respectively.
We find no correlation between the disc radii and the star formation
activity. Overall the sizes of discs in passive and star-forming systems are identical over all the redshift range probed by our data. This goes along with the fact that no relation between the disc structure and the bulge components is measured. 

\section{Discussion}
\label{sec:discussion}


\subsection{The mechanisms of bulge growth}

For the first time, we have explored the morphologies of galaxies hosting massive bulges with the aim of putting constraints on bulge formations mechanisms.

We observe that $\sim40\%$ of massive classical bulges are embedded in disc-dominated galaxies (figure~\ref{fig:nd}). This suggests that either the discs survive bulge formation or are (re)built from the surrounding gas, which ultimately leads to a decrease of the bulge-to-total ratio (\citealp{Hammer2009}, \citealp{Hopkins2009a}) (given the abundance of bulges embedded in discs, both mechanisms have to be relatively common processes). This result is supported by other works that have also found that newly quenched galaxies at $z\sim2$ have disc components (e.g. \citealp{Gargiulo2017,Huertas2016}). 

We find that bulges in galaxies of all morphologies ($B/T>=$0.2) have similar sizes at fixed stellar mass independently of the mass of the disc. This may be due to a variety of processes capable of forming bulges, e.g. mergers (e.g. \citealp{Toomre1977, Hammer2005, Hopkins2010}), disc instabilities (\citealp{Bournaud2016}) or rapid gas compaction (e.g. \citealp{Tacchella2017}), which could wash out clear signatures visible in the mass-size relation.

Moreover, the disc scale length does not depend on the bulge prominence (see figure~\ref{fig:gamma_evol_D}), suggesting a scenario, a mechanism of bulge formation in which the disc component is either not affected or rapidly rebuilt. The presence of a disc does not rule out mergers as main driver for bulge formation in disc systems: several numerical simulations have shown that large discs can survive merger events (e.g. \citealp{Hopkins2009a}).

Several works have speculated on the presence of two main quenching modes (fast and slow). Bulge-dominated galaxies could be formed by more violent events than bulges in disc-dominated systems especially at $z>2$ (e.g. \citealp{Huertas2015, Huertas2016}).
\cite{Barro2013, Barro2017} also introduced a fast and a slow track to build bulges. Dynamical and photometric analysis in the local universe (\citealp{Cappellari2011,Emsellem2011}) have pointed out that the so-called slow rotators, which should be close to our pure bulge dominated population, have different assembly histories than fast-rotators. Namely the slow rotator are believed to follow an assembly history more dominated by (dry) mergers. Our results suggest that different bulge formation/assembly histories, if present, should conspire towards similar size distributions, within the precision of our measurements. 

In semi-analytical models, bulges grow either through mergers or disc instabilities. \cite{Guo2011} computed sizes of bulges, formed through disc instabilities, by transferring a mass equal to the one that is necessary to stabilize the disc. This assumption predicts that bulges produced by disc instabilities (more common in late-type morphologies) are smaller at fixed stellar mass. This is because disc-instability driven bulges are assumed (in \cite{Guo2011} model) to be less efficient in growing the mass and sizes of bulges. We also checked \cite{Shankar2013}, variant of the \cite{Guo2011} model which includes gas dissipation and neglects the orbital energy in mergers (see Figure 1B in \citealp{Shankar2013}).  This model predicts that galaxies that grew their bulges via disc instability, have sizes, at fixed stellar mass, smaller than those of classical bulges, irrespective of their B/T. 

\begin{figure}
$\begin{array}{cc}
\includegraphics[width=0.44\textwidth]{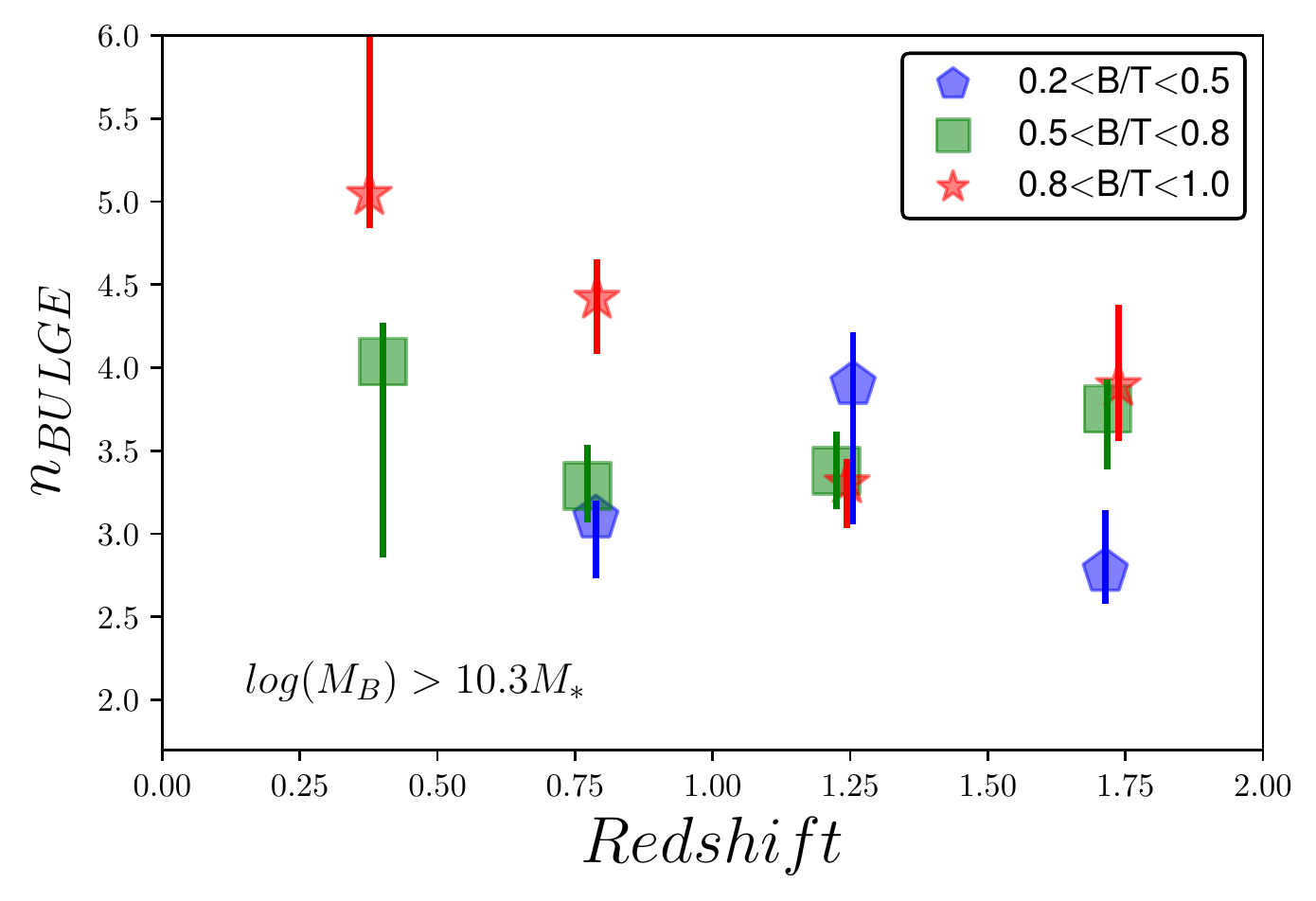} \\
\includegraphics[width=0.44\textwidth]{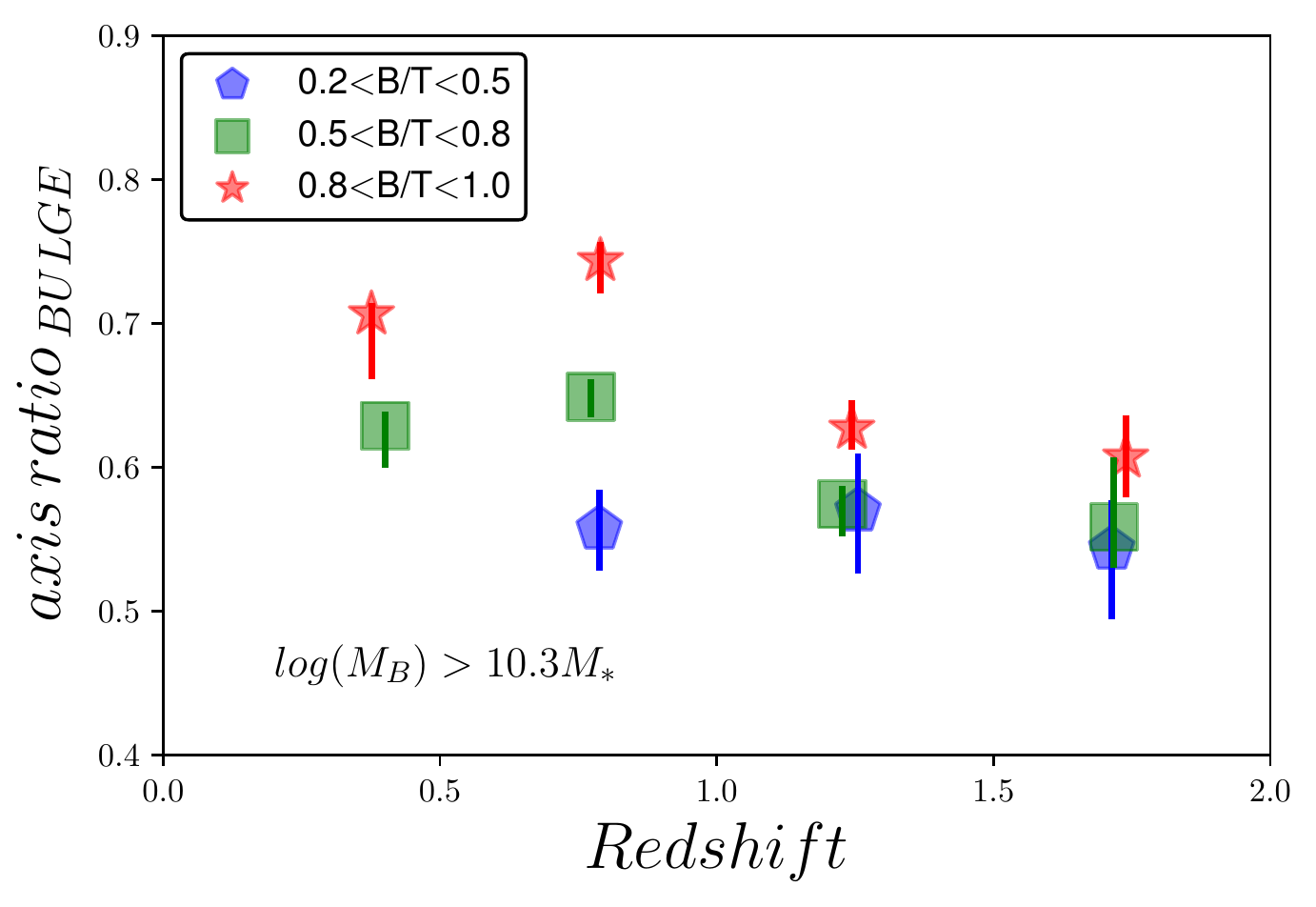}
\end{array}$
\caption{Top panel: Evolution of the median \Sersic index of the bulges more massive than $3\times10^{10} M_{\odot}$  in $B/T$ bins as labeled. Bottom panel: Median values of the axis ratio for the bulge component in redshift bins. Errors bars are estimated through bootstrapping $1000$ times in both cases.}
\label{fig:n_evol}
\end{figure}

In order to get more insights we explore in figure~\ref{fig:n_evol} the median \Sersic indices of the two bulge populations. Numerical simulations predict indeed that the \Sersic index is very sensitive to merger events (e.g \citealp{Nipoti2012}), since it is expected to increase significantly due to dry minor mergers. If galaxies with different morphologies have experienced different merger histories, it might be reflected in the \Sersic index of the bulge component, if not in the size. \\
Figure~\ref{fig:n_evol} shows the redshift evolution of the median \Sersic index of the bulge component (taken from the 5000 \AA\, rest frame model) for bulges with $log(M_{*,B}/M_\odot)>10.3$. The average \Sersic indices is always larger that $\sim2$, consistent with being \emph{classical bulges}. Moreover, the \Sersic values are compatible for a large range of morphologies. However, we measure a trend for larger \Sersic values as well as  stronger evolution, for pure bulges ($B/T>0.8$), in agreement with the previous findings by \cite{Shankar2018}. We also find that pure bulges are marginally rounder ($b/a\sim0.8$ vs. $b/a\sim0.6$) than bulges hosted by galaxies with $B/T<0.8$. This could be an indication of a more merger dominated history for these galaxies. 

\subsection{The effect of quenching on the internal structure}
\label{sec:disc_Q}

The second key question our results can put constraints on is how bulge growth is related to the shutdown of star formation. 
The analysis of structural properties and scaling relations provide some clues in this direction. In Dimauro et al. (2019, in prep.) we will explore the topic in more detail.\\
\\
We found that bulges in star-forming galaxies (at fixed stellar mass), are slightly larger than in passive systems (figure~\ref{fig:gamma_evol_B_SF}), while the disc component shows no dependence with the star-formation activity (figure~\ref{fig:gamma_evol_D_SF}). At the same time we found no difference in the size of bulges in galaxies of different B/T (morphologies). These results seem to be in contradiction. However, they only reflect the non-perfect match between morphology and star-formation. There is even a population of pure blue bulges as pointed out by previous works (e.g. \citealp{Barro2013}) as well as passive discy systems (e.g. \citealp{Huertas2016}) in the CANDELS survey. The exact fractions and abundances will be discussed in a forthcoming work.

Our finding  suggests that in our sample of mostly field galaxies, any structural changes that could result from quenching processes are reflected on the bulge component rather than in the disc. 
Mechanisms like strangulation (e.g. \citealp{Peng2015, Carollo2016}) can predict such a signature. This shows very clearly that morphologies (B/T) and star formation activity are not interchangeable properties. Not all star forming galaxies are disc-dominated and not all passive galaxies are bulge-dominated. While the paradigm of a bimodal distribution breaks down in dense environments (\citealp{Kuchner2017}), we clearly also find a signature of this in the field sample used here. In clusters, discs are slightly smaller in quiescent galaxies than in star forming systems (\citealp{Kuchner2017}). The environmental quenching, offers a number of mechanisms that are more relevant for this class of galaxies than for the field counterpart. 

However, assuming that the star-forming galaxies are the progenitors of the passive one with similar mass, our results of figure~\ref{fig:gamma_evol_B_SF} suggest that an additional increase of the central density, after the bulge formation, would be required during or after quenching to explain the different sizes. This signature is compatible with the \emph{blue nugget} galaxies, passing trough a compaction phase before quenching (\citealp{Tacchella2017}).
\begin{figure}
$\begin{array}{cc}
\includegraphics[width=0.44\textwidth]{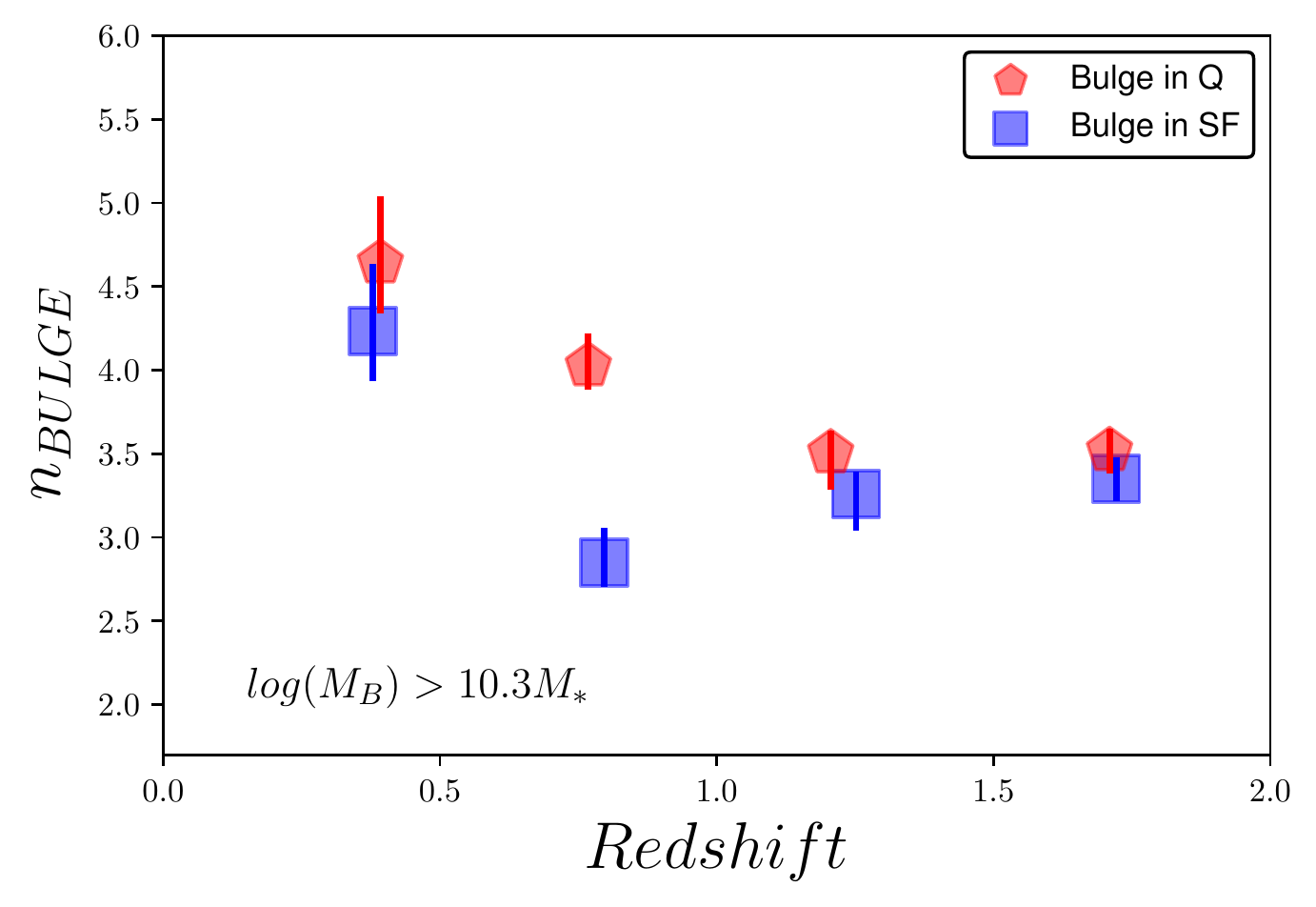} \\
\includegraphics[width=0.44\textwidth]{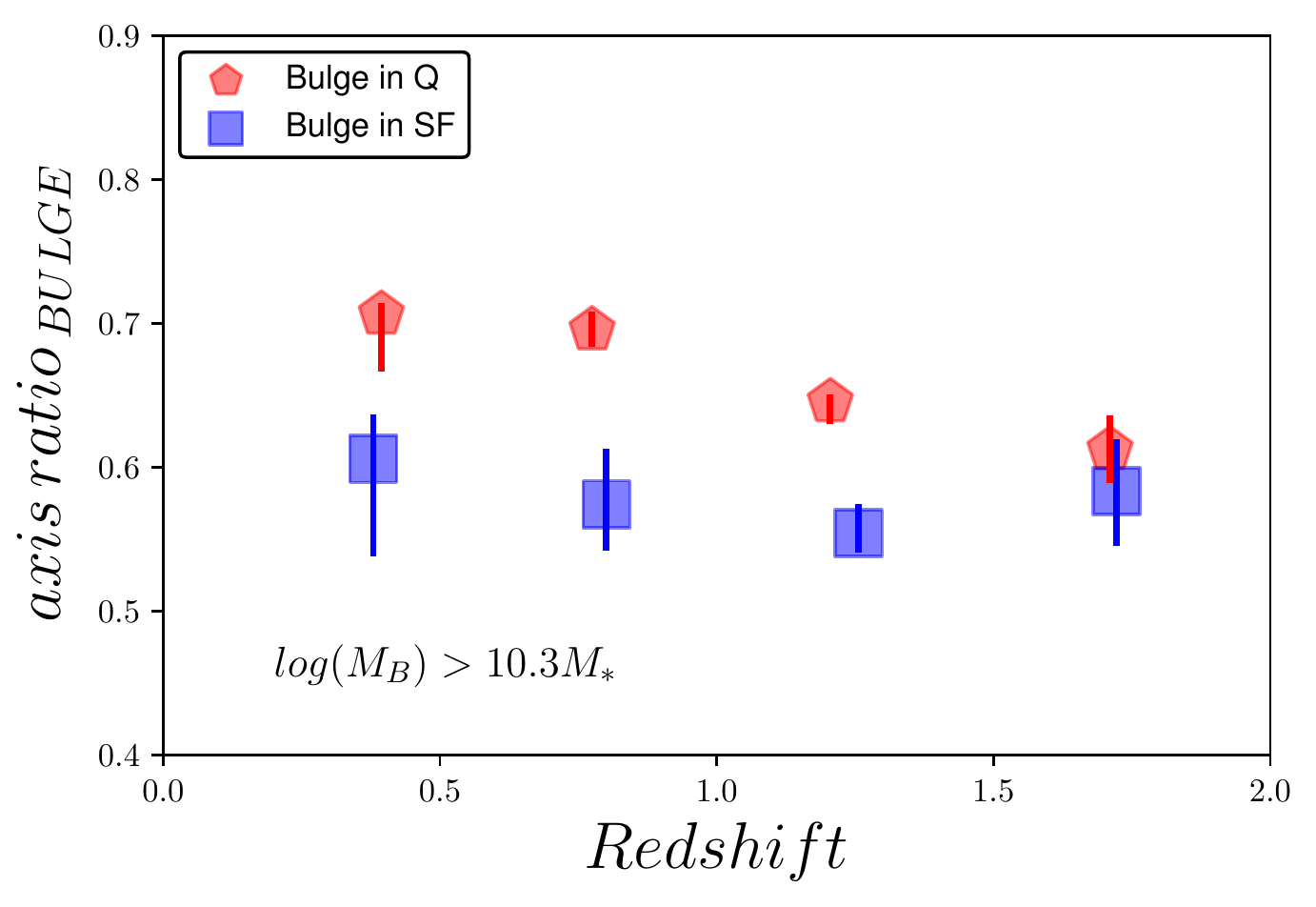}
\end{array}$
\caption{Top panel: Evolution of the median \Sersic index of the bulges more massive than $3\times10^{10} M_{\odot}$  in star forming and quiescent galaxies as labeled. Bottom panel: Median values of the axis ratio for the bulge component in redshift bins. Errors bars are estimated through bootstrapping $1000$ times in both cases.}
\label{fig:n_evol_SF}
\end{figure}

 The measurement is also unavoidably affected by progenitor bias. Bulges in passive galaxies at a given redshift are most probably the descendants of star-forming galaxies at higher redshifts (\citealp{Carollo2013, Lilly2016}). Since, as shown in table~\ref{tbl:fit_mass_size_bulge} the effective radii of bulges increase by a factor $\sim2.5$ from $z\sim2$, those high redshift bulges in star-forming galaxies were smaller and therefore the difference we measure can reflect different formation times instead of morphological transformations. In other words, bulges in star-forming galaxies are formed at later epochs than bulges in passive galaxies, at the epoch of observation. Analysis of the stellar mass density of $z\sim2$ massive galaxies (\citealp{Tacchella2015, Tacchella2017}), show indeed that the central cores are already in place at this epoch. This result would be still compatible with the fact that no size difference is observed in the disc component of passive and star-forming galaxies given the mild size evolution of discs in the redshift range probed by our work. 

Figure \ref{fig:n_evol_SF} shown the evolution of the median of the \Sersic index and the axis ratio. While the \Sersic index progressively increases through cosmic time, no significant difference detected between the two populations of bulges. This can be again referred back to the non-perfect overlap between morphology and star formation. However, bulges in passive galaxies appear rounder than the bulge hosted in star-forming systems. This trend is in agreement with our results on the sizes. Indeed, the size difference we measured is also the results of the fact that bulges in star forming galaxies are more elongated. 

Determining the ages of bulges can certainly provide invaluable information to disentangle between different scenarios and also shed light on the nature of the difference. This is the main purpose of our future work with the present dataset.

\section{Summary and conclusions}
\label{sec:conclusion}

We have analyzed the mass-size relation of a complete sample of bulges and discs more massive than $2\times10^{10}$ solar masses from $z\sim2$ in the CANDELS fields. We used the bulge-to-disc decompositions from DM18. To our knowledge, this is the first systematic analysis of the structural properties of a mass selected sample of bulges and discs at high redshift with such a big sample. \\
 
Our main results are:
\begin{itemize}
\item Star-forming galaxies and discs follow similar mass-size relations while bulges show a less steep relation than quiescent galaxies.
\item $\sim40\%$ of massive bulges are surrounded by discs. We interpret this as a signature that disc survival/regrowth after bulge formation is a common process.
\item Bulges and discs follow different mass-size relations at all redshifts. The relation for discs has a typical slope of $\alpha\sim0.2$ with a slight decrease at high redshift. At $z\sim2$ they are $30\%$ smaller than today. Bulges follow a mass-size relation with a typical slope of $\alpha\sim0.5$
and an intrinsic scatter of $\sim0.2$. The zeropoint increases by a factor of 2.5 from $z\sim2$. 
\item  The mass-size relation of bulges and discs weakly depends on the morphology of the host galaxy (probed by the $B/T$ ratio). Indeed, bulges and discs hosted by  galaxies with $B/T<0.5$ have consistent sizes within the 15-20\% ($\pm$ 2-$\sigma$) and a similar median \Sersic index for the bulge (for a given stellar mass). It suggests that the mechanisms of bulge growth are similar for most massive galaxies. 
\item Pure bulges ($B/T>0.8$) have similar sizes than bulges surrounded by discs ($0.2<B/T<0.8$). Despite that, they are marginally rounder ($b/a\sim0.8$ vs. $b/a\sim0.6$) and have larger \Sersic values compared to bulges hosted by galaxies with lower B/T. An assembly history more dominated by merger can explain these trends.
\item The size of the disc component weakly depends on the bulge-to-total ratio. It suggests that the disc structure is not affected by the central bulge. However, discs, hosted by galaxies with $B/T>0.5$, have larger sizes compared to pure disc systems. This result is in agreement with theoretical prediction. At a given disc mass, higher B/T implies a larger total mass and thus larger halo mass. Hence, the disc size is larger since it is directly proportional to the virial radii.   
\item Bulges of star-forming galaxies are found to be $\sim20\%$ larger than same stellar mass bulges of quiescent systems. This signal can be an indication that additional central \emph{compaction} is required after or during quenching, although it can also be a consequence of a different formation epoch without morphological transformation (progenitor bias). 
\item The size of the disc component at fixed stellar mass does not show strong dependencies with the star formation activity of the host. Sizes are compatible within $\pm$ 2-sigma range of 12-15\%. This result suggest that the quenching does not affect the disc structure when it survives.
\end{itemize}

\section*{Acknowledgments}
This work has been essentially funded by to a French national PhD scholarship. MHC is also thankful to Google for their unrestricted gift to explore deep-learning techniques for galaxy evolution.
PGP-G acknowledges support from the Spanish Government grant AYA2015-63650-P.
FS acknowledges partial support from a Leverhulme Trust Research Fellowship

\bibliographystyle{mnras}
\bibliography{/biblio}

\appendix
\section{Mass-size fit: the contamination factor}
\label{sec:appendix1}

The mass-size fit method from \cite{vanderWel2014} includes possible misclassification between early and late type galaxies. This is done applying a contamination factor (C) that takes into account of misclassification probability. 
$$L_{ET}=\Sigma\,ln[W((1-C).P_{ET} + C.P_{LT}  +0.01)] $$
$$L_{LT}=\Sigma\,ln[W((1-C)P_{LT} + C.P_{ET}  +0.01)] $$

\begin{figure*}
\begin{center}
$\begin{array}{c c}
\includegraphics[width=0.4\textwidth]{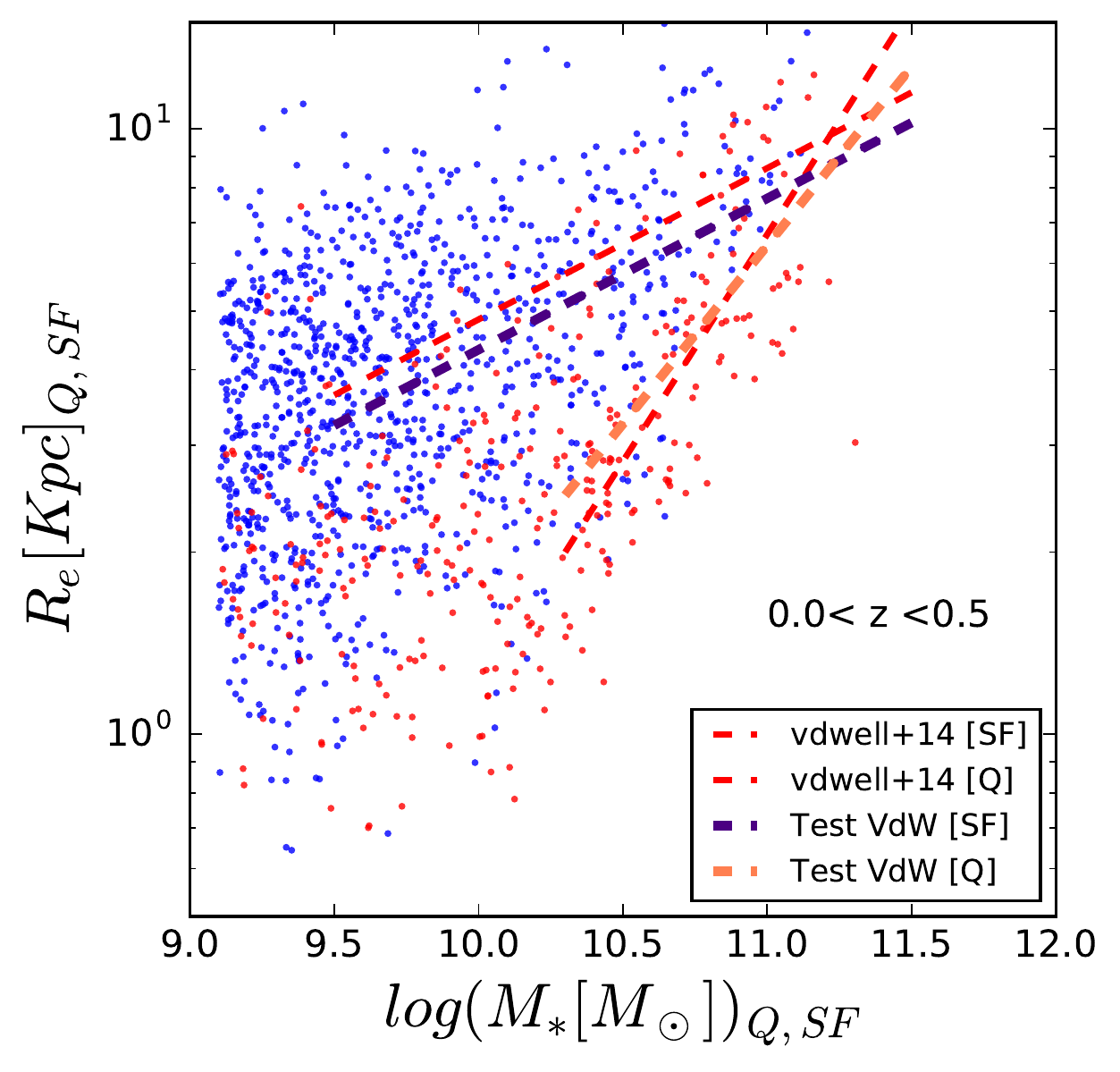} & \includegraphics[width=0.4\textwidth]{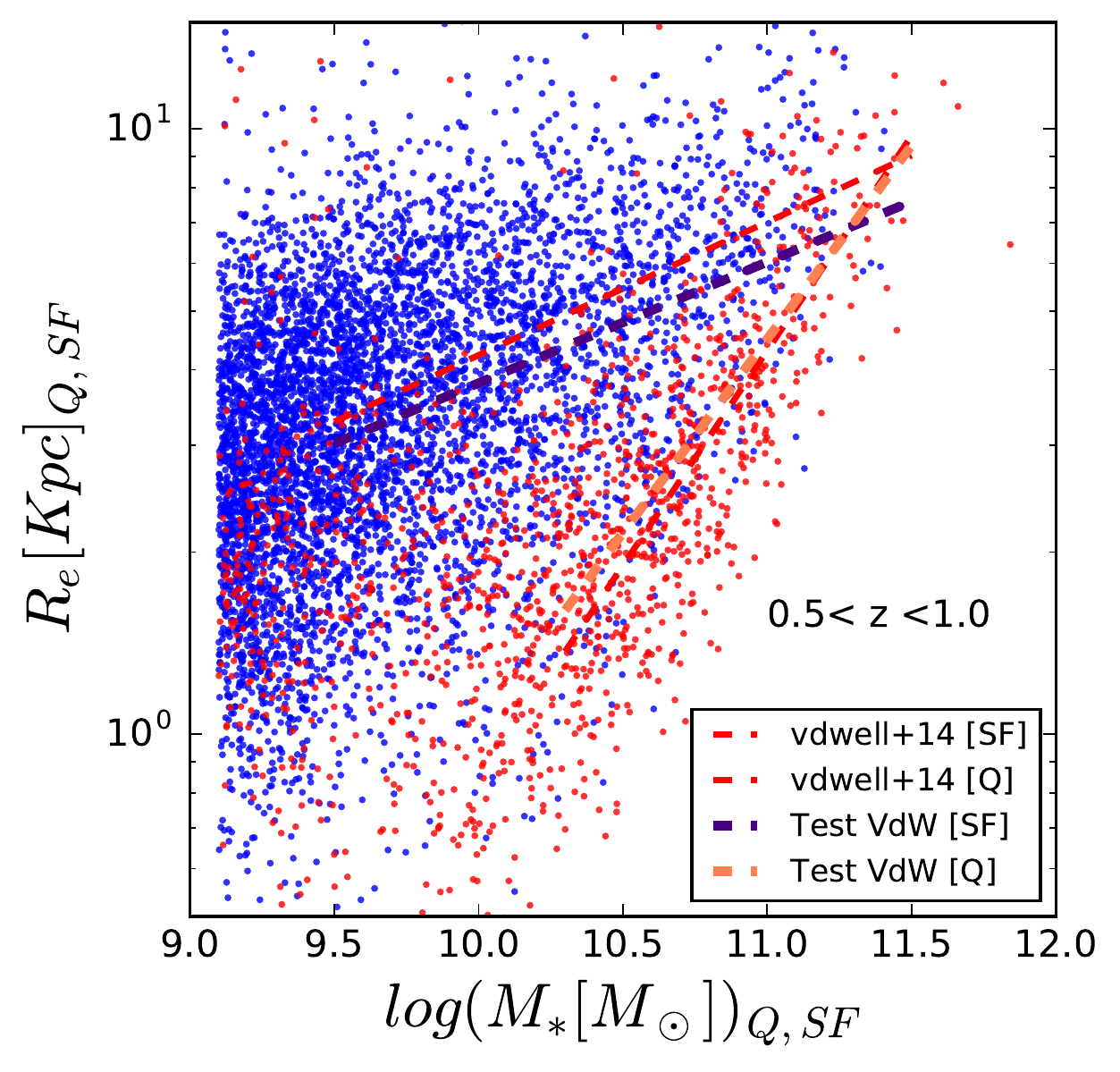} \\
\includegraphics[width=0.4\textwidth]{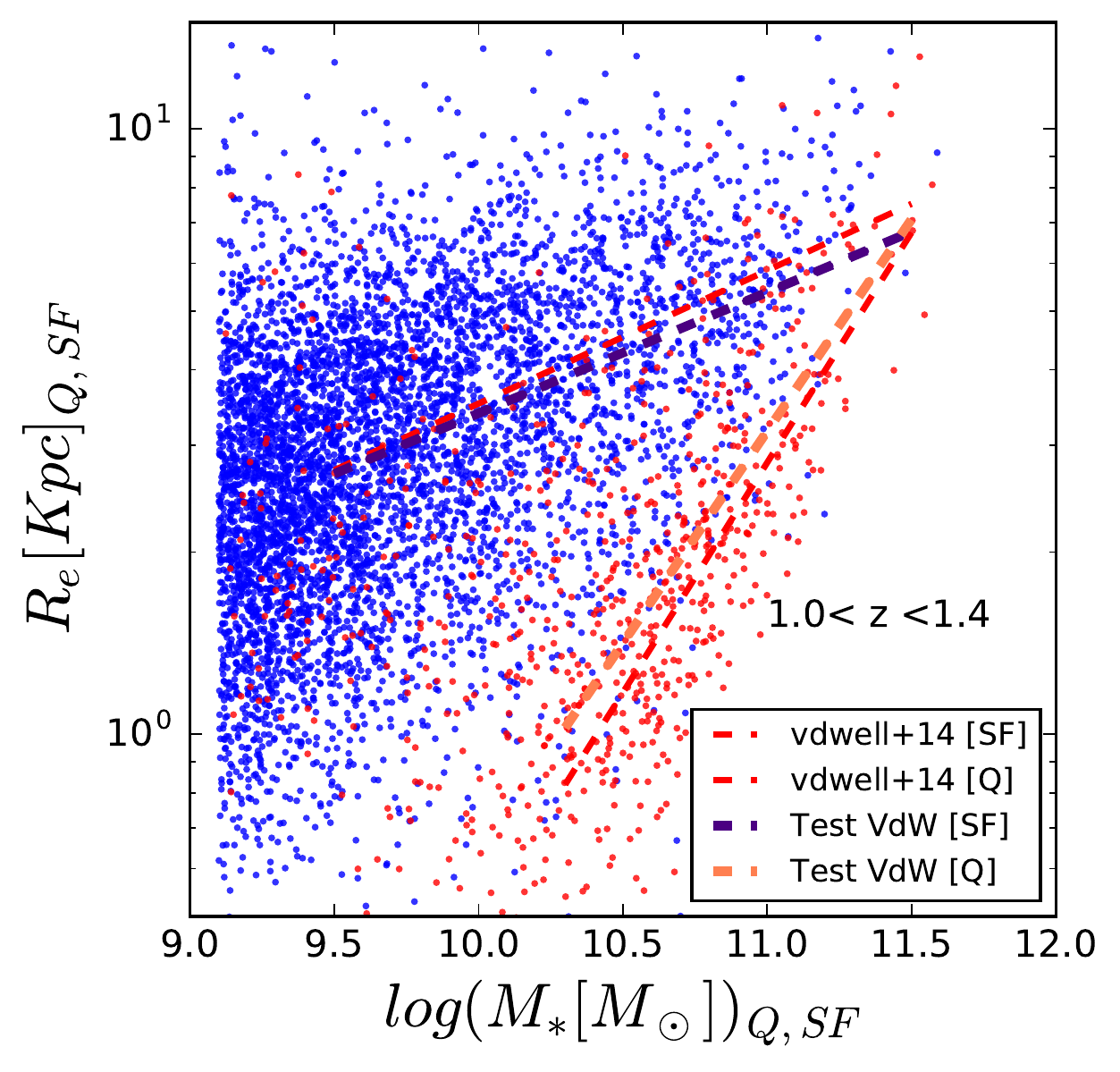} & \includegraphics[width=0.4\textwidth]{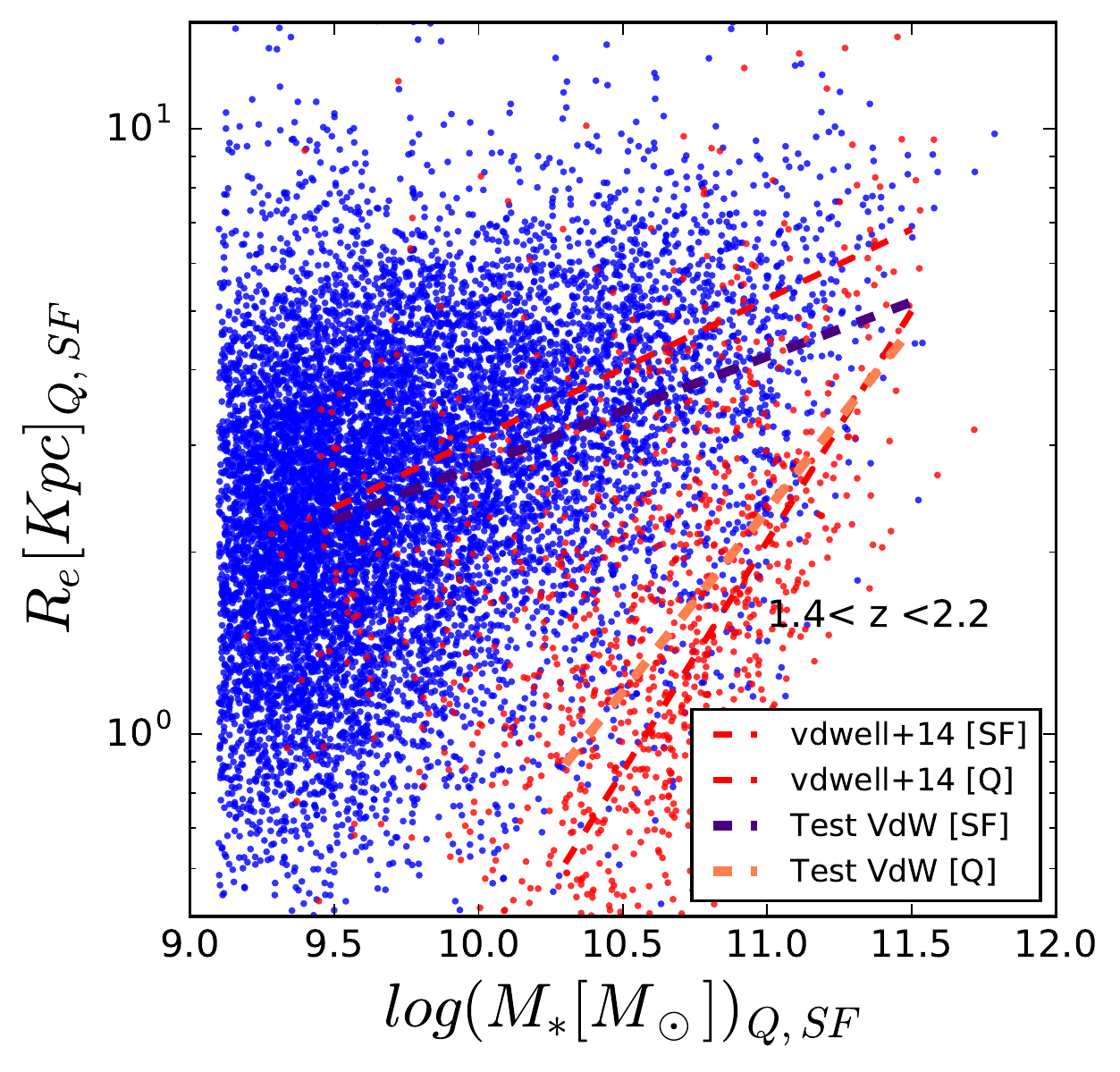} \\
\end{array}$
\caption{Mass-size relations of star forming and quiescent galaxies done using the Van der Well 2012 catalog. The red dotted lines are the best fit from Van der Well 2014, while the blue and orange dotted lines are the results from our fit method that does not take into account of the contamination factor.}
\label{fig:test_mass_size_VdW}
\end{center}
\end{figure*}

Moving from the early-late type galaxies to the bulge-disc populations, the same correction cannot be applied. For that purpose we tested the impact of this correction applying the mass-size fit without this factor. The sample used is selected from the \cite{vanderWel2014} catalog, using the same selection as it is done in his work. Results are shown in figure: \ref{fig:test_mass_size_VdW}. The mass-size fits are compatible. 

\section{Similarities and differences between the two catalogs}
\label{sec:appendix2}
\begin{figure*}
\begin{center}
$\begin{array}{c c}
\includegraphics[width=0.44\textwidth]{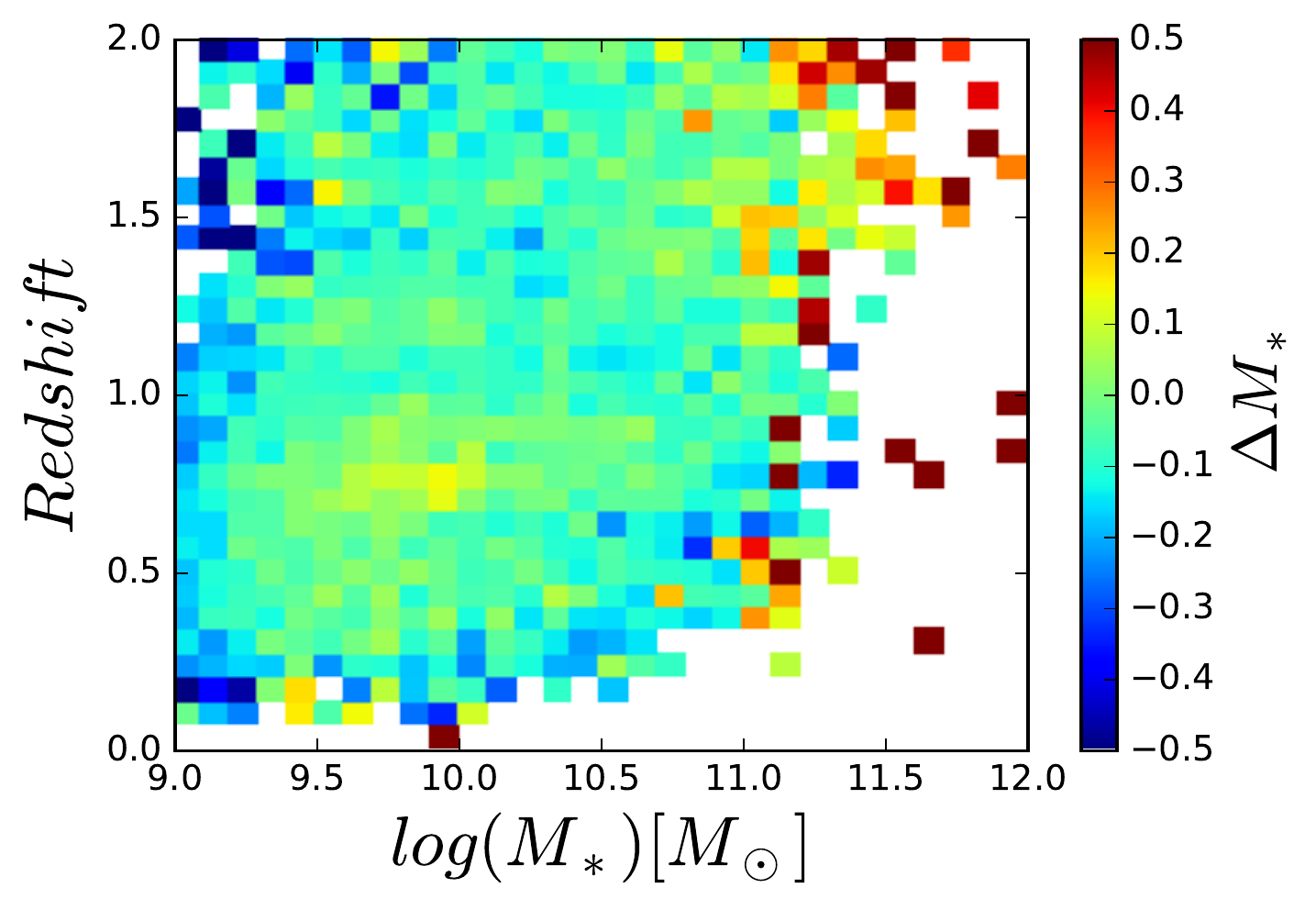} &
 \includegraphics[width=0.44\textwidth]{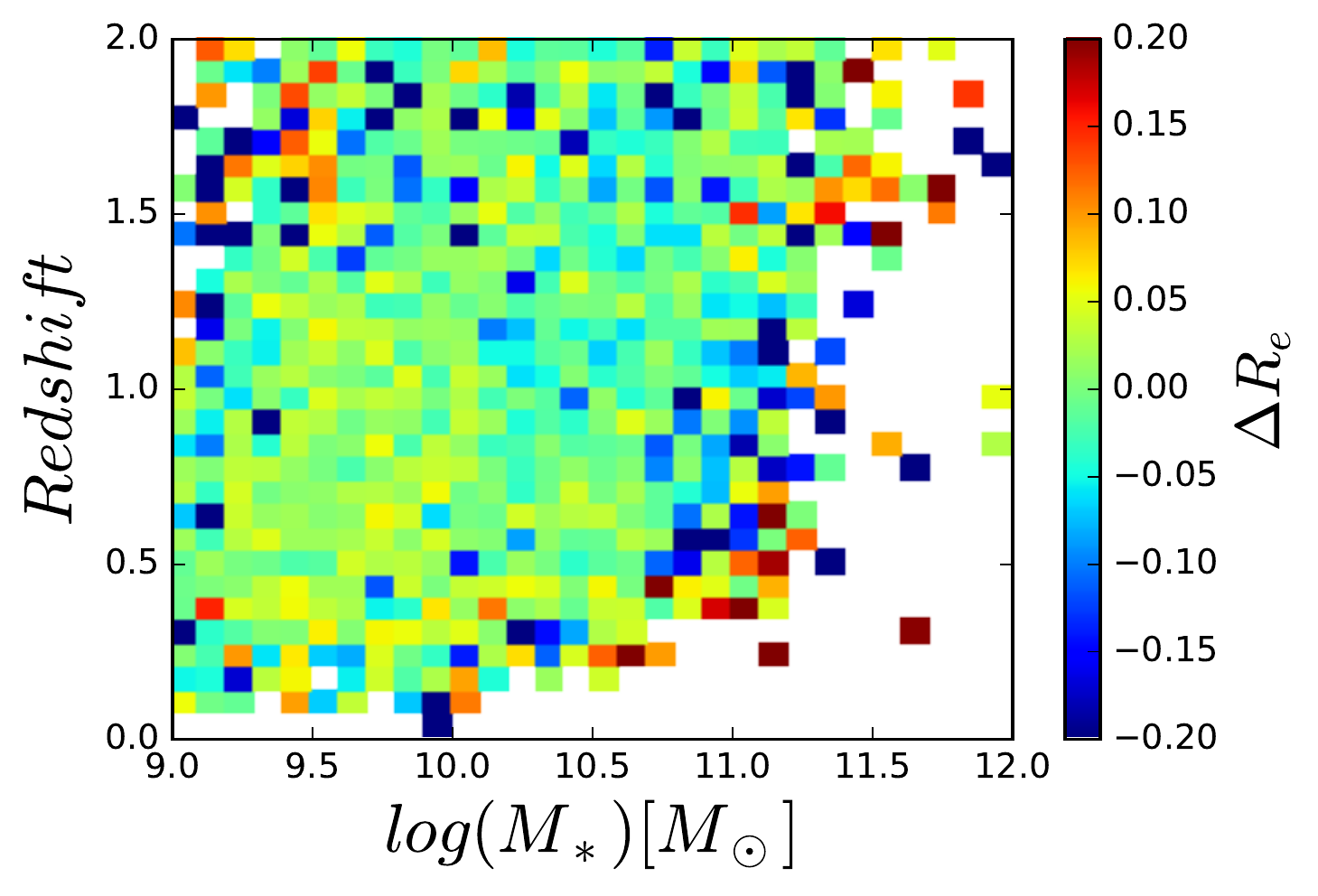} \\
\includegraphics[width=0.44\textwidth]{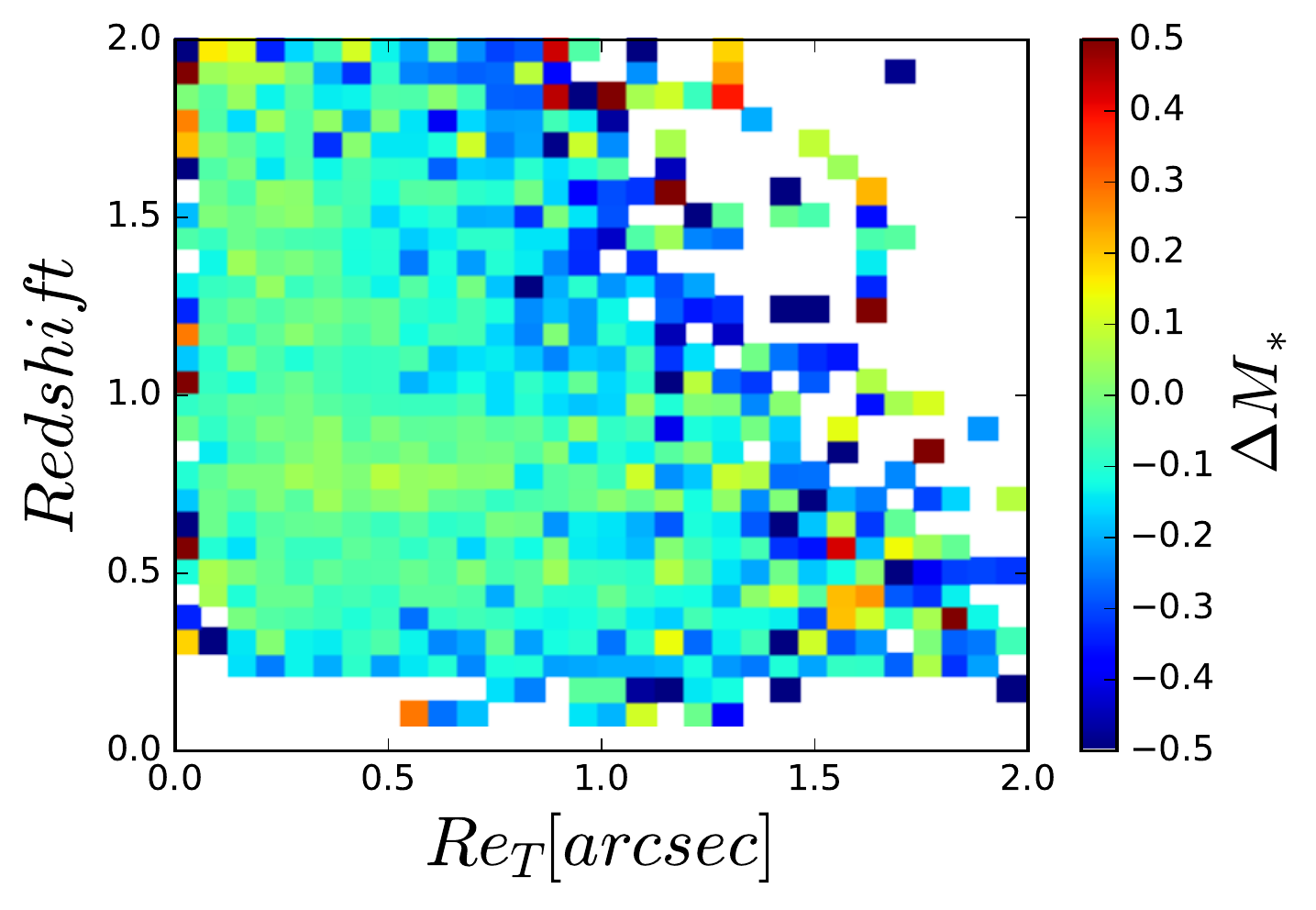} & 
\includegraphics[width=0.44\textwidth]{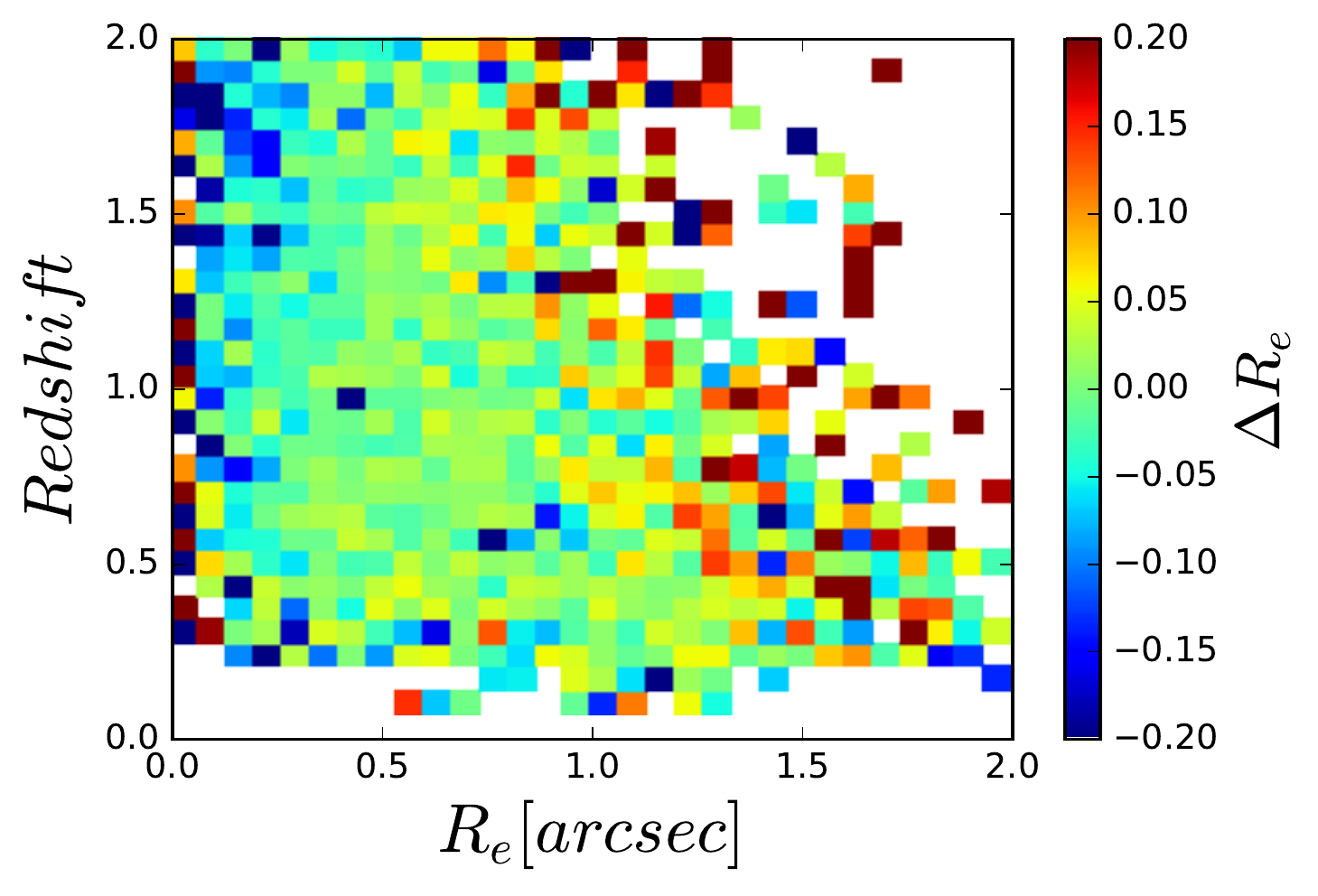} \\
\end{array}$
\caption{Colormaps that shown the distribution of $\Delta M_* = M_* - M_{VdW}$ and $\Delta R_e = R_e - R_{e,VdW}$ in the redshift - $M_* (R_e)$ plane. No clear gradient is observed, discarding the possibility of a bias in the measure. }
 \label{fig:dist_size_mass}
\end{center}
\end{figure*}

Figure \ref{fig:mass_size_SF_Q} compares the mass-size fits done from \cite{Dimauro2018} catalog and the best fits from \cite{vanderWel2014}. The method used in both cases similar hence similar results are expected. The reason of this disagreement can be related to a combination of many factors. It might be a consequence of the different approach used to estimate sizes. Just to recall, in this work a multi wavelength analysis is used while \cite{vanderWel2014} modeled profiles independently in each band. However, in \cite{Dimauro2018} is it shown that the two measurements are in agreement with no bias and a scatter of the order of $\sim10\%$. Figure \ref{fig:dist_size_mass} shown the distribution of the difference in mass and in size between the two catalogs in the mass(size) - redshift plane. We do not observe clear gradient with the redshift as well as galaxy properties like the mass and the size. This result discards the possibility of a bias in the final model.  

A second possible reason is related on how errors are computed. In this work they are estimated by comparing the results from two different settings while \cite{vanderWel2014} uses two identical runs of Galfit on the same object but with different S/N data.
Finally the difference in completeness between the two samples can also affect the results. In the present work a selection was done for $m_{F160}$ < 23 while in the van der Wel et al. [2014] the magnitude limit is fixed at $m_{F160}$ < 24.5. Consequently the diversity in the selection can affect the distribution of masses and sizes which can eventually explain the difference in the fits. 

\begin{figure*}
\begin{center}
$\begin{array}{c c}
\includegraphics[width=0.4\textwidth]{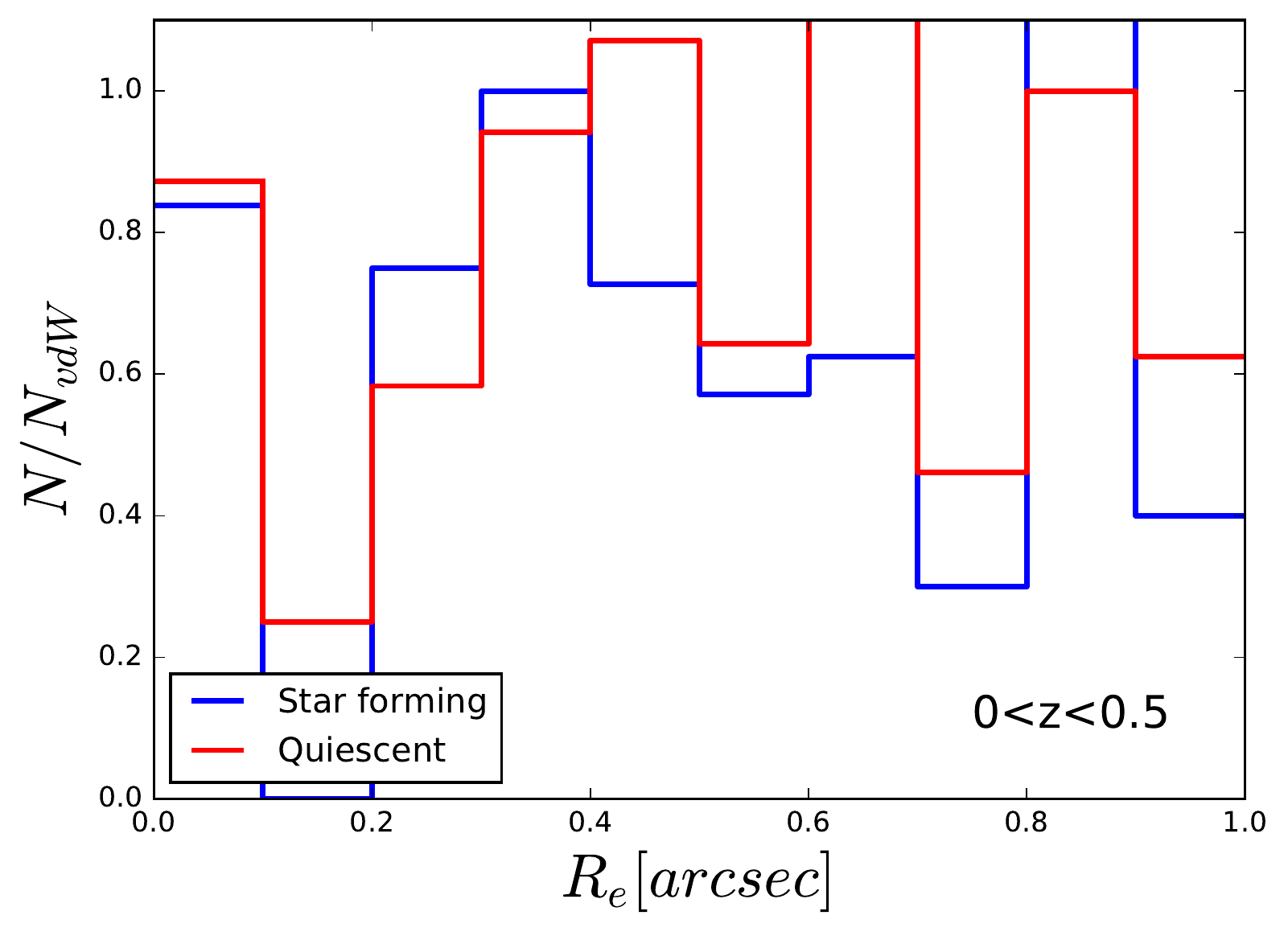} & \includegraphics[width=0.4\textwidth]{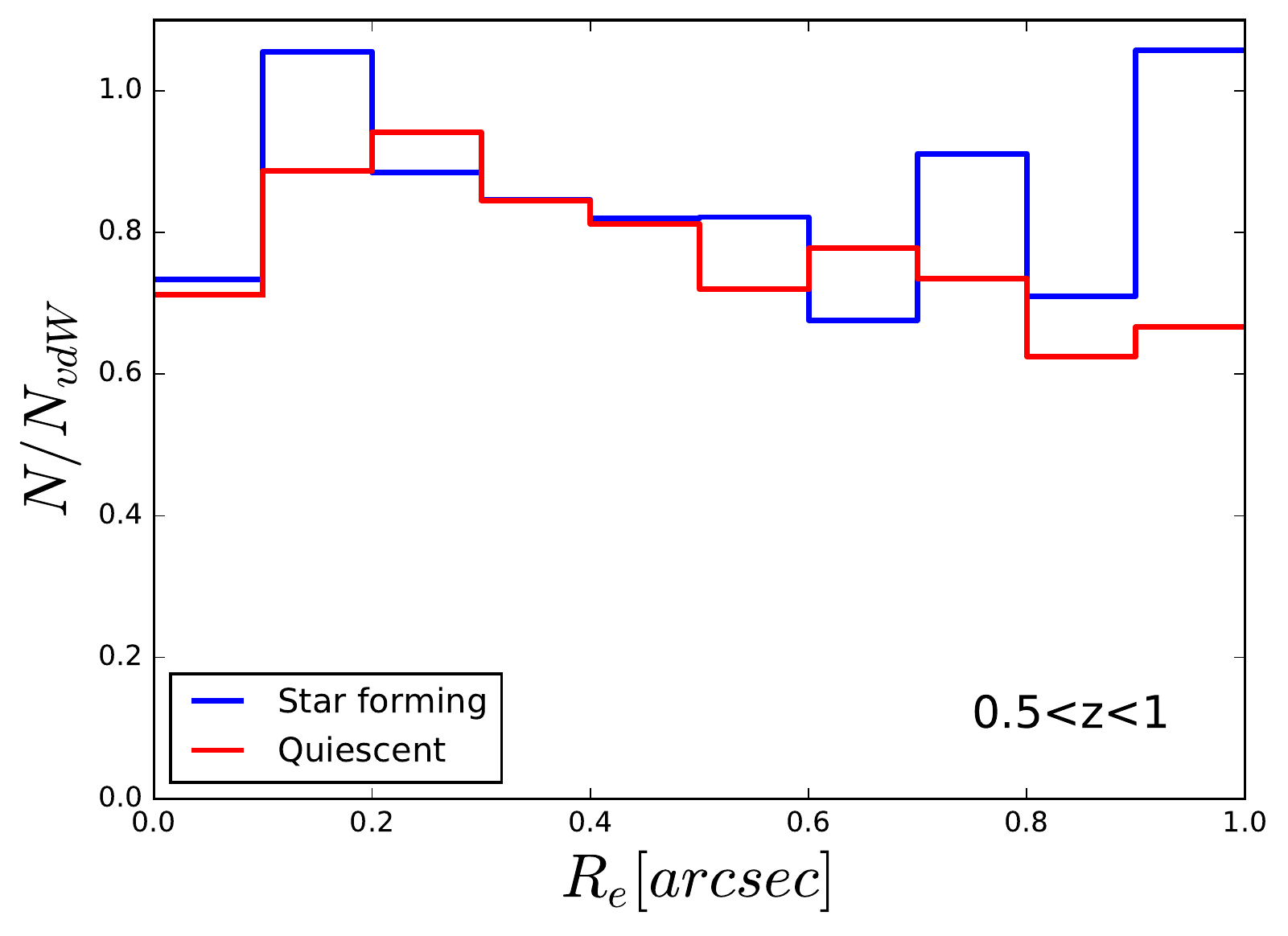} \\
\includegraphics[width=0.4\textwidth]{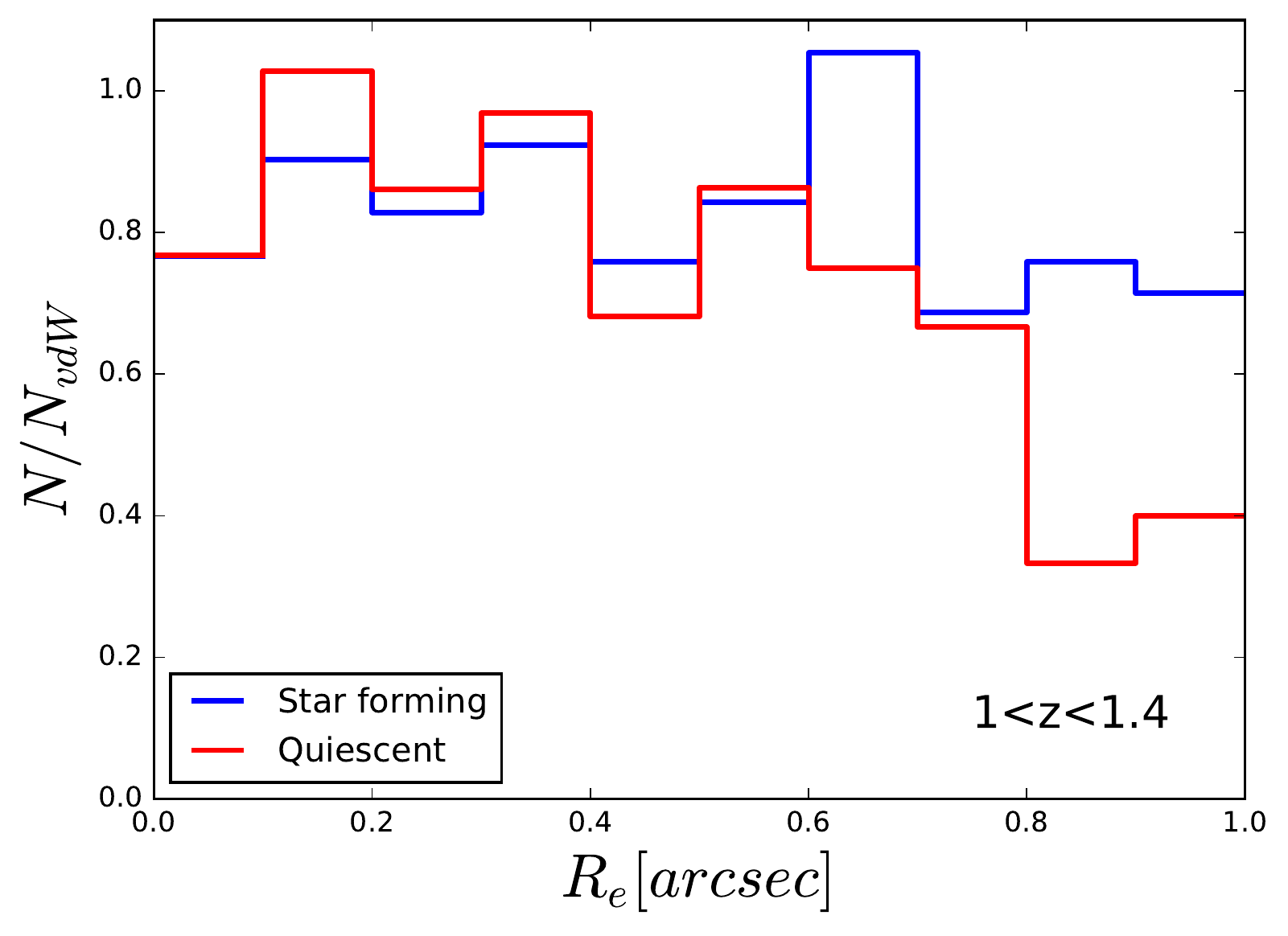} & \includegraphics[width=0.4\textwidth]{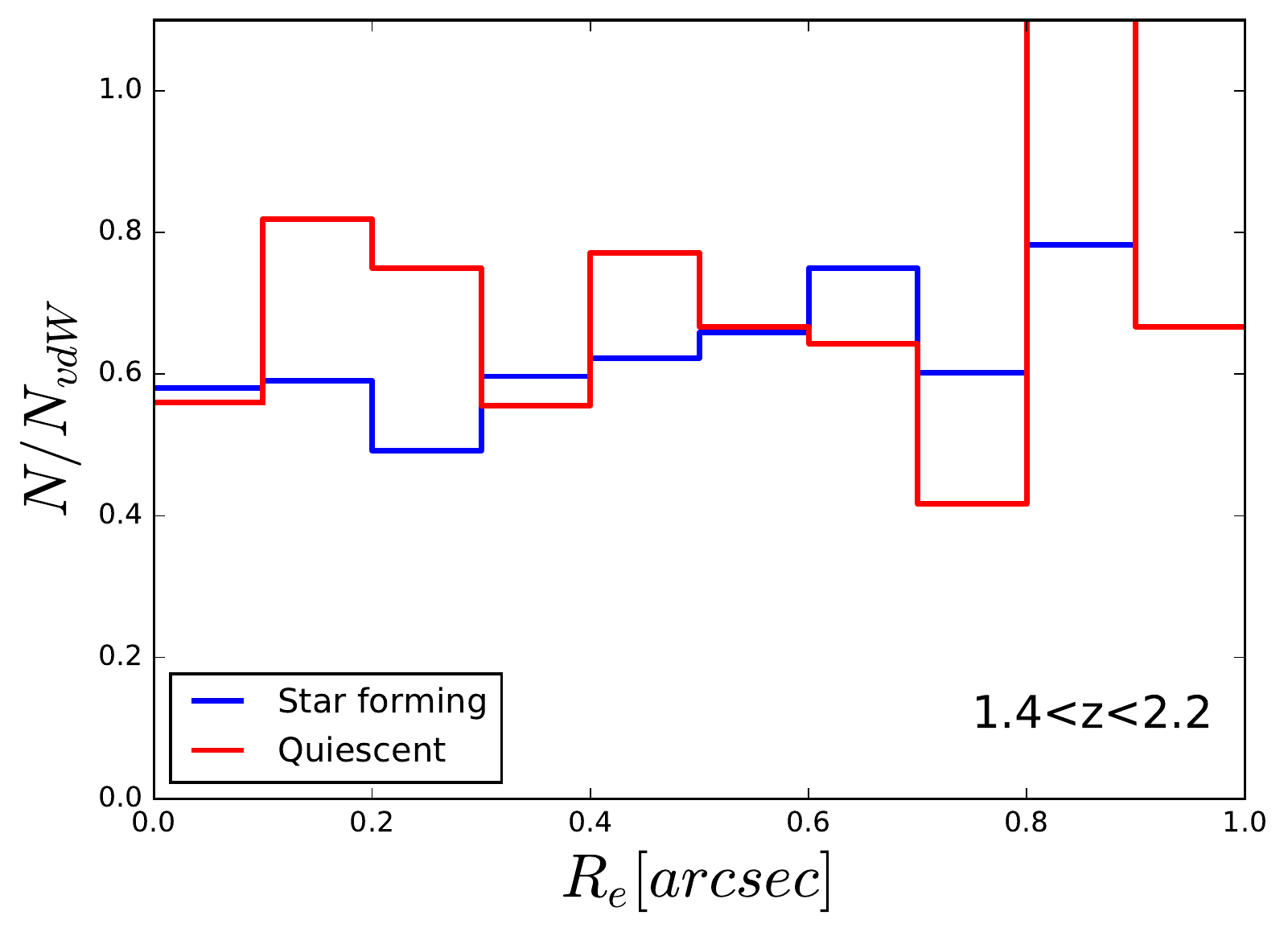} \\
\end{array}$
\caption{Comparison of sizes distribution between the catalog presented in this work and the one from van der Wel et al., 2014. In the y-axis is reported the fraction of galaxies for a given size bin.}
 \label{fig:comp_size_VdW}
\end{center}
\end{figure*}

In order to test this effect, figure \ref{fig:comp_size_VdW} shows the ratio between the number of galaxies in our catalog and \cite{vanderWel2014} for a given size range. As expected, a fraction of galaxies is lost in each bin. However no significant trend with size is observed. No evidence of a strong bias. Although for the quiescent populations there is a drop at larger radii that increases at z > 1 that could explain the deviation from the best fit in this range of z.

\section{Sizes comparison between passive galaxies and bulges}
\label{sec:appendix3}
\begin{figure*}
\begin{center}
$\begin{array}{c c}
\includegraphics[width=0.45\textwidth]{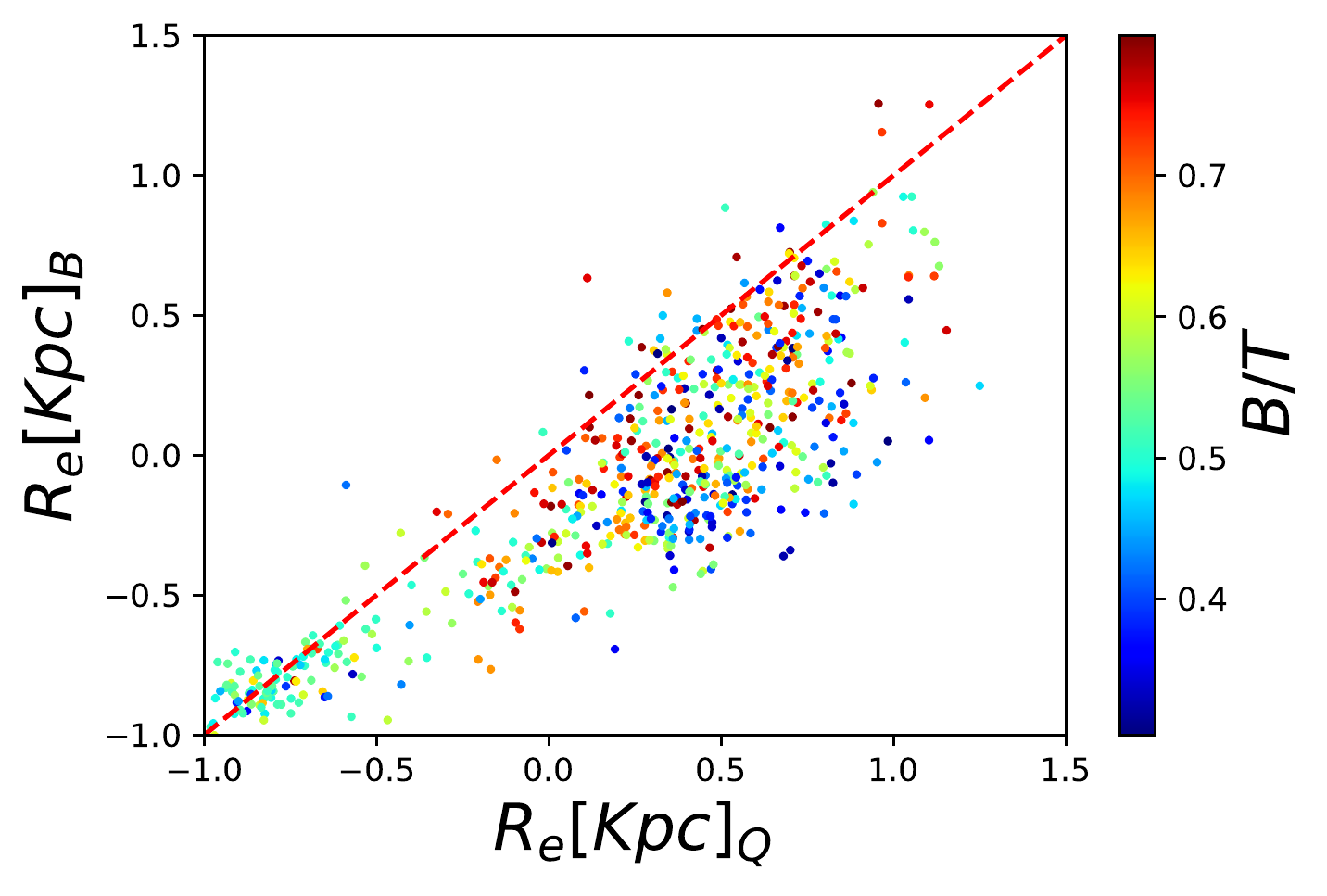} & 
\includegraphics[width=0.45\textwidth]{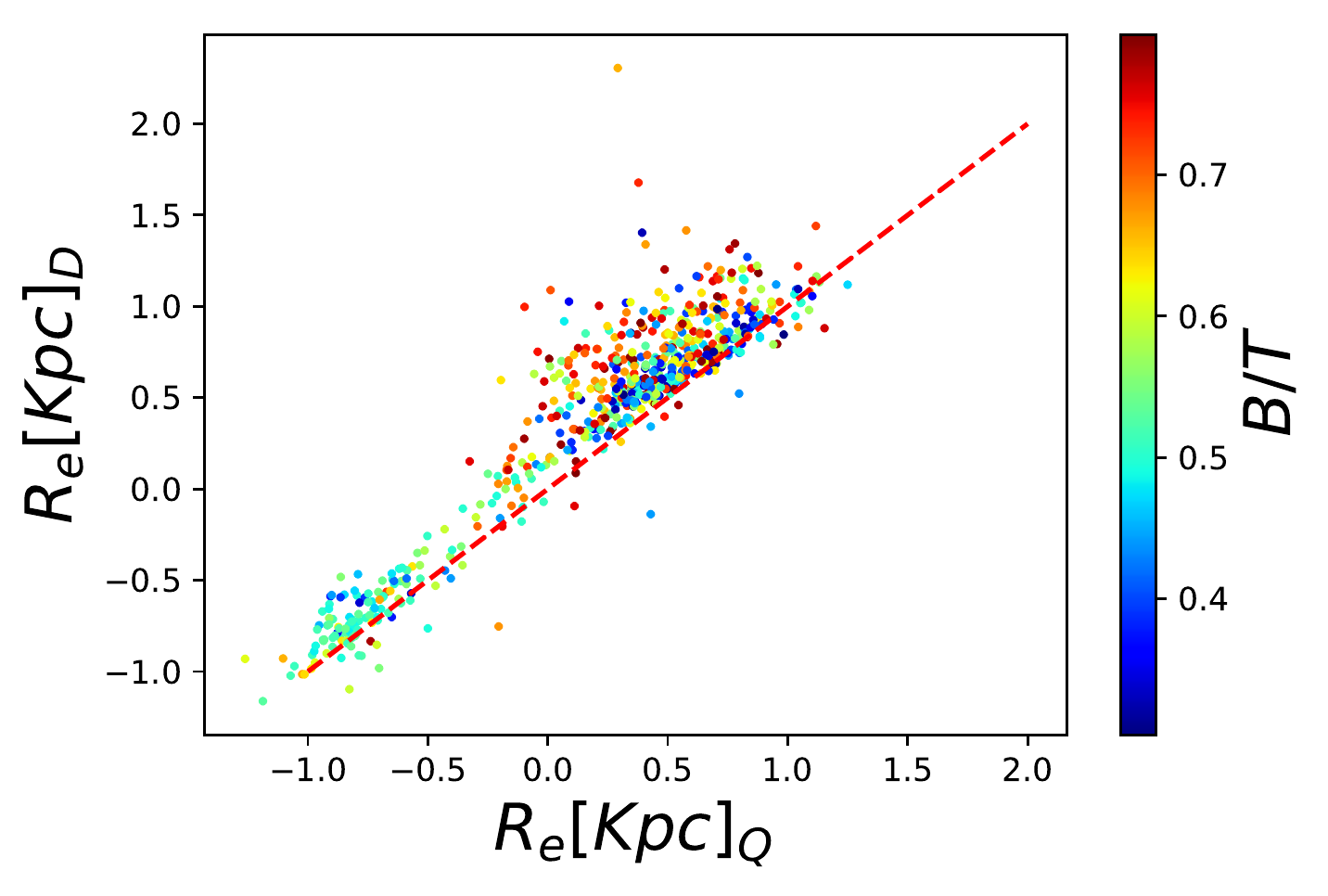} \\
\end{array}$
\caption{Comparison between total and bulges sizes for quiescent galaxies with $0.2\leq B/T\leq0.8$. The color code is showing $B/T$. } 
\label{fig:mass_size_Q_reB}
\end{center}
\end{figure*}
The mass-size relation of bulges is flatter then the one of passive galaxies. Possible reason resides in the measure of the half light radii. The size estimation from the 1-comp \Sersic model accounts also of the presence of the disc. Indeed, while the two sizes (total and bulges sizes), are similar for bulge-dominated galaxies, they are quite different when an extended disc is present. 
In order to check this effect we focus the following analysis only on galaxies with $0.2\leq B/T\leq0.8$. 
Figure \ref{fig:mass_size_Q_reB} shows the comparison between the total half light radii and the size of bulges and discs in the left and the right panel respectively. The color code is done accounting of the $B/T$. It can be seen that the scatter increases, moving from compact galaxies and galaxies hosting a relevant bulge, to galaxies with a different range of $B/T$. Such subsample of galaxies that in this plot show a large scatter, are the ones that in the massive end of the mass-size plane will move to lower sizes affecting the slope of the best fit.

\section{Medium values}
\label{sec:appendix4}

\begin{table*}
\tiny
\centering
\begin{tabular}{ccccccccccccc}
\toprule
\toprule
   \multicolumn{7}{c}{BULGE } &  \multicolumn{6}{c}{DISC }\\
    \cline{1-13}
    \multicolumn{1}{c|}{z}  &\multicolumn{1}{c}{ } & \multicolumn{5}{c}{  $M_{*}$    }& \multicolumn{1}{c}{ } & \multicolumn{5}{c}{  $M_{*}$   }  \\
    \cline{3-7}
    \cline{9-13}
   & &  10.35 &10.65 &10.95 &11.25 &11.55 & &10.35 &10.65 &10.95 &11.25 &11.55\\
   \cline{1-7}
   \cline{9-13}
   0-0.5 & SF & 0.364  & 0.485  & 0.718 & - & - &SF &  0.860  & -  & - & - & -  \\ 
	  & Q& 0.385 & 0.341  & 0.541 & - & -&Q& 0.864 & -  & - & - & - \\  
     \cline{1-13}
   0.5-1.0  &SF & 0.402  & 0.470  & 0.545 & 0.622 & -&SF & 0.734  & 0.809  & 0.849 & - & -  \\  
	  &Q & 0.155 & 0.374  & 0.511 & 0.538 & - &Q & 0.721 & 0.765  & 0.765 & - & -\\  
	  \cline{1-13}
1.0-1.5  &SF & 0.191  & 0.329  & 0.417 & 0.326 & - &SF & 0.687  & 0.726  & 0.759 & 0.790 & -\\   
	  &Q & 0.109 & 0.184  & 0.302 & 0.583 & - &Q & 0.650 & 0.660  & 0.974 & - & -\\    
  \cline{1-13}
1.5-2.0  &SF & 0.178  & 0.238  & 0.129 &  & - &SF & 0.659  & 0.687  & 0.701 & 0.731 & - \\ 
	 &Q & 0.014 & 0.089  & 0.185 & 0.215 & - &Q & 0.403 & 0.724  & 0.540 & 0.689 & - \\  
\bottomrule
\bottomrule
\end{tabular}
\caption{Median sizes values [kpc] for bulges/discs hosted in star forming(SF) or quenched galaxies(Q) in mass bin as shown in figures \ref{fig:mass_size_B_SQ}, \ref{fig:mass_size_D_SQ} }
\label{tbl:tb_size_bulge_sf}
\end{table*}

\begin{table*}
\tiny
\centering
\begin{tabular}{cccccccccccccc}
\toprule
\toprule
   \multicolumn{7}{c}{BULGE } &  \multicolumn{7}{c}{DISC }\\
    \cline{1-7}
     \cline{9-14}
    \multicolumn{1}{c|}{z}  &\multicolumn{1}{c}{BT} & \multicolumn{5}{c}{  $M_{*}$ } &\multicolumn{2}{c}{BT} & \multicolumn{5}{c}{  $M_{*}$ } \\
     \cline{3-7}
    \cline{10-14}
   & &  10.3 &10.5 & 10.7 & 10.9 & 11.1 & && 10.3 &10.5 & 10.7 & 10.9 & 11.1\\
   \cline{1-7}
    \cline{9-14}
   0-0.5 & 0.2-0.5 & -  & -  & - & - & -&&0.0-0.2 & 0.899 & -  & - & - & - \\
	  & 0.5-0.8& 0.167 & 0.357  & - & - & -& &0.2-0.5 & 0.789 & -  & - & - & - \\
	  &0.8-1.0 & 0.390 & 0.447  & 0.677 & - && - &0.5-0.8 & 0.917 & -  & - & - & - \\
     \cline{1-14}
   0.5-1.0  &0.2-0.5 & 0.247  & 0.421  & - & - & - & &0.0-0.2 & 0.695  & 0.763  & 0.819 & - & - \\
	  &0.5-0.8 & 0.204 & 0.356  & 0.512 & - & - &&0.2-0.5 & 0.744 & 0.832  & 0.908 & - & -  \\
	  &0.8-1.0 & 0.241 & 0.409  & 0.510 & 0.592 && - & 0.5-0.8 & 0.800 & 0.930  & - & - & -   \\
	  \cline{1-14}
1.0-1.5  &0.2-0.5 & 139  & 0.132  & - & - & - & &0.0-0.2 & 0.681  & 0.713  & 0.765 & 0.637 & -\\
	  &0.5-0.8 & 0.125 & 0.234  & 0.207 & - & - & &0.2-0.5 & 0.674 & 0.708  & 0.783 & - & - \\
	  &0.8-1.0 & 0.161 & 0.245  & 0.365 & 0.398 & -& &0.5-0.8 & 0.732& 0.809  & 0.987 & - & -  \\
  \cline{1-7}
1.5-2.0  &0.2-0.5 & 0.03  & 0.161  & 0.061 & - & - &&0.0-0.2 & 0.665  & 0.687  & 0.687 & 0.697 & - \\
	 &0.5-0.8 & 0.11 & 0.104  & 0.124 & 0.191 & - &&0.2-0.5 & 0.518 & 0.681  & 0.638 & - & -\\
	 &0.8-1.0 & 0.178 & 0.177  & 0.176 & 0.296 & 0.464 & &0.5-0.8 & 0.679 & 0.768  & 0.739 & - & -  \\
\bottomrule
\bottomrule
\end{tabular}
\caption{Median sizes values [kpc] of bulges and discs in bins of mass bin and B/T as shown in figures \ref{fig:mass_size_B_BT},\ref{fig:mass_size_D_BT}}
\label{tbl:med_tb_size_bulge_bt}
\end{table*}

\begin{table*}
\tiny
\begin{tabular}{c|c|ccccc}
\toprule
\toprule
    \multicolumn{1}{c|}{z}  &\multicolumn{1}{c}{ } & \multicolumn{5}{c}{  $M_{*}$    } \\
        \cline{3-7}
   & &  10.3 &10.5 & 10.7 & 10.9 & 11.1\\
\hline
0-0.5  &B & 0.253  & 0.263  & 0.425 & 0.511 & 0.55 \\
	  &D & 0.732 & 0.762  & 0.826 & 0.840 & 0.808 \\
\hline	  
0.5-1.0 &B & 0.253  & 0.263  & 0.425 & 0.511 & 0.55 \\
	   &D & 0.732 & 0.762  & 0.826 & 0.840 & 0.808 \\
\hline
1.0-1.5 &B & 0.170 & 0.169 & 0.232 & 0.304 & 0.372 \\
	   &D & 0.660 & 0.725 & 0.722 & 0.780 & 0.783 \\
\hline
1.5-2.0 &B& 0.213 & 0.122  & 0.161 & 0.155 & 0.234  \\
            &D& 0.648 &  0.681 & 0.696 & 0.697 & 0.654 \\
\bottomrule
\bottomrule
\end{tabular}
\caption{Median sizes values [kpc] of bulges and discs in mass bin as shown in figure \ref{fig:mass_size_B_D} }
\label{tbl:tb_size_bulge_disk_sf}
\end{table*}

\bsp	
\label{lastpage}
\end{document}